\documentstyle[12pt,axodraw,epsf,amssymb]{article}
\begin{document}
\baselineskip 0.6cm
\newcommand{\gsim}{ \mathop{}_{\textstyle \sim}^{\textstyle >} }
\newcommand{\lsim}{ \mathop{}_{\textstyle \sim}^{\textstyle <} }
\newcommand{\vev}[1]{ \left\langle {#1} \right\rangle }
\newcommand{\bra}[1]{ \langle {#1} | }
\newcommand{\ket}[1]{ | {#1} \rangle }
\newcommand{\Dsl}{\mbox{\ooalign{\hfil/\hfil\crcr$D$}}}
\newcommand{\EV}{ {\rm eV} }
\newcommand{\KEV}{ {\rm keV} }
\newcommand{\MEV}{ {\rm MeV} }
\newcommand{\GEV}{ {\rm GeV} }
\newcommand{\TEV}{ {\rm TeV} }
\def\diag{\mathop{\rm diag}\nolimits}
\def\Spin{\mathop{\rm Spin}}
\def\SO{\mathop{\rm SO}}
\def\O{\mathop{\rm O}}
\def\SU{\mathop{\rm SU}}
\def\U{\mathop{\rm U}}
\def\Sp{\mathop{\rm Sp}}
\def\SL{\mathop{\rm SL}}
\def\tr{\mathop{\rm tr}}

%%%%%%%%%%
%%%%%%%%%%      title page
%%%%%%%%%%

\begin{titlepage}

\begin{flushright}
CERN-TH/2002-185\\
UT-02-44
\end{flushright}

\vskip 2cm
\begin{center}
{\large \bf  A Solution to the Doublet-Triplet Splitting
 Problem \\ in the Type IIB Supergravity}

\vskip 1.2cm
T.~Watari$^{a,b}$ and T.~Yanagida$^{a,b,}$\footnote{
On leave from University of Tokyo, Japan}

\vskip 0.4cm
$^{a}$ {\it Theory Division, CERN, CH-1211 Geneva 23, Switzerland} \\
$^{b}$ {\it Department of Physics, University of Tokyo, \\
          Tokyo 113-0033, Japan}\\

\vskip 1.5cm
\abstract{The doublet--triplet mass splitting problem 
is one of the most serious problems in supersymmetric 
grand unified theories (GUTs).
A class of models based on a product gauge group, 
such as the SU(5)$_{\rm GUT} \times$U(3)$_{\rm H}$ or 
the SU(5)$_{\rm GUT} \times$U(2)$_{\rm H}$,
realize naturally the desired mass splitting 
that is protected by an unbroken R symmetry.
It has been pointed out that various features in the models suggest 
that these product-group unification models are embedded 
in a supersymmetric brane world.
We show an explicit construction of those models 
in the supersymmetric brane world based on the Type IIB supergravity 
in ten dimensions. 
We consider ${\bf T}^6/({\bf Z}_{12}\times{\bf Z}_2)$ orientifold for 
the compactified six extra dimensions.
We find that all of the particles needed for the GUT-symmetry-breaking
sector are obtained from the D-brane fluctuations. 
The three families of quarks and leptons are introduced at an orbifold
singularity, although their origin remains unexplained. 
This paper includes extensive discussion on anomaly cancellation 
in a given orbifold geometry. 
Relation to the Type IIB string theory, realization
 of R symmetry as a rotation of extra-dimensional space, 
and effective superpotential at low energies are also discussed.}

\end{center}
\end{titlepage}

%%%%
%%%%\include{introduction}
%%%%
%%%%%%%%%%%%%%%%%%%%%%%%%%%%%%%%%%%%%%%%%%%%%%%%%%%%%%%%%%%%%%%%%%%%%%
\section{Introduction}
%%%%%%%%%%%%%%%%%%%%%%%%%%%%%%%%%%%%%%%%%%%%%%%%%%%%%%%%%%%%%%%%%%%%%%
Supersymmetric (SUSY) Grand Unified Theories (GUTs) have attracted many
people for a long time because of a number of theoretical beauties \cite{GG}. 
They have become even more attractive recently 
since the precise measurements of the standard-model gauge coupling
constants support their SUSY SU(5) unification \cite{DRW,unify}. 

However, there are serious problems in the SUSY SU(5) models.
The most severe problem is to provide coloured-triplet 
Higgs multiplets with masses of the order of the GUT scale, 
$\sim 10^{16}$ GeV, keeping two Higgs doublets almost massless.
Another problem is the absence of the dimension-5 proton decay
\cite{dim5,dim5-phen}. 
This problem is closely related to the previous problem,
because both indicate a particular structure of 
the mass matrix of the coloured Higgs multiplets and 
a symmetry behind it.

A class of models of the SUSY GUTs
in Refs. \cite{Yanagida-ss,HIY-ss,HY-ss,IY-ss} is 
one of the solutiions to these problems.
A discrete R symmetry plays a crucial role there \cite{IY-ss}.
This R symmetry forbids both  large masses of the Higgs doublets and 
dimension-5 operators for proton decay, simultaneously.
Coloured Higgs multiplets are provided with mass terms 
through a kind of missing-partner mechanism \cite{mpm}.
Their mass partners (triplets without doublets)
emerge as composite states, at the price of introducing a new gauge group
above the GUT scale.
Thus, we call it the product-group unification\footnote{
Various unification models based on the SU(5)$\times$SU(5) gauge group have
been proposed to explain naturally the doublet--triplet mass splitting
\cite{BDS,IY-55,WittenG2}.
However, in this paper, we mean by the ``product-group unification'' 
the class of models developed in Refs. \cite{Yanagida-ss,HIY-ss,HY-ss,IY-ss}, 
which was referred to as ``semi-simple unification'' 
in \cite{IWY-ss,WY-ss,FW-ss,FW-ss2}.}.

This class of models has to give up the ``unification'' by a simple
group. 
The gauge coupling constants are not universal either; 
the SU(5) gauge group has weak coupling, while the newly introduced
group(s) has(have) to have relatively large coupling(s). 
However, we do not consider these features as ugly, because 
they are quite naturally explained along with a number of other
features of the models, 
if the models are embedded in a SUSY brane world 
\cite{IWY-ss,WY-ss}\footnote{A similar model has been proposed 
in \cite{Csaki}.}.
Here, the extra dimensions are assumed to be smaller than the inverse 
of the GUT scale.
Qualitative arguments in Ref. \cite{IWY-ss} show that 
the SUSY brane-world structure behind the models 
is quite a natural possibility. 
The present authors briefly show in the previous letter \cite{WY-ss} 
an explicit construction of the brane-world model by adopting the D3-D7
system of the ten-dimensional supergravity.
This article provides an extensive and more detailed construction.
Theoretical consistency and relation to string theories 
are also discussed.  

This article is organized as follows.
The original product-group unification in the four-dimensional space-time
is briefly  reviewed in section \ref{sec:4D}.
Motivation for extending the original models to 
the SUSY brane world are summarized in section \ref{sec:IIB}.
Section \ref{sec:principle} explains our principle of the brane-world 
model construction in the Type IIB supergravity. 
The geometry of the particular orbifold compactification we adopt 
is explained in section \ref{sec:geometry}. 
Sections \ref{sec:u2} and \ref{sec:u3} are devoted 
to explicit constructions of two different models 
of the product-group unification. 
We describe the D-brane configurations on the orbifold and orbifold
projection conditions. 
The whole sector relevant to the SU(5)-symmetry breaking is
perfectly obtained from massless modes on D-branes, although 
we cannot find the origin of quarks and leptons.
Anomaly cancellation on the orbifold is discussed for both models.
Relation to the Type IIB string theory is also briefly mentioned in
subsubsection \ref{subsubsec:u3-anomaly-rel}.
Particles that have a definite origin in extra-dimensional space,
i.e. particles obtained from massless modes on D-branes and
from the supergravity multiplet in the bulk, cannot have arbitrary 
R charges.
We show that the R-charge assignment in the extra-dimensional
construction of the models can coincide with the one required 
for successful phenomenology.
Effective superpotentials below the Kaluza--Klein scale 
are discussed at the ends of both sections.
Section \ref{sec:fine} provides a summary of this paper 
and a brief discussion of phenomenological consequences 
of the present models.

%%%%
%%%%\include{review}
%%%%
%%%%%%%%%%%%%%%%%%%%%%%%%%%%%%%%%%%%%%%%%%%%%%%%%%%%%%%%%%%%%%%%%%%%%%
\section{Product-Group Unification Models  \\ in Four Dimensions}
\label{sec:4D}
%%%%%%%%%%%%%%%%%%%%%%%%%%%%%%%%%%%%%%%%%%%%%%%%%%%%%%%%%%%%%%%%%%%%%%

The symmetry that governs the product-group unification models
is a discrete R symmetry: (mod 4)-R symmetry \cite{IY-ss}.
The R charges of all the fields in the minimal SUSY standard model 
(MSSM) are given in Table \ref{tab:mssm-R4}. 
Higgsino does not have a mass of the order of the fundamental scale 
because of this symmetry, and the Giudice--Masiero mechanism \cite{GM}
provides the SUSY-invariant mass term ($\mu$-term) for Higgs multiplets 
of the order of the electroweak (TeV) scale after 
the (mod 4)-R symmetry is broken by the non-vanishing vacuum value of
the superpotential. 
The (mod 4)-R is actually the unique symmetry in the MSSM compatible
with the SU(5)$_{\rm GUT}$ that satisfies the above two properties and 
that may have \cite{KMY-da} a vanishing mixed anomaly \cite{I-da,BD-da} 
with the SU(5)$_{\rm GUT}$ gauge group. 

An immediate consequence of this symmetry is the absence 
of the dimension-5 proton decay. 
At the same time, this symmetry implies that additional 
SU(3)$_C$-triplets (without SU(2)$_L$-doublets) should be  
introduced as mass partners of the coloured Higgs multiplets.
They should have R charge 2 rather than 0.
The other possibility\footnote{The only way to avoid providing only
triplets or only doublets in SU(5) unified theories is to introduce
two sets of infinite number of the SU(5)-({\bf 5}+{\bf 5}$^*$), where
one set has R charge 0 and the other has R charge 2.
This is what is done in \cite{Witten,HN}.} 
is that the SU(5)$_{\rm GUT}$ covariant fields, $H^i({\bf 5})$ and 
$\bar{H}_i({\bf 5}^*)$, contain only doublets as one-particle
degrees of freedom without triplets in the SU(5)$_{\rm GUT}$-breaking phase.
The product-group unification we discuss in this paper
is a framework \cite{Yanagida-ss} that provides explicit models 
for the above two possibilities.

%%%%%%%%%%%%%
%% U(3) model
%%%%%%%%%%%%
Let us first explain a model based on a product gauge group 
SU(5)$_{\rm GUT} \times$U(3)$_{\rm H}$ \cite{IY-ss}.
Quarks and leptons are singlets of the U(3)$_{\rm H}$ gauge group 
and form three families of {\bf 5}$^*$+{\bf 10} of the SU(5)$_{\rm GUT}$.
Higgs multiplets that contain two Higgs doublets are $H({\bf 5})^i$ 
and $\bar{H}({\bf 5}^*)_i$, which are also singlets of the U(3)$_{\rm
H}$.
Fields introduced for the SU(5)$_{\rm GUT}$ breaking are given as follows:
$X^{\alpha}_{\;\;\beta}(\alpha ,\beta =1,2,3)$ transforming 
as (${\bf 1}$,{\bf adj.}=${\bf 8}+{\bf 1}$) under 
the $\SU(5)_{\rm GUT}\times \U(3)_{\rm H}$ gauge 
group, and $Q^{\alpha}_{\;i}(i=1,...,5)+Q^{\alpha}_{\;6}$
and $\bar{Q}^i_{\;\alpha}(i=1,...,5)+\bar{Q}^6_{\;\alpha}$
transforming as ({\bf 5}$^*$+{\bf 1},{\bf 3}) and 
({\bf 5}+{\bf 1},{\bf 3}$^*$).
Indices  $i$ are for the $\SU(5)_{\rm GUT}$ and $\alpha$ or $\beta$ 
for the $\U(3)_{\rm H}$.
The chiral superfield $X^\alpha_{\;\;\beta}$ is also written as 
$X^c (t_c)^\alpha_{\;\;\beta}(c=0,1,...,8)$, where $t_a(a=1,...,8)$ are 
Gell-Mann matrices of the SU(3)$_{\rm H}$ gauge group\footnote{
The normalization condition $\tr(t_a t_b) = \delta_{ab}/2$ is understood. 
Note that the normalization of the following $t_0$ is determined 
such a way that it also satisfies $\tr(t_0t_0)=1/2$.} and 
$t_0 \equiv {\bf 1}_{3 \times 3}/\sqrt{6}$,
where U(3)$_{\rm H}\simeq $ SU(3)$_{\rm H}\times$U(1)$_{\rm H}$.
The R charges (mod 4) of these fields are given in Table \ref{tab:u3-R4}.
The mixed anomaly (R mod 4)[SU(3)$_{\rm H}$]$^2$ happens to vanish
\cite{KMY-da}.
The most general superpotential is given \cite{IY-ss} by 
\begin{eqnarray}
W &=&\sqrt{2} \lambda_{\rm 3H} \bar{Q}^i_{\;\alpha} X^a(t_a)^{\alpha}_{\;\beta}
                             Q^{\beta}_{\;i} 
+ \sqrt{2} \lambda_{\rm 3H}' \bar{Q}^6_{\;\alpha} X^a(t_a)^{\alpha}_{\;\beta}
                             Q^{\beta}_{\;6}   \nonumber \\
& & \!\!\!\!\! 
+ \sqrt{2} \lambda_{\rm 1H} \bar{Q}^i_{\;\alpha} X^0(t_0)^{\alpha}_{\;\beta}
                             Q^{\beta}_{\;i} 
+ \sqrt{2} \lambda_{\rm 1H}' \bar{Q}^6_{\;\alpha} X^0(t_0)^{\alpha}_{\;\beta}
                             Q^{\beta}_{\;6}  \nonumber \\
& &  \!\!\!\!\!
- \sqrt{2}\lambda_{\rm 1H}  v^2 X^\alpha_{\;\alpha}    \label{eq:super}  \\
  & & \!\!\!\!\! + h' \bar{H}_i \bar{Q}^i_{\;\alpha} Q^{\alpha}_{\;6} 
      + h \bar{Q}^6_{\;\alpha} Q^{\alpha}_{\;i}H^i   \nonumber \\
  & & \!\!\!\!\! + y_{\bf 10} {\bf 10} \cdot {\bf 10} \cdot H 
      + y_{\bf 5^*} {\bf 5}^* \cdot {\bf 10} \cdot \bar{H} + \cdots,
    %  + f   {\bf 5}^*_i \frac{\bar{Q}^i_{\;\alpha}Q^{\alpha}_{\;j}}{M_*^2}
    %           {\bf 10}^{jk} \bar{H}_k, 
\nonumber
\end{eqnarray}
where the parameter $v$ is taken to be of the order of the GUT scale, 
$y_{\bf 10}$ and $y_{\bf 5^*}$ are Yukawa coupling constants 
of the quarks and leptons, and $\lambda_{\rm 3H},\lambda_{\rm 3H}',
\lambda_{\rm 1H},\lambda_{\rm 1H}',h'$ and $h$ are dimensionless coupling 
constants. 
Ellipses stand for neutrino mass terms and non-renormalizable terms. 
The fields $Q^{\alpha}_{\;i}$ and $\bar{Q}^i_{\;\alpha}$ in the
bifundamental representations acquire vacuum expectation
values (VEV's), $\vev{Q^\alpha_{\;\;i}}$ = $v \delta^\alpha_{\;\; i}$ and 
$\langle\bar{Q}^i_{\;\;\alpha}\rangle$ = $v \delta^i_{\;\;\alpha}$, 
because of the second and third lines in (\ref{eq:super}). 
Thus, the gauge group $\SU(5)_{\rm GUT}\times \U(3)_{\rm H}$ 
is broken down to that of the standard model.
The mass terms of the coloured Higgs multiplets arise from 
the fourth line in (\ref{eq:super}) in the GUT-breaking vacuum. 
No unwanted particle remains massless after the gauge group is broken
down to that of the standard model.
In other words, the above model is constructed so that 
the U(3)$_{\rm H}$ gauge interactions leave only two composite massless 
fields (moduli), $(\bar{Q}^i_{\;\; \alpha}Q^\alpha_{\;\; 6})$ and 
$(\bar{Q}^6_{\;\; \alpha}Q^\alpha_{\;\; i})$, 
after they are integrated out.
These two composite fields have R charge 2 (see Table \ref{tab:u3-R4}), 
and contain only SU(3)$_C$-triplets (without SU(2)$_L$-doublets) 
as one-particle degrees of freedom.
Thus, they can be the mass partners of the coloured Higgs multiplets.
Therefore, the doublet--triplet mass splitting problem is naturally solved.

Fine structure constants of the $ % \U(3)_{\rm H} \simeq
\SU(3)_{\rm H}\times\U(1)_{\rm H}$ must be larger than that of the
$\SU(5)_{\rm GUT}$. 
This is because the gauge coupling constants $\alpha_C, \alpha_L$ 
and $\alpha_Y$ of the MSSM are given by
\begin{equation}
 \frac{1}{\alpha_C} = \frac{1}{\alpha_{\rm GUT}} + \frac{1}{\alpha_{\rm 3H}},
\end{equation}
\begin{equation}
 \frac{1}{\alpha_L} =  \frac{1}{\alpha_{\rm GUT}}, \quad \quad \quad
\end{equation}
and 
\begin{equation}
\frac{3/5}{\alpha_Y} = \frac{1}{\alpha_{\rm GUT}} + 
                                                 \frac{2/5}{\alpha_{\rm 1H}},
\end{equation} 
at tree level, where $ \alpha_{\rm GUT}, 
\alpha_{3\rm H}$ and $\alpha_{1\rm H}$ are fine structure constants 
of $\SU(5)_{\rm GUT}$, 
$\SU(3)_{\rm H}$ and $\U(1)_{\rm H}$, respectively. 
The values of $1/\alpha_{\rm 3H}$ and $1/\alpha_{\rm 1H} $ 
must be within a few percent of the $1/\alpha_{\rm GUT}$ 
at the GUT scale to reproduce the approximate unification 
of  $\alpha_C, \alpha_L$ and $5 \alpha_Y /3$.

The large coupling constant of the SU(3)$_{\rm H}$ required above, 
however, is not stable in the renormalization group running 
% during the one order of magnitude between the GUT scale and 
% the cut-off scale $M_*$ 
\cite{FW-ss}.
Although its beta function is zero at 1-loop, renormalization 
at higher-loop levels is not negligible; $n$-th loop effects arise with  
$(3\times \alpha_{\rm 3H}/(4\pi))^n \sim 1$. 
At 2-loop level, it is easy to see that the SU(3)$_{\rm H}$ coupling 
becomes infinity immediately above the GUT scale.
In other words, the SU(3)$_{\rm H}$ gauge interactions immediately
become weak below the cut-off scale.
Thus, the approximate SU(5)$_{\rm GUT}$ relation of the MSSM
gauge couplings is not a natural consequence
unless the large coupling constant of the SU(3)$_{\rm H}$ 
is stable against the radiative corrections.

An interesting way to solve this problem is to impose a specific
relation,
\begin{equation}
 \frac{(\lambda_{\rm 3H})^2}{4\pi} \simeq \frac{(\lambda_{\rm 3H}')^2}{4\pi}
 \simeq \alpha_{\rm 3H}. 
\label{eq:N=2relation}
\end{equation}  
This relation is stable under the
renormalization group, because there is a symmetry in the limit of
$g_{\rm GUT},h,h',y_{\bf 10},y_{\bf 5^*} \rightarrow 0$: 
an ${\cal N}$ = 2 SUSY.
The matter contents of the SU(5)$_{\rm GUT}$-breaking sector have 
a multiplet structure of the ${\cal N}$ = 2 SUSY \cite{HY-ss}:
the $\U(3)_{\rm H}$  vector multiplet and the $\U(3)_{\rm H}$-{\bf adj.}
chiral multiplet, $X^\alpha_{\;\;\beta}$, form an ${\cal N}$ = 2 vector 
multiplet, and the vector-like pairs 
$(Q^\alpha_{\; k},\bar{Q}^{k}_{\;\;\alpha})$ ($k=1,...,6$) in this sector 
form ${\cal N}$ = 2 hypermultiplets.  
The superpotential (\ref{eq:super}) 
from the first to the third line exhibits the form of interactions of 
the ${\cal N}$ = 2 SUSY gauge theories \cite{IWY-ss}. 
The approximate ${\cal N}$ = 2 SUSY exists when the ${\cal N}$ = 2 relation
in Eq. (\ref{eq:N=2relation}) is satisfied.
This relation, in turn, is stable because of the symmetry.
Then, the perturbative renormalization to the gauge coupling 
is 1-loop-exact in this ${\cal N}$ = 2 SUSY limit. 
Higher-loop renormalization to the SU(3)$_{\rm 3H}$ gauge coupling 
only appears by involving weak couplings,  
$g_{\rm GUT}, h,h',y_{\bf 10}$ and $y_{\rm 5^*}$, and hence 
the large gauge coupling can be preserved under the renormalization group.
This is the main reason why we impose the approximate ${\cal N}$ = 2 SUSY in
the SU(5)$_{\rm GUT}$-breaking sector.
We also impose 
$(\lambda_{\rm 1H})^2/(4\pi)\simeq (\lambda'_{\rm 1H})^2/(4\pi) 
\simeq \alpha_{\rm 1H}$ so that the approximate ${\cal N}$ = 2 SUSY is
maintained in the full SU(5)$_{\rm GUT}$-breaking sector.

The gauge coupling constant of the U(1)$_{\rm H}$ is asymptoticcally non-free. 
The coupling, which is already strong at the GUT scale, 
becomes infinity below the Planck scale, 
$M_{\rm Pl}\simeq 2.4 \times 10^{18}$ GeV. 
Even the ${\cal N}$ = 2 SUSY does not solve this problem.
Thus, the cut-off scale (in other words, the fundamental scale) $M_*$ 
of this model should lie below the Planck scale. 
On the other hand, the fundamental scale should be higher than the GUT
scale by at least one order of magnitude, so that 
the SU(5)$_{\rm GUT}$-breaking corrections to the gauge coupling
constants, through non-renormalizable interactions such as
\begin{equation}
 W = \tr \left(\left(\frac{1}{g^2} + \frac{\vev{\bar{Q}Q}}{M_*^2}\right)
{\cal W}^{\alpha,SU(5)} {\cal W}_\alpha^{SU(5)} \right),
\label{eq:tree-SU(5)breaking}
\end{equation}
are suppressed below $10^{-2}$.

%%%%%%%%%%%%%%%%%%%%%%%
%%%%   U(2) Model
%%%%%%%%%%%%%%%%%%%%%%%

The other model of the product-group unification is based on 
an SU(5)$_{\rm GUT} \times$U(2)$_{\rm H}$ gauge group,
where U(2)$_{\rm H}\simeq$ SU(2)$_{\rm H}\times$U(1)$_{\rm H}$. 
This model provides two Higgs doublets without triplets
as one-particle degrees of freedom in massless composite fields, 
after the U(2) gauge group is integrated out.
This model realizes the other possibility discussed at the beginning 
of this section.

Matter contents of this model are 
$X^\alpha_{\;\;\beta}$ ({\bf 1},{\bf adj.}={\bf 3}+{\bf 1}), 
$Q^\alpha_{\;\; i}+Q^\alpha_{\;\; 6}$({\bf 5}$^*$+{\bf 1},{\bf 2}) and 
$\bar{Q}^i_\alpha+\bar{Q}_\alpha^6$({\bf 5}+{\bf 1},{\bf 2}$^*$) 
($\alpha,\beta = 4,5$;$i=1,...,5$) 
in addition to the three families of quarks and leptons, 
({\bf 5}$^*$+{\bf 10},{\bf 1}). 
The ordinary Higgs fields $H^i({\bf 5})$ and $\bar{H}_i({\bf 5}^*)$ 
are not introduced.
The R charges of those fields are given in Table \ref{tab:u2-R4}.
Mixed anomaly (R mod 4)[SU(2)$_{\rm H}$]$^2$ happens to vanish again.
The superpotential is given by
\begin{eqnarray}
W &=&\sqrt{2} \lambda_{\rm 2H} \bar{Q}^i_{\;\alpha} X^a(t_a)^{\alpha}_{\;\beta}
                             Q^{\beta}_{\;i} 
+ \sqrt{2} \lambda_{\rm 2H}' \bar{Q}^6_{\;\alpha} X^a(t_a)^{\alpha}_{\;\beta}
                             Q^{\beta}_{\;6}   \nonumber \\
& & \!\!\!\!\! 
+ \sqrt{2} \lambda_{\rm 1H} \bar{Q}^i_{\;\alpha} X^0(t_0)^{\alpha}_{\;\beta}
                             Q^{\beta}_{\;i} 
+ \sqrt{2} \lambda_{\rm 1H}' \bar{Q}^6_{\;\alpha} X^0(t_0)^{\alpha}_{\;\beta}
                             Q^{\beta}_{\;6}  \nonumber \\
& &  \!\!\!\!\!
- \sqrt{2}\lambda_{\rm 1H}  v^2 X^\alpha_{\;\alpha}    \label{eq:super2}  \\
& & \!\!\!\!\! 
+ c_{\bf 10} {\bf 10}^{i_1 i_2} {\bf 10}^{i_3 i_4} 
                   (\bar{Q}Q)^{i_5}_{\;\; 6} 
     + c_{\bf 5^*} (\bar{Q}Q)^6_{\;\; i} \cdot {\bf 10}^{ij}\cdot {\bf 5}^*_j  
      + \cdots,
    %  + f   {\bf 5}^*_i \frac{\bar{Q}^i_{\;\alpha}Q^{\alpha}_{\;j}}{M_*^2}
    %           {\bf 10}^{jk} \bar{H}_k, 
\nonumber
\end{eqnarray}
where $t_a$($a=1,2,3$) is now one half of the Pauli matrices.
The SU(5)$_{\rm GUT} \times$U(2)$_{\rm H}$ symmetry is broken down to
that of the standard model through the expectation values 
$\vev{Q^\alpha_{\;\;i}}$ = $v \delta^\alpha_{\;\; i}$ and 
$\langle\bar{Q}^i_{\;\;\alpha}\rangle$ = $v \delta^i_{\;\; \alpha}$. 
When the U(2)$_{\rm H}$ gauge interactions are integrated out, 
two moduli remain massless in addition to the chiral quarks and
leptons, which are $(\bar{Q}^i_{\;\;\alpha}Q^\alpha_{\;\;6})$ and 
$(\bar{Q}^6_{\;\;\alpha}Q^\alpha_{\;\;i})$.
These two composite fields contain only SU(2)$_L$-doublets as 
one-particle degrees of freedom in the SU(5)$_{\rm GUT}$ breaking phase. 
Thus, they play the role of the two Higgs doublets in the MSSM. 
Their R charges are 0 (see Table \ref{tab:u2-R4}) as required. 
There is no unwanted massless particle in this model either.

The gauge coupling constants of the SU(2)$_{\rm H}\times$U(1)$_{\rm H}$
should also be relatively large, for  the same reason 
as in the SU(5)$_{\rm GUT}\times$U(3)$_{\rm H}$ model.
An ${\cal N}$ = 2 SUSY relation:  
\begin{equation}
 \frac{(\lambda_{\rm 2H})^2}{4\pi} \simeq 
 \frac{(\lambda_{\rm 2H}')^2}{4\pi} \simeq \alpha_{\rm 2H}, \qquad \qquad
 \frac{(\lambda_{\rm 1H})^2}{4\pi} \simeq 
 \frac{(\lambda_{\rm 1H}')^2}{4\pi} \simeq \alpha_{\rm 1H},  
\label{eq:N=2relation2}
\end{equation}
also stabilizes the large coupling constant of the SU(2)$_{\rm H}$
gauge group.
The cut-off scale should lie below the Planck scale, as explained in the
previous model. 

Here, we summarize five remarkable features that are common to the two
models described above.
First of all, the gauge groups of these ``unification theories'' have 
a product-group structure, and 
secondly, the SU(5)$_{\rm GUT}$ gauge coupling constant is small 
while the rest of the gauge couplings are relatively large.  
Third, there is an approximate ${\cal N}$ = 2 SUSY in the
$\SU(5)_{\rm GUT}$-breaking sector. 
The ${\cal N}$ = 2 SUSY is crucial in maintaining 
the approximate SU(5)$_{\rm GUT}$ unification 
of the MSSM gauge couplings at the GUT scale.
It is quite remarkable that the matter contents and interactions 
support the approximate ${\cal N}$ = 2 SUSY.
Fourth, the cut-off scale should lie below the Planck scale 
because of the asymptotically non-free running of the U(1)$_{\rm H}$ gauge
coupling constant.
Finally, the discrete R symmetry that governs these models should be
preserved in an accuracy better than the $10^{-14}$ level 
to keep the two Higgs doublets almost massless.

%%%%%%%%%%%%%%%%%%%%%%%%%%%%%%%%%%%%%%%%%%%%%%%%%%%%%%%%%%%%%%%%%%%%%
\section{Motivations of Product-Group Unification \\ 
in Type IIB Supergravity}
\label{sec:IIB}
%%%%%%%%%%%%%%%%%%%%%%%%%%%%%%%%%%%%%%%%%%%%%%%%%%%%%%%%%%%%%%%%%%%%%

\subsection{Product-Group Unification in Supersymmetric Brane World}
\label{subsec:IIB-braneworld}
The five features listed at the end of the previous section 
are understood quite naturally when we embed the models 
in a SUSY brane world \cite{IWY-ss,WY-ss},
as we briefly explain in this subsection. %\ref{subsec:IIB-braneworld}.
This is the reason why we develop explicit construction of models in
a SUSY brane world in this paper.

% why SUSY brane world? to break extended SUSY in UV theory through geometry!
It is quite reasonable to think of supersymmetric higher dimensions 
when one considers the approximate ${\cal N}$ = 2 SUSY 
in the SU(5)$_{\rm GUT}$-breaking sector as an indication of an extended
SUSY at short distances (rather than as an accident); 
any extended SUSY can be easily broken down to the
${\cal N}$ = 1 SUSY through extra-dimensional geometry, while it is
difficult to obtain successful models using the partial SUSY-breaking 
mechanisms in the four-dimensional space-time \cite{partial}.
The SU(5)$_{\rm GUT}$-breaking sector should be localized in
an extra-dimensional manifold, or otherwise the low energy matter contents
would be chiral and have the multiplet structure of only the ${\cal N}$ = 1
SUSY. 
This is the primary reason why we embed the original models into
a SUSY brane world.

% How the N=2 SUSY kept? How is the whole geometry?
There is a possibility that a localized sector has ${\cal N}$ = 2 SUSY (eight
SUSY charges), when the short-distance physics possesses ${\cal N}$ = 4 SUSY 
(sixteen SUSY charges)\footnote{
The ${\cal N}$ = 4 SUSY is understood as (1,1) SUSY in six dimensions, 
and is understood as ${\cal N}$ = 1 SUSY in more than six dimensions.};
sixteen SUSY charges are necessary to realize the ${\cal N}$ = 2 SUSY on
a localized sector, because the localized sector itself breaks 
translational symmetry in the extra dimensions  
and hence SUSY is broken by at least half.
Partial breaking of the translational symmetry on the localized sector
{\it can} leave half of the original SUSY charges unbroken 
when the localized sector satisfies BPS conditions (e.g. see appendix of
\cite{IWY-ss}). 
Then, the ${\cal N}$ = 2 SUSY is preserved if we impose sixteen SUSY charges in
extra-dimensional space-time and if the local geometry
around the localized sector are not the extra source of 
further breakings of SUSY charges.
Then, the approximate ${\cal N}$ = 2 SUSY (the third feature) 
is no longer an accident.

The whole geometry of the compactified manifold, on the other hand,
is chosen so that it preserves only the ${\cal N}$ = 1 SUSY 
of the four-dimensional space-time. 
It is only the local geometry around the localized SU(5)$_{\rm GUT}$-breaking 
sector that is required to preserve the ${\cal N}$ = 2 SUSY.

% SU(5) vector multiplet in the bulk
On the other hand, the SU(5)$_{\rm GUT}$ vector multiplet 
should propagate in the extra space dimensions.
Indeed, gauge fields of the SU(5)$_{\rm GUT}$ should 
propagate around the localized sector, since the hypermultiplets in the
bifundamental representation  
($\bar{Q}^{i}_{\;\; \alpha}$,$Q^\alpha_{\;\; i}$)  are charged under the
SU(5)$_{\rm GUT}$ gauge group, 
while the SU(5)$_{\rm GUT}$ vector multiplet should not be confined 
in the localized sector 
since we should not have chiral multiplets of the SU(5)$_{\rm GUT}$-{\bf
adj.} representation (i.e. ${\cal N}$ = 2 SUSY partner) 
in the models explained in the previous section.

% size of the extra dimensions 
When the compactified manifold has a moderately large volume in the
$M_*$ units, the effective four-dimensional Planck scale $M_{\rm Pl}$ 
is higher than the fundamental scale, which is given by
\begin{equation}
 M_{\rm pl}^2 \simeq M_*^2 (M_*^\delta \times {\rm volume}),
\label{eq:weak-grav}
\end{equation}
where $\delta$ is the number of sligtly large dimensions 
of the compactified manifold.
The cut-off scale lies below the effective Planck scale
of the four-dimensional gravity. 
Therefore, the fourth feature is translated into the moderately large 
volume of the extra dimensions.
Now the gauge coupling constant of the $\SU(5)_{\rm GUT}$ Kaluza--Klein 
zero mode becomes weak with respect to that of the $\U(3)_{\rm H}$ 
$(\U(2)_{\rm H})$; this is because only the SU(5)$_{\rm GUT}$ gauge field
propagates in the extra dimensions and its gauge coupling is suppressed as 
\begin{equation}
 \frac{1}{\alpha_{\rm GUT}} \simeq 
 \frac{1}{\alpha_*}(M_*^\delta \times {\rm volume}), \qquad \quad
 \frac{1}{\alpha_{\rm 3H,2H,1H}} \simeq \frac{1}{\alpha_*}.
\label{eq:weak-SU(5)}
\end{equation}
Thus, the disparity in the gauge coupling constants (the second feature)
follows naturally. We impose\footnote{The volume for the gravity in
Eq. (\ref{eq:weak-grav}) and the volume for the SU(5)$_{\rm GUT}$ gauge
field in Eq. (\ref{eq:weak-SU(5)}) are not necessarily the same, in
general.
However, we have no motivations to consider such a situation.}
\begin{equation}
 (M_*^\delta \times {\rm volume}) \sim 10^2
\label{eq:vol-required}
\end{equation}
to maintain the approximate SU(5)$_{\rm GUT}$ unification.
Then, in turn, $M_* \simeq 10^{-1} M_{\rm pl}$ follows\footnote{
This relation is independent of the number of extra dimensions $\delta$.} 
from Eq. (\ref{eq:weak-grav}), which is also a desirable value for the
cut-off scale.

% discrete R symmetry
Another benefit of higher dimensions
is that R symmetries % (required from successful phenomenology) 
can be realized as discrete gauged symmetries 
below the compactification scale \cite{gauged-R}.
A discrete rotational symmetry of the compactified manifold is in general
recognized as an R symmetry below the compactification scale. 
The rotational symmetry is a gauge symmetry, since it is a subgroup 
of the extra-dimensional Lorentz group.
The R symmetry is thus exact, unless broken spontaneously.
The fifth feature finds its natural explanation when 
the (mod 4)-R symmetry is identified with a suitable rotational symmetry
of the compactified manifold. 

% product-group structure 
As we have seen so far, an effective field theory with a localized
sector and with an higher-dimensional SUSY
is able to explain basic structures of the product-group unification models.
Both the extended SUSY and the localization of 
gauge fields are natural ingredients of higher-dimensional supergravities. 
In fact, a number of indications have been obtained,  
which suggest that SUSY gauge theories are localized 
on solitonic solutions of the higher-dimensional supergravities 
\cite{HS} called D-branes\footnote{Those solitonic solutions were
formerly called ``black $p$-branes'' \cite{HS}. 
We make no distinction between ``D-branes'' 
in supergravities and ``black $p$-branes''.}.
Once we adopt this picture, then the product-group structure of the 
``unified gauge group'' (the first feature) is quite a natural
consequence since each stack of D-branes provides each factor 
of the product group.

\subsection{In Type IIB Supergravity}

Therefore, it is quite interesting to consider that 
the product-group unification
models are realized on D-branes in higher-dimensional supergravities.
References \cite{IWY-ss,WY-ss} identify the D3--D7 system (bound states of
D3- and D7-branes) of the Type IIB supergravity, in ten-dimensional
space-time, with the origin of the SU(5)$_{\rm GUT} \times$U(3)$_{\rm H}$
(U(2)$_{\rm H}$) gauge group.

% What is Type IIB ?
The Type IIB supergravity (in ten-dimensional space-time) 
has the maximally extended SUSY and highly restricted multiplet structure.
There are thirty-two SUSY charges (eight times those of the four-dimensional
${\cal N}$ = 1 SUSY), which are combined into two  
SUSY generators ${\cal Q}$ and ${\cal Q}'$, irreducible under the SO(9,1).
Both the ${\cal Q}$ and ${\cal Q}'$ are Weyl and Majorana spinors 
of the SO(9,1), and hence each one contains sixteen SUSY charges. 
The Type IIB supergravity allows only one SUSY multiplet, 
the supergravity multiplet. 
No other multiplet is allowed as a massless representation of 
the SUSY generators and the Lorentz symmetry SO(9,1).
The supergravity multiplet consists of one hundred and twenty-eight 
bosonic states and one hundred and twenty-eight fermionic states. 
The one hundred and twenty-eight bosonic states are described 
by ten-dimensional metric, which contains thirty-five on-shell states, 
two real scalar fields $\phi$ called dilaton and $C_{(0)}$, 
two 2-form fields $B_{\mu\nu}$ and $C_{(2)}$, 
both containing twenty-eight states, and a self-dual 4-form field 
$C_{(4)}^{+}$ containing thirty-five states.
The one hundred and twenty-eight fermionic states consist of two Weyl and
Majorana gravitinos (fifty-six states each) and two Weyl and Majorana
spinors (eight states each).

% What are D-brane solutions?
D-branes are soliton solutions made of those fields. 
D7-branes are soliton solutions that extend in seven spatial dimensions.
There are two codimensions in the ten-dimensional space-time.
D7-branes are made\footnote{
The D7-brane is a multi-valued solution of $C_{(0)}$, as is evident from 
the formula of 7-brane charges for D7-branes, Eq. (\ref{eq:7brane-charge}).
The multi-valued soliton solution  is a natural possibility 
because the D7-branes have only two codimensions. 
The $\tau \equiv (C_{(0)}+i e^{-\phi})$ of D7-brane solutions 
is considered as single-valued 
up to SL$_2$({\bf Z}) monodromy around them, 
where this SL$_2$({\bf Z}) is a subgroup of the SL$_2$({\bf R})
symmetry of the Type IIB supergravity that acts on 
$\tau$. (See, e.g. \cite{Schwarz}.)} 
of the metric, the dilaton $\phi$ and the $C_{(0)}$, 
whose discretized 7-brane charges are measured by 
\begin{equation}
 \oint dC_{(0)},
\label{eq:7brane-charge}
\end{equation}
where the closed path is taken so that it winds around the D7-branes.
D3-branes are solitons that extend in three spatial dimensions made of 
the metric and the $C_{(4)}^+$. 
The 3-brane charges of the D3-branes are measured by
\begin{equation}
 \int_{S_5} dC_{(4)}^+,
\end{equation}
where the 5-form is integrated over a 5-sphere that wraps 
the D3-branes.
The 5-brane charges of the D5-branes are measured by 
\begin{equation}
 \int_{S_3} dC_{(2)}
\end{equation}
on a 3-sphere surrounding the D5-branes, although this solution is not
relevant to our construction.
What is called the D3--D7 system
%% \footnote{Explicit solution of this bound state is not obtained.
%% % yet their existence and its stability (BPS nature) are expected.
%% } 
is a bound state of the D3- and D7-branes.

%%% SUSY of D7-branes, and D3--D7 system
The D7-branes are BPS solutions of the Type IIB supergravity, 
on which half of SUSY charges (sixteen SUSY charges, i.e. 
the ${\cal N}$ = 1 SUSY in eight dimensions) 
are realized linearly \cite{HS,Polchinski,GK}. 
The SUSY charges preserved in the presence of 
D7-branes are\footnote{$\Gamma^{98}\equiv\Gamma^9 \Gamma^8$. A similar
notation is used throughout this paper.} 
\begin{equation}
 {\cal Q} -  \Gamma^{98} {\cal Q}'.
\label{eq:SUSY-D7}
\end{equation}
The SUSY charges preserved in the presence of D3-branes are
\begin{equation}
  {\cal Q} + \Gamma^{987654} {\cal Q}'.
\label{eq:SUSY-D3}
\end{equation}
The SUSY charges on the D3--D7 system, i.e. the SUSY charges 
that belong both to (\ref{eq:SUSY-D7}) and to (\ref{eq:SUSY-D3}), 
are equivalent to (\ref{eq:SUSY-D7}) that satisfy an additional constraint 
\begin{equation}
 -\Gamma^{7654} ( {\cal Q} -  \Gamma^{98} {\cal Q}') =  
              ( {\cal Q} -  \Gamma^{98} {\cal Q}').
\label{eq:D3-D7}
\end{equation}
Since the $\Gamma^{7654}$ has two eigenvalues\footnote{
Note that $(-\Gamma^{7654})^2=1$. This is why four extra
dimensions, transverse to the localized SU(5)$_{\rm GUT}$-breaking sector, 
are necessary.}, 1 and $-1$, both with the same multiplicity,
the SUSY charges are further broken by half on the D3--D7 system
\cite{Polchinski,GK}. 
Eight SUSY charges (SUSY charges in the eigenspace 
$(-\Gamma^{7654})=1$)  are left unbroken among the sixteen SUSY charges 
in (\ref{eq:SUSY-D7}).

Now we see that the D3--D7 system preserves the ${\cal N}$ = 2 SUSY, 
which is necessary to the SU(5)$_{\rm GUT}$-breaking sector. 
Therefore, we % consider that the higher-dimensional effective field theory 
% with localized SU(5)$_{\rm GUT}$-breaking sector (naturally?) originates
% from the D3--D7 system
consider the D3--D7 system as the origin of the SU(5)$_{\rm GUT}$-breaking
sector\footnote{
There is another system that possesses the ${\cal N}$ = 2 SUSY: i.e. 
NS5--D4--D6 system \cite{GK}. 
We do not discuss that system in this paper.
Degrees of freedom on the NS5-brane are also relevant because  
the extra dimensions are compactified, and moreover the weak coupling 
limit is not applicable in the case that interests us.}.
The U(3)$_{\rm H}$ or the U(2)$_{\rm H}$ gauge group is expected to
arise on D3-branes and the SU(5)$_{\rm GUT}$ gauge group on D7-branes.

% discrete R symmetry from 89th plane
Discrete rotational symmetry of the plane transverse to the D7-branes 
is identified \cite{IWY-ss,WY-ss} with an origin of 
the (mod 4)-R symmetry, which is crucial in the product-group 
unification models.
This is the primary reason why we construct models in {\em ten}-dimensional
space-time. 

\subsection{Purpose of This Paper}

% explicit model construction
The discussion given above does not go beyond qualitative arguments. 
Explicit models should be constructed to examine theoretical
consistencies in the Type IIB supergravity, 
which is the purpose of this paper.

% orbifold approximation
It is quite difficult to handle a six-dimensional compactified manifold, 
unless it has a simple geometry.
We only adopt an orbifold of a six-dimensional torus as the compactified
manifold\footnote{We use the word ``manifold'' even if it has
singularities. It is an abuse of terminology, though.}. 
The purpose of the following sections is to see 
whether the basic idea is consistently realized in
the orbifold compactification of the Type IIB supergravity.
Namely, we want to see if the SU(5)$_{\rm GUT}$-breaking sector of 
the product-group unification is naturally realized on the D3--D7 system.

% What we are going to do
Once the orbifold geometry is fixed, we can calculate the low-energy
spectrum on a set of D-brane configuration and orbifold projection
conditions.
We show that the whole SU(5)$_{\rm GUT}$-breaking sector is obtained
from fluctuations localized on D-branes.
Although not all the particles of the whole theory are obtained 
as D-brane fluctuations in our construction (quarks and leptons are missing), 
we consider missing particles that do not arise as D-brane
fluctuations arise as fields at fixed points (see also discussion in
section \ref{sec:principle}).  
Once particles are obtained from the D-brane fluctuations, 
then their R charges are far from arbitrary.
Now that the (mod 4)-R symmetry is identified with a rotational symmetry
of the compactified manifold, we can determine how each particle
transforms under the rotation and equivalently under the R symmetry.
Therefore, the model building in such a higher-dimensional space-time
is subject to a stringent consistency check.
Anomaly cancellation at orbifold fixed points is also discussed.
This also serves as a non-trivial consistency check.

%%%%
%%%%\include{preparation}
%%%%
%%%%%%%%%%%%%%%%%%%%%%%%%%%%%%%%%%%%%%%%%%%%%%%%%%%%%%%%%%%%%%%%%%%%%%
\section{Our Principle of Model Construction \\ in Type IIB Supergravity}
\label{sec:principle}
%%%%%%%%%%%%%%%%%%%%%%%%%%%%%%%%%%%%%%%%%%%%%%%%%%%%%%%%%%%%%%%%%%%%%

%  Not IIB string
Our discussion is based on supergravity, and we do not assume the Type
IIB string theory.
It is true that the Type IIB string theory is one of the candidates
of quantum gravity that effectively provides the Type IIB
supergravity below the fundamental scale (i.e. the string scale),
but there is no proof that it is the only one.
The Type IIB string theory has a definitely fixed spectrum 
that extends up to infinity above the fundamental scale.
We have no strong motivation that directly suggests
to us to impose such a stringent restriction\footnote{
Even if one does not assume the exact spectrum of the Type IIB string
theory, the SL$_2$({\bf Z}) symmetry is necessary in the UV spectrum,
since this symmetry is crucial for the existence of the D7-brane solution.}.
Our study keeps genericity and is independent of the ultraviolet (UV)
spectrum above the cut-off scale of the supergravity.
Thus, we have more freedom to construct realistic models, 
since we do not specify the UV physics.
In particular, we do not ask the question of whether the UV physics 
required for our model 
is contained in the vast variety of vacua of what is called the M-theory
\cite{M-theory}.
This is not a question to be solved at present, since we do not have a
precise definition of the M-theory.

% matter content on D-branes as in string theories
Massless matter contents and interactions on D-branes are known very well 
if one assumes the Type IIB string theory.
On the other hand, localized massless sectors on D-branes are not 
well understood when one assumes only the Type IIB
supergravity\footnote{
T.W. thanks M.~Nishimura for useful discussion.}.
However, various studies on the AdS/CFT correspondence \cite{AdS/CFT1} 
seem to suggest that those sectors in supergravity 
are the same as those predicted by string theories 
in particular cases\footnote{
Most of the studies on the AdS/CFT does not aim 
at showing a correspondence between supergravities and
gauge theories, but rather a correspondence between string theories and
gauge theories \cite{AdS/CFT1}. 
However, most of the evidences of this correspondence have been obtained
for large 't Hooft couplings, where string corrections are not important 
(i.e. only the supergravity is relevant).
Such results, which are rather independent of the UV spectrum, are
just what we are interested in here.}; 
there are a number of evidences \cite{AdS/CFT4}, when the 't Hooft coupling 
$g^2_{\rm Yang-Mills} N$ is large, that U($N$) gauge theories 
with sixteen SUSY charges are localized on coincident $N$ D3-branes 
of the supergravity, which is the same as the predictions of the Type
IIB string theory. 
Although such studies have not yet given {\em proof} that 
the gauge theories localized on {\em any} D-brane configuration 
in supergravities with {\em arbitrary} 't Hooft coupling 
are the same as those in the Type IIB string theory,
we assume this to be the case \cite{GPTT}.

Since string theories provide massless matter contents and 
their interactions that are known to obey all consistency conditions 
of the higher-dimensional supergravity, 
it is convenient to adopt those string predictions 
as the starting point of our model construction.
This is another reason why we assume the same massless matter contents 
and interactions on D-branes 
as in the string theories% \footnote{
%%  Once the matter contents and SUSY charges are determined on D7-branes 
%%  (sixteen SUSY charges) and D3-D7 system (eight SUSY charges),
%%  then these are almost sufficient in determining interactions of these
%%  fields.}
.
% We might have lost genericity by assuming in that way, but we do not
% care about it.

% SU(5) from five D7-branes and......
%%%% {\em This paragraph might not be necessary here.}
We consider that the SU(5)$_{\rm GUT}$ gauge group comes from 
five D7-branes and the U(3)$_{\rm H}$ or U(2)$_{\rm H}$ gauge
group from three or two D3-branes, respectively \cite{IWY-ss, WY-ss}.
The gauge group would be U(5) on D7-branes rather than SU(5)$_{\rm
GUT}$, but it is not a problem as will be shown in subsections
\ref{subsec:u2-phen} and \ref{subsec:u3-phen}. 
We show that all the matter contents of the SU(5)$_{\rm GUT}$-breaking
sector are obtained from the fluctuations of the D3-D7 system.
We try to understand as many particles of the models as possible 
as the D-brane fluctuations (i.e. massless fields on D-branes), 
but quarks and leptons are not obtained. 

% Fields at orbifold singularities
The Type IIB string theory also has definite predictions on the
massless matter contents and interactions at orbifold singularities.
Those massless matter contents are called the twisted sector. 
However, we do not restrict ourselves to the matter contents of the
twisted sector determined by the Type IIB string theory.

The twisted sectors of the Type IIB string theory play important roles
in the following two aspects:
first, in restoring the modular invariance of the string world-sheet 
and, second, in keeping the unitarity of the theory.
The first aspect, the modular invariance of world-sheets, is crucial 
to make the string theories UV-finite \cite{Polchinski}.
Theories with modular invariance are constructed so that the 1-loop
amplitude with UV momentum, where an infinite number of massive particles
are in the loop, is equivalent to the 1-loop amplitude with infrared
(IR) momentum, where only massless particles are in the loop. 
This equivalence between the UV and IR amplitudes enables one
to cut out the UV part from theories, rendering the theories UV-finite.
This clearly shows that modular invariance is a constraint between
UV physics and IR physics, not a constraint on purely IR physics.
In particular, this means that the necessary massless contents at the
orbifold singularities are different 
when the UV spectrum in the bulk is different.
Moreover, we are not sure whether the modular invariance in string theories  
is the only way to make a theory UV-finite.
Therefore, we have no reason to adopt the massless matter contents
at orbifold singularities predicted by the Type IIB string theory;
our study is based on a generic Type IIB supergravity and 
we do not specify the theory above the cut-off scale.
Thus, our framework is not restrictive enough to determine 
the massless contents at orbifold singularities in a top-down way.

The second aspect, the unitarity, is a well-defined notion within field
theories. 
Anomalies appear at orbifold singularities, even if the theory is
consistent in the {\em ten}-dimensional spce-time. 
Some of the massless fields in the twisted sector in the Type IIB string
theory provide massless degrees of freedom necessary to cancel 
the pure gravitational anomalies, and some of them 
realize the (generalized) Green--Schwarz mechanism \cite{GS,DSW} 
at orbifold singularities, cancelling the anomalies.
Therefore, we also require in our model 
that the massless matter contents at singularities are such 
that anomalies are all cancelled.

% Since there are a number of mechanisms known so far that provide massless
% particles at singularities \cite{massless}, 
We expect that suitable particles can be
supplied at singularities by a certain mechanism of a fundamental theory 
if they are required for theoretical consistency there.
We do not specify the mechanism in this paper.
General interactions localized at singularities are also considered
as long as they satisfy the symmetries around there.
Since we do not know the origin of particles that arise at
singularities, there is no way of determining their interactions other 
than theoretical consistencies and symmetries.

There are consistency conditions called Ramond--Ramond tadpole
cancellation \cite{GP} in (string-based) Type IIB orientifolds 
\cite{IIB-orbifold}.
% These conditions are the principle in the (string-based) model constructions 
% on the Type IIB orientifolds. 
These conditions are closely related to the matter contents and
interactions of the twisted sector. 
These conditions are {\em generically} equivalent to conditions for 
vanishing non-Abelian anomalies at fixed-point singularities
\cite{RR-tadpoleI,RR-tadpoleII},  
and at the same time they ensure automatically
that the mixed anomalies arising at fixed points 
are cancelled through the fixed-point interactions 
(generalized Green--Schwarz mechanism) 
predicted by the Type IIB string theory 
\cite{IIB-orbifold,anomalousU(1)-IIB,RR-tadpoleII}. 

However, it is also known that the Ramond--Ramond tadpole cancellation
is sometimes more stringent than the triangle anomaly cancellation 
\cite{RR-tadpoleII}.
We clarify the relation between the Ramond--Ramond tadpole cancellation
and the anomaly cancellation in subsubsection
\ref{subsubsec:u3-anomaly-rel}, 
in addition to the arguments on the anomaly cancellation in subsections
\ref{subsec:u2-anomaly} and \ref{subsec:u3-anomaly}.
We conclude, there, that various anomalies vanish or can be cancelled 
through the (generalized) Green--Schwarz mechanism, 
while one cannot argue some of the Ramond--Ramond tadpole cancellation 
conditions in our models, since we do not specify 
the UV spectrum and the matter contents at singularities.

%%%%%%%%%%%%%%%%%%%%%%%%%%%%%%%%%%%%%%%%%%%%%%%%%%%%%%%%%%%%%%%%%%%%%%%
\section{Geometry of the ${\bf T}^6/({\bf Z}_{12}\vev{\sigma}\times
{\bf Z}_2\vev{\Omega R_{89}})$ Orientifold}
\label{sec:geometry}
%%%%%%%%%%%%%%%%%%%%%%%%%%%%%%%%%%%%%%%%%%%%%%%%%%%%%%%%%%%%%%%%%%%%%%

%%% introduction of orientifold planes
%%%
The extra dimensions should be compactified to obtain realistic  
low-energy physics.
Then, the fluxes of gauge fields $C_{(0)}$ and $C_{(4)}^+$ 
have nowhere to escape in the compactified extra dimensions.
This implies that the totality of 7-brane charges and 
that of 3-brane charges scattered within the compact manifold 
should be zero. 
However, it is impossible for D7-branes alone (D3-branes alone) to have
vanishing total 7-brane charges (3-brane charges) without breaking SUSY,
because the unbroken half-SUSY on each BPS D-brane is determined by  
the sign of its 7- (3-)brane charge \cite{OPplane1}.
This is a well-known phenomenon for the BPS monopoles and instantons 
\cite{OPplane1}. 
Therefore, in order for the 7- (3-)brane charges to cancel out within the
compactified manifold, 
there should be new objects with 7- (3-)brane charge whose
sign is opposite to the charge of D7- (D3-)branes and which also
preserve the same SUSY as D7- (D3-)branes do.
Such candidates are known in string theories:
orientifold 7-planes (O7-planes) and orientifold 3-planes (O3-planes)
\cite{OPplane2}.  
Orientifold planes emerge in string theories 
when the world-sheet parity ($\Omega$)
(flipping of the orientation of strings) is gauged. One can gauge 
an arbitrary combination of the world-sheet parity and 
order-2 space transformation ($g^2=1$;$g \in $SO(6)) 
rather than a simple world-sheet parity, i.e. one can gauge $\Omega g$. 
Orientifold planes are (loci of) fixed points 
of such order-2 space transformations.

Although the orientifold planes are well-formulated in string
theories \cite{Polchinski,GK}, their origin is not clearly understood 
in general field theories\footnote{An O6-plane is realized 
as a solitonic solution of eleven-dimensional supergravity 
(in the context of M-theory) in \cite{Sen}. 
This O6-plane solution is also associated to a ${\bf Z}_2$
projection and is found to have the same 6-brane charge as that
predicted by string theories.}. 
However, we assume that there exist ``orientifold planes'' 
even in the Type IIB supergravity, since there is no clear obstruction
against this assumption. 
Furthermore, we assume that the ``orientifold planes'' have almost 
the same properties as those in the Type IIB string theories.
Namely, O$p$-planes have $p$-brane charges opposite to that of the
D$p$-branes, and they are always (loci of) ${\bf Z}_2$-fixed points 
as explained above.
Here, an orientifold $p$-plane carries $- 2^{p-4}$ times the 
$p$-brane charge of a single D$p$-brane, 
as in string theories \cite{Polchinski,GK}\footnote{
It is not clear at all whether we have to impose this $p$-brane charge
of the orientifold planes for the consistency of models in a field
theory. 
Our construction of models, nevertheless, does not change so much
even if the $p$-brane charges of the orientifold planes 
are different from the predictions of the string theories. 
There is a related discussion at the beginning of subsection 
\ref{subsec:u2-config}.}.

%%%  geometry of ${\bf T^2}/({\bf Z}_{12}\vev{\sigma} \times {\bf
%%%  Z}_2\vev{\Omega R_{89}})$ orientifold

The D7-branes, O7-planes, D3-branes and O3-planes are put in a
six-dimensional torus ${\bf T}^6$. 
We will obtain most of matter contents of the product-group 
unification models
by gauging the orientifold projection, and by gauging
the discrete rotational symmetry of the ${\bf T}^6$.
The geometry we adopt in this paper is 
${\bf T}^6/({\bf Z}_{12}\vev{\sigma} \times {\bf Z}_2\vev{\Omega R_{89}})$.
Let us first explain the geometry of this manifold.
The geometry is important when we discuss the D-brane configuration, 
anomaly cancellation, discrete R symmetry and effective superpotential.
(However, the reader can skip the rest of this section for now, 
and come back to it when necessary.)

${\bf T}^6$ denotes a six-dimensional torus $({\bf
C}^3=\{(z_1,z_2,z_3)|z_1,z_2,z_3 \in {\bf C}\})/\Gamma_0$ in which 
the lattice $\Gamma_0$ is spanned by six base vectors 
${\bf e}_4$, ${\bf e}_5$, ${\bf e}_6$, ${\bf e}_7$, ${\bf e}_8$, and 
${\bf e}_9$.
In other words, two points ${\bf y},\tilde{{\bf y}} \in {\bf C}^3$ 
are identified with each other if and only if 
\begin{equation}
 \tilde{{\bf y}} = {\bf y} + n_m  {\bf e}_m \qquad \qquad 
  (n_m \in {\bf Z}, \quad m = 4,\cdots, 9).  
\end{equation}
The base vectors ${\bf e}_{4,5,6,7}$ will be chosen so that 
they span the first two complex planes ${\bf C}^2=\{(z_1,z_2)| z_1,z_2 \in
{\bf C}\}$, and ${\bf e}_{8,9}$ 
for the last complex plane, whose coordinate is $z_3$, in the orbifold we
adopt in this paper.

We assume D7-branes stretched in the first two complex planes, and hence we
need O7-planes parallel to the D7-branes to cancel those 7-brane charges.
We gauge the combination $\Omega R_{89}$, where $R_{89}$ denotes
the angle-$\pi$ rotation in the last complex plane, 
having loci of fixed points that extend parallel to the D7-branes.
The loci of the $R_{89}$-fixed points are the O7-planes.
Although this orientifold projection breaks half of the 
thirty-two SUSY charges of the Type IIB supergravity, the unbroken 
sixteen SUSY charges are the same as those preserved 
on the D7-branes (i.e. (\ref{eq:SUSY-D7})), 
since the O7-planes are parallel to the D7-branes. 
There are four O7-planes within the six-dimensional torus, 
whose $z_3$-coordinates are given by
\begin{equation}
 z_3 = \frac{1}{2}n_{m''} {\bf e}_{m''}, 
\qquad ({\rm mod~} n_{m''}  {\bf e}_{m''}|_{m''=8,9} 
\quad  {\rm where~} n_{m''} \in {\bf Z}).
\end{equation}
The total 7-brane charge of the O7-planes is $-32$ because each
O7-plane carries\footnote{Some works state that each O7-plane
carries the 7-brane charge = $-4$, and that the total 7-brane charge of
O7-planes is cancelled by sixteen D7-branes. This discrepancy comes
from adopting two different descriptions:
counting 7-brane charges either in projected space ${\bf T}^6/{\bf
Z}_2\vev{\Omega R_{89}}$, or in the covering space ${\bf T}^6$.
We adopt the latter counting throughout this paper.} 
the 7-brane charge $=-8$. 
Thus, there should be thirty-two D7-branes within the six-dimensional
torus {\bf T}$^6$.

We also need O3-planes parallel to the D3-branes so that 
the totality of the 3-brane charges vanishes in the six-dimensional torus.
This implies that $\Omega R_{456789}$ should also be gauged,
where $R_{456789}$ reverses all six extra dimensions. 
In other words, $R_{4567}$ should also be gauged, where $R_{4567}$ 
reverses all ${\bf e}_{4,5,6,7}$.
Indeed, gauging ${\bf Z}_2\vev{R_{4567}}$ is equivalent to gauging 
${\bf Z}_2\vev{\Omega R_{456789}}$,
under the condition that $\Omega R_{89}$ is already gauged because of
the isomorphism  
${\bf Z}_2\vev{R_{4567}}\times {\bf Z}_2\vev{\Omega R_{89}} \simeq
{\bf Z}_2\vev{\Omega R_{456789}}\times {\bf Z}_2\vev{\Omega R_{89}}$.
Thus, the orbifold group should contain ${\bf Z}_2\vev{R_{4567}}$ as a
subgroup. 
The SUSY charges broken by this orbifold projection 
${\bf Z}_2\vev{R_{4567}}$ are the same as those broken 
in the presence of the D3-branes.
Indeed, the SUSY charges on which $-\Gamma^{7654}$ acts trivially 
are not twisted by the $R_{4567}$ since 
the $-\Gamma^{7654}$ is the same as the spinor representation 
of $R_{4567}=\exp ((\pi \Gamma^{45}-\pi \Gamma^{67})/2)$. 
Hence the eight SUSY charges (${\cal N}$ = 2 SUSY in four-dimensional
spce-time) are preserved in the D3--D7 system along with the O7- and O3-planes. 
There are sixty-four $R_{456789}$-fixed points, whose coordinates are
\begin{equation}
 {\bf y}' = \frac{1}{2}n_{m'} {\bf e}_{m'}|_{m'=4,5,6,7} \quad {\rm ~and~} 
\quad  z_3 = \frac{1}{2}n_{m''}{\bf e}_{m''} |_{m''=8,9} 
\end{equation}
mod $n_m {\bf e}_m|_{m=4,...,9}$, where $n_m \in {\bf Z}$. 
Thus, the total 3-brane charges from those O3-planes are $-32$ because
each O3-plane carries the 3-brane charge $= -1/2$.
Therefore, thirty-two D3-branes are required in the six-dimensional torus.
The orbifold group should be much larger, so that the SUSY of the whole
geometry preserves only ${\cal N}$ = 1 SUSY of the four-dimensional spce-time,
i.e. only four SUSY charges.
Thirteen ${\bf Z}_n$-type orbifold groups\footnote{We restrict 
our attention only to orbifolds of the form ${\bf T}^6/{\bf Z}_n$.
One can, in principle, consider orbifolds such as ${\bf T}^6/({\bf Z}_n
\times {\bf Z}_m)$ or more complicated ones.
However, the variety of higher-dimensional constructions does not become
very much richer by considering such possibilities.}  
are listed in \cite{orbifold-II}
that can be imposed on the six-dimensional torus, keeping the ${\cal N}$ = 1
SUSY. Four of them preserve even ${\cal N}$ = 2 SUSY, and 
two others do not contain ${\bf Z}_2\vev{R_{4567}}$ as their subgroup.
We adopt the ${\bf Z}_{12}\vev{\sigma} \equiv \{ \sigma^k |
k=0,...,11\}$ 
orbifold\footnote{We follow the notation of \cite{IIB-orbifold}.} 
among the seven remaining candidates. 
There are a couple of reasons why we choose this group, each of 
which is explained in the course of the following discussion.

The generator $\sigma$ of the present orbifold group, 
${\bf Z}_{12}\vev{\sigma}$, acts on the ${\bf C}^3$ as
\begin{equation}
 \sigma : {\bf y}\equiv (z_b)|_{b = 1,2,3} \in {\bf C}^3 \longmapsto 
 \sigma \cdot {\bf y} \equiv (e^{2 \pi i v_b} z_b)|_{b = 1,2,3} \in {\bf C}^3 
\label{eq:generator}
\end{equation}
with $(v_b)|_{b=1,2,3} = (1/12,-5/12,4/12)$. 
Note that $\sigma^6 = R_{4567}$, and hence the ${\bf
Z}_{12}\vev{\sigma}$ contains ${\bf Z}_2\vev{\sigma^6 = R_{4567}}$ 
as required, and hence there surely exist O3-planes 
(and D3-branes) in the orbifold.
On the other hand, since the $\sigma^6$ is the only order-2 element in
the ${\bf Z}_{12}\vev{\sigma}$, only the $\Omega R_{89}$ and $\Omega R_{89}
\sigma^6$ in 
${\bf Z}_{12}\vev{\sigma} \times {\bf Z}_2\vev{\Omega R_{89}}$
are the elements that lead to the existence of orientifold planes.
Therefore, the O7- (O3-)planes transverse to the ${\bf e}_{8,9}$
(${\bf e}_{4,5,6,7,8,9}$) directions 
are the only orientifold planes, and no other types of
orientifold planes exist in this manifold. 
%Note also that the $\{ \Omega R_{89} ,\sigma^6 \cdot \Omega R_{89}\}$
%are the only order-2 elements among the entire orbifold group 
%${\bf Z}_{12}\vev{\sigma}\times {\bf Z}_2\vev{\Omega R_{89}}$.
%This implies that O7-planes stretched in the first two complex planes
%and O3-planes are the only orientifold planes in our manifold, and hence
%we do not have to introduce other species of D-branes to cancel 
%the charges of unwanted orientifold planes.

The six-dimensional torus ${\bf T}^6$ should be chosen so that it has 
the ${\bf Z}_{12}\vev{\sigma}$ symmetry. Thus, the $\Gamma_0$, which
determines the six-dimensional torus, is chosen as 
\begin{eqnarray}
&& {\bf e}_4 = (1,1,0)L_4,  \qquad 
   {\bf e}_5 = (\zeta,\zeta^{-5},0)L_4,  \\
&& {\bf e}_6 = (\zeta^2,\zeta^2,0)L_4, \qquad 
   {\bf e}_7 = (\zeta^3,\zeta^{-3},0)L_4, \\
&& {\bf e}_8 = (0,0,1)L_2, \qquad 
   {\bf e}_9 = (0,0,\omega)L_2,
\label{eq:Gamma-0}
\end{eqnarray}
where $\zeta = e^{2 \pi i /12}$, $ \omega = e^{2 \pi i /3}$ and
$L_4,L_2$ are two independent length scales of the six-dimensional torus; 
$L_4$ corresponds to the size of the first four-dimensional torus
and $L_2$ to that of the remaining two-dimensional torus.

The D7-branes are stretched in the ${\bf e}_{4,5,6,7}$ directions, on
which the SU(5)$_{\rm GUT}$ gauge fields are expected to propagate. 
Therefore, we require that the $L_4$ be slightly larger 
than the fundamental-scale inverse, 
because the volume of these extra four dimensions should be large enough
to account for the disparity between gauge couplings of the SU(5)$_{\rm
GUT}$ and U(3)$_{\rm H}$ or U(2)$_{\rm H}$.
Equations (\ref{eq:weak-SU(5)}) and (\ref{eq:vol-required}) imply that 
\begin{equation}
 \frac{1}{4}(M_* L_4)^4 \sim 10^{2},
\label{eq:length}
\end{equation}
where $(1/12) \times 3 (L_4)^4$ is the volume of extra dimensions where 
the SU(5)$_{\rm GUT}$ gauge field propagates. 
Here, the D7-branes that provide the SU(5)$_{\rm GUT}$ are assumed to 
reside at a ${\bf Z}_{12}\vev{\sigma}$-fixed\footnote{The orientifold
projection ${\bf Z}_2\vev{\Omega R_{89}}$ does not reduce the volume on
which SU(5)$_{\rm GUT}$ propagates because it acts only in the transverse
directions to the D7-branes.} locus.
On the other hand, the $L_2$ is considered to be of the order of 
the inverse of the fundamental scale.

The generator $\sigma$ in (\ref{eq:generator}) rotates 
three complex planes separately,
and satisfies $\sum_{b=1}^3v_b =0$. 
Thus, it belongs to an $\SU(3)$ subgroup of the six-dimensional
rotational symmetry $\SO(6) \simeq \SU(4)$. 
The $\sigma$ is regarded as an $\SU(4)$ element, which 
is written as\footnote{
Section(s) 3 (and 5) and appendices of Ref. \cite{IWY-ss} might be useful 
to understand this paragraph and the following sections. 
SO(6) $\simeq$ SU(4) transformation properties of gauge fields, 
scalars, fermions and SUSY charges are explicitly written there.} 
\begin{equation}
 \sigma = e^{- 2 \pi i \diag(\tilde{v}_a)_{a=0,1,2,3}} 
 = e^{- 2\pi i \diag(0,\frac{1}{12},
\frac{-5}{12},\frac{4}{12})} \in \SU(4),  
\end{equation}
where $\tilde{v}_0\equiv (v_1+v_2+v_3)/2$ and $\tilde{v}_b\equiv v_b -
\tilde{v}_0$ for b=1,2,3, or equivalently,
\begin{equation}
 \diag (\tilde{v}_a)|_{a=0,1,2,3} \equiv
\diag (\frac{v_1+v_2+v_3}{2},\frac{v_1-v_2-v_3}{2},
       \frac{-v_1+v_2-v_3}{2},\frac{-v_1-v_2+v_3}{2}). 
\end{equation}
Note that $\tilde{v}_0=0$ and $\tilde{v}_b=v_b$ (for $b=1,2,3$) when
$\sum_{b=1}^3 v_b =0$. The $\sigma$ belongs to an SU(3) 
subgroup at the lower-right corner of the SU(4).
This $\tilde{v}_0=0$, or equivalently an eigenvalue 
$e^{-2 \pi i \tilde{v}_0}=1$, implies that SUSY charges 
are partially preserved in the orbifold geometry \cite{CHSW}.
The unbroken SUSY charges, which correspond to the first entry of the
{\bf fund.} representation of the SU(4) 
(i.e. $\sigma = e^{-2\pi i \tilde{v}_0}$ eigenspace), 
are half of the ${\cal N}$ = 2 SUSY of the D3--D7 system, 
because $((-\Gamma^{7654}) = \sigma^6 = e^{-2\pi i \diag(\tilde{v}_a) \times
6}) = 1 $ eigenspaces are the first and fourth entries of the {\bf
fund.} representation of SU(4) ($\simeq$ {\bf spinor} representation 
of SO(6)).
That is, the ${\cal N}$ = 1 SUSY, which is one half of the ${\cal N}$ = 2 SUSY 
preserved in the D3--D7 system, is also preserved in the entire
orbifold geometry.

% The geometry of the orientifold ${\bf T}^6/({\bf Z}_{12}\vev{\sigma}
%\times {\bf Z}_2\vev{\Omega R_{89}})$ is obtained 
%as ${\bf C}^3/(\Gamma_0 \ltimes ({\bf Z}_{12}\vev{\sigma}\times{\bf
%Z}_2\vev{R_{89}})) = ({\bf C}^3/\Gamma_0)/({\bf
%Z}_{12}\vev{\sigma}\times {\bf Z}_2\vev{R_{89}})$. 

There are several types of singularities on this geometry. 
Loci of points fixed under the $\sigma^6$ or $\sigma^3$ 
form two-dimensional singularities in the {\bf T}$^6$ 
that extend in the directions spanned
by ${\bf e}_8$ and ${\bf e}_9$ (i.e. singularities with (five+one) spce-time
dimensions). 
This is because $v_3 \times 3 \in {\bf Z}$ or, in other words, 
the $\sigma^3$ does not rotate the last complex plane. 
The coordinates of these singularities in the first two complex planes, 
${\bf y}'=(z_1,z_2) \in {\bf C}^2$, are given by
\begin{eqnarray}
&\sigma^6\mbox{-}{\rm fixed} & {\bf y'}= n_{m'} \frac{{\bf e}_{m'}}{2}|_{ 
m' = 4,\cdots, 7}, \qquad ({}^\forall z_3),  \label{eq:sigma6-fixed}\\
&\sigma^3\mbox{-}{\rm fixed}&  {\bf y'}=
n_4 \frac{{\bf e}_4+{\bf e}_5 + {\bf e}_6}{2}
+ n_5 \frac{{\bf e}_4 + {\bf e}_7}{2} 
+ n_6 {\bf e}_6 + n_7 {\bf e}_7, \qquad ({}^\forall z_3), 
\label{eq:sigma3-fixed}
\end{eqnarray}
where $n_{m'}|_{m'=4,5,6,7}$ are integers. 
There are sixteen loci of $\sigma^6$-fixed points of within the covering
space ${\bf T}^6$, four of which are loci of points fixed under 
the $\sigma^3$.
One locus (${\bf y}'$ = {\bf 0}) among the latter four 
is fixed under the $\sigma^1$ as a locus.
Three remaining loci among the four in {\bf T}$^6$ 
form an orbit of ${\bf Z}_{12}\vev{\sigma}/{\bf Z}_4\vev{\sigma^3}$, 
and become a single singularity in ${\bf T}^6/{\bf Z}_{12}\vev{\sigma}$.
Twelve remaining loci of $\sigma^6$-fixed points in {\bf T}$^6$ 
form two distinct orbits of 
${\bf Z}_{12}\vev{\sigma}/{\bf Z}_2\vev{\sigma^6}$ and become 
two distinct singularities in ${\bf T}^6/{\bf Z}_{12}\vev{\sigma}$.
Thus, there are four distinct two-dimensional singularities 
in ${\bf T}^6/{\bf Z}_{12}\vev{\sigma}$.
The ${\bf Z}_2\vev{\Omega R_{89}}$ projection acts only within each
two-dimensional singularity, and hence there are four two-dimensional
singularities in 
${\bf T}^6/({\bf Z}_{12}\vev{\sigma}\times {\bf Z}_2\vev{\Omega R_{89}})$.
The isotropy group is ${\bf Z}_4\vev{\sigma^3}$ at a generic point of the
first two singularities and is ${\bf Z}_2\vev{\sigma^6}$ at a generic
point of the last two singularities.

Other singularities are points in the ${\bf T}^6$ (i.e. singularities
with only (three+one) spce-time dimensions), whose coordinates 
${\bf y}=({\bf y}',z_3)\in {\bf C}^3$ are 
\begin{eqnarray}
\sigma^4\mbox{-}{\rm fixed}&& {\bf y'}=
n_4 \frac{{\bf e}_4+{\bf e}_6}{3}
+ n_5 \frac{{\bf e}_5 + {\bf e}_7}{3}
+ n_6 {\bf e}_6 + n_7 {\bf e}_7, 
\label{eq:sigma4-fixed}\\
\sigma^{2,1}\mbox{-}{\rm fixed}&& {\bf y'}=
n_{m'} {\bf e}_{m'}|_{m' = 4,...,7}
\end{eqnarray}
in the first two complex planes, where $n_{m'}|_{m'=4,...,7}$ are
integers, and
\begin{equation}
\sigma^{4,2,1}\mbox{-}{\rm fixed} \quad z_3 = {\bf 0},
\pm \frac{{\bf e}_8+2 {\bf e}_9}{3}  \qquad 
({\rm mod~}{\bf e}_{8,9} \quad {\rm where~}n_{8,9} \in {\bf Z})
\end{equation}
in the last complex plane. Note that all fixed points of $\sigma^2$
are fixed also under the $\sigma^1$ in this geometry.

Both four-dimensional loci in the {\bf T}$^6$ determined by 
$z_3 = ({\bf e}_8+2{\bf e}_9)/3$ and $z_3 = - ({\bf e}_8+2{\bf e}_9)/3$) 
are fixed under the $\sigma^1$ as a four-dimensional locus, respectively. 
Although each point on those loci are moved by $\sigma^1$ except 
${\bf y}'=0$, they move only within each four-dimensional locus.
These two four-dimensional ${\bf Z}_{12}\vev{\sigma}$-fixed loci, which
are (seven+one)-dimensional sub-spce-time, are distant from 
the loci of ${\bf Z}_2\vev{\Omega R_{89}}$ fixed points, 
where O7-planes reside.
The existence of this (seven+one)-dimensional fixed loci away from O7-planes is
one of the reasons why we chose the ${\bf Z}_{12}\vev{\sigma}$ orbifold.
We put D7-branes on these (seven+one)-dimensional fixed loci in the next
section.
D7-branes should be put on fixed loci; otherwise orbifold projection
conditions would not be imposed.
Incidentally, unwanted massless matter contents would remain in the
spectrum when a D3--D7 system coincides with orientifold planes.
Therefore, we need a fixed locus away from O7-planes.
Only the ${\bf Z}_6$, ${\bf Z}'_6$ and ${\bf Z}_{12}$
orbifolds\footnote{Notations are based on \cite{IIB-orbifold}.} 
have such a fixed locus among the seven candidates that reduce 
the higher-dimensional SUSY down to four-dimensional ${\cal N}$ = 1 SUSY 
accommodating the D3--D7 system.  

%%%%
%%%%\include{u2}
%%%%
%%%%%%%%%%%%%%%%%%%%%%%%%%%%%%%%%%%%
\section{SU(5)$_{\rm GUT} \times$ U(2)$_{\rm H}$ Model}
\label{sec:u2}
%%%%%%%%%%%%%%%%%%%%%%%%%%%%%%%%%%%%%

In the following two sections we explicitly construct the product-group
unification models in the Type IIB supergravity.
We begin with the construction of the SU(5)$_{\rm GUT} \times$U(2)$_{\rm
H}$ model, because its structure is simpler in some aspects.
The SU(5)$_{\rm GUT} \times$U(3)$_{\rm H}$ model is discussed in 
section \ref{sec:u3}.

\subsection{D-brane Configuration and Orbifold Projection}
\label{subsec:u2-config}

Matter contents below the Kaluza--Klein scale depend on 
the D-brane configuration (locations of D-branes on the orbifold geometry) 
and the orbifold projection conditions.
Let us first describe how %(the SU(5)$_{\rm GUT}$-breaking sector of) 
the SU(5)$_{\rm GUT} \times$U(2)$_{\rm H}$ model is obtained.

%%%%%%%%%%%%%%%%%%%%%%%%%%%%%%%%%%%%%%%%%%%%%%%%%%%%%%%%%%%
%%% Matter contents before the orbifold projection
%% Massless matter contents 
We assume that the gauge theory on $N$ coincident D7-branes 
(distant from any O7-planes) 
consists of a U($N$) vector multiplet ($\Sigma^k_{\;\;l}$) ($k,l = 1,...,N$)
of the ${\cal N}$ = 1 SUSY of eight-dimensional space-time. 
$M$ coincident D3-branes are also expected to have a U($M$) vector multiplet
($X^\alpha_{\;\;\beta}$) ($\alpha,\beta=1,...,M$) of the ${\cal N}$ = 4 SUSY 
of four-dimensional space-time. 
There would be a hypermultiplet 
($\bar{Q}^{k}_{\;\; \alpha}$,$Q^\alpha_{\;\; k}$) of 
the four-dimensional ${\cal N}$ = 2 SUSY 
when the $M$ D3-branes are on the $N$ D7-branes.
Here, the ${\cal N}$ = 1 chiral multiplets $\bar{Q}^k_{\;\; \alpha}$ and 
$Q^\alpha_{\;\; k}$ are in the bifundamental representation 
under the U($N$)$\times$U($M$) gauge group, transforming 
(${\bf N},{\bf M^*}$) and (${\bf N}^*$,${\bf M}$), respectively. 
These are the particles we assume at the starting point of our model
construction in addition to the Type IIB supergravity multiplet in the
ten-dimensional bulk. These are exactly the same massless
modes as in the D3--D7 system of the Type IIB string theory.
D7-branes are labelled by indices $k,l$, and D3-branes by
$\alpha,\beta$.

We are concerned only with those massless modes in the following, 
except for massless modes that are required at orbifold singularities.
It is clear that those matter contents on D-branes satisfy all consistency 
conditions of the supergravity.
%There is no anomaly in gauge theories on D7-branes and D3-branes 
%with the Type IIB supergravity. 
%The pentagonal anomaly on D7-branes and the triangle anomaly on 
%D3-branes vanish because there are only vector-like representations.
Anomalies would arise only at orbifold singularities. 
These issues are discussed in subsection \ref{subsec:u2-anomaly}.

%% No particular assumption on massive spectrum
We do not specify the spectrum of massive particles above 
the cut-off scale. Those particles are not relevant to the physics at
the GUT scale, or at low energies.
It is true that the theoretical consistency conditions can be modified 
if there are infinite numbers of massive particles. 
However, we do not consider such theoretical conditions involving massive
sectors, since they are highly dependent on the UV spectrum above 
the cut-off scale.

%% No winding mode

In particular, string theories predict so-called winding modes. 
These are massive excitations of a ``string'' in which 
a ``string'' winds around the circle of a torus.  
We do not require theoretical consistency conditions 
that involve the contributions from winding modes\footnote{
Winding modes are equivalent to Kaluza--Klein modes through the
T-duality in string theories. 
Thus, the gauge theories on D7-branes or on D3-branes are 
essentially ten-dimensional gauge theories, in some aspects, in the
presence of the winding modes. 
In particular, the hexagonal anomalies of ten-dimensional
gauge theories are required to vanish, which is the case when both 
D7-branes and D3-branes have ``ten-dimensional'' SO(32) gauge theory 
with sixteen SUSY charges (or its spontaneous breakdown) \cite{SO-32,GP}.
This provides an independent reason for the total number of D7-branes
and D3-branes to be thirty-two within the flat six-dimensional torus 
${\bf T}^6$.
We do not respect this reason, since it is heavily dependent on the
presence of the winding modes, but we still respect the total number of
D-branes since we borrow ``orientifold planes'' from string theories,
which might not be fully justified within pure supergravity, as we  
discussed in section \ref{sec:geometry}. 
However, the total number of D-branes is not an important constraint 
for our model construction.}. 
This is because those conditions depend on the UV spectrum above the cut-off
scale, which we do not specify, and are not conditions of 
the physics below the cut-off scale.

%%%%%%%%%%%%%%%%%%%%%%%%%%%%%%%%%%%%%%%%%%%%%%%%%%%%%%%%%%%%%%%
%%% D7 configuration and orbifold projection
%% 12 D7-branes on the two fixed loci away from O7
We use twelve out of thirty-two D7-branes 
to provide the SU(5)$_{\rm GUT}$ vector multiplet.
They should be put on (seven+one)-dimensional loci fixed under the 
${\bf Z}_{12}\vev{\sigma}$ orbifold group, so that the
vector multiplet with sixteen SUSY charges is projected out except for
a four-dimensional ${\cal N}$ = 1 SU(5)$_{\rm GUT}$ vector multiplet (without
chiral multiplets in the SU(5)$_{\rm GUT}$-{\bf adj.} representation).
Six of them are put at a fixed locus $z_3=\frac{{\bf e}_8+2 {\bf
e}_9}{3}$ and six remaining D7-branes are at 
the ${\bf Z}_2\vev{\Omega R_{89}}$-image of the fixed locus, 
i.e. at $z_3 = -\frac{{\bf e}_8+ 2{\bf e}_9}{3}$.
The reason why we put six D7-branes at each fixed locus 
rather than five will become clear later in this subsection
\ref{subsec:u2-config}  
(it is because we require that $\bar{Q}^6_{\;\; \alpha}$ and 
$Q^\alpha_{\;\; 6}$ are obtained from the D3--D7 system).

The twenty D7-branes that have not been used are placed at the other fixed
locus $z_a = {\bf 0}$ or are floating in the bulk.
Their existence is irrelevant to the dynamics of the 
SU(5)$_{\rm GUT}$-symmetry breaking, while they provide a room 
for constructing the SUSY-breaking sector, 
inflation sector and some other gauge theories we do not know yet.

%%% relation to F-theory
Once the configuration of 7-branes is fixed, we can calculate
the behaviour of the dilaton VEV through equations of motion.
In particular, the F-theory \cite{F-theory} implies that  
$\tau \equiv (C_{(0)}+ie^{-\phi})$ goes to $i \infty$ at the fixed loci 
where we put six D7-branes \cite{D3-probe}.
However, equations of motion derived from the Type IIB supergravity is not
reliable in the vicinities of D7-branes within $1/M_*$ (since we do not specify
short-distance physics above the cut-off scale $M_*$), and hence 
it does not make sense to discuss the short-distance behaviour of the
dilaton VEV.
Moreover, a precise relation is not known in supergravities between 
the dilaton VEV around the D7-branes 
and the effective coupling constant of the gauge theories 
on D7-branes\footnote{See also the question raised 
in the introduction of \cite{Seiberg}.}.
Therefore, we cannot discuss what the natural values are 
for the effective coupling constants $1/\alpha_*$ and 
$(M_*^\delta {\rm volume})/\alpha_*$ in (\ref{eq:weak-SU(5)}).
The relative ratio, however, is reasonable, since gauge theories 
on D7-branes become non-dynamical in the large-volume limit.

%% orientifold projection
Now, the fields on the D7-branes at $z_3 = \frac{{\bf e}_8+2{\bf
e}_9}{3}$ are identified with their image at 
$z_3 = - \frac{{\bf e}_8+2{\bf e}_9}{3}$ under the projection condition 
associated with the $\Omega R_{89}$.
Thus, we only need discuss the fields on one of these images.
Those fields are a U(6) vector multiplet $\Sigma^k_{\;\; l}(x,{\bf y'})$ 
($x\in {\bf R}^{3,1}$, ${\bf y}'=(z_1,z_2) \in {\bf C}^2$) 
of the ${\cal N}$ = 1 SUSY in eight-dimensional space-time.
Fields contained in this multiplet are also described in terms of 
four-dimensional ${\cal N}$ = 1 SUSY: one vector multiplet 
$(\Sigma_0)^k_{\;\; l}(x,{\bf y'},\theta)\equiv 
{\cal W}_\alpha^{\rm U(6)} (x,{\bf y'},\theta)$ and three chiral multiplets 
$(\Sigma_b)^k_{\;\; l}(x,{\bf y'},\theta)$ ($b=1,2,3$).

%% rotational property
Let us see how those fields transform\footnote{Ref. 
\cite{IWY-ss} might be useful again.} under the rotational symmetry 
of ${\bf C}^3$ before we discuss the orbifold projection conditions.
Let us consider three independent rotations $z_b \mapsto e^{i \alpha_b}
z_b$ for ($b=1,2,3$). 
The four-dimensional gauge-field strength in $\Sigma_0$ 
is a singlet under the rotation of extra-dimensional space. 
The complex scalars of $\Sigma_{b=1,2}$ receive the phase factors 
$e^{i \alpha_b}$  since they originate from polarizations 
of gauge fields in higher-dimensional space pointing at extra dimensions. 
Four fermions (Weyl in four-dimensional space-time) in $\Sigma_a$'s 
($a=0,...,3$) receive phase factors $e^{i \tilde{\alpha}_a}$ due to the
Lorentz transformation of the higher-dimensional space-time 
in the spinor representation, where $\tilde{\alpha}_0 \equiv
(\alpha_1+\alpha_2+\alpha_3)/2$ and $\tilde{\alpha_b}\equiv\alpha_b -
\tilde{\alpha}_0$ for ($b=1,2,3$). Thus, in terms of 
four-dimensional ${\cal N}$ = 1 superfields, they transform as
\begin{eqnarray}
 \Sigma_0(x,{\bf y}',\theta) &\mapsto &e^{i\tilde{\alpha}_0}
 \Sigma_0(x,\tilde{{\bf y}'},e^{-i\tilde{\alpha}_0}\theta), 
 \label{eq:rot-D7-0}\\
 \Sigma_{b}(x,{\bf y}',\theta) &\mapsto & e^{i \alpha_b}
 \Sigma_{b}(x,\tilde{{\bf y}'},e^{-i\tilde{\alpha}_0}\theta) \quad {\rm for~} 
b=1,2,
 \label{eq:rot-D7-12}
\end{eqnarray}
where $\tilde{{\bf y}'}$ is a point that is moved to the point ${\bf
y}'$ under the rotation, i.e. 
$\tilde{{\bf y}'}=(e^{-i\alpha_1}z_1,e^{-i\alpha_2}z_2)$ and  
${\bf y}'$ = ($z_1$,$z_2$). The complex scalar component of $\Sigma_3$
receives the phase factor $e^{i\alpha_3}$, so that 
\begin{equation}
 \Sigma_3(x,{\bf y}',\theta) \mapsto e^{i \alpha_3}
 \Sigma_3(x,\tilde{{\bf y}'},e^{-i\tilde{\alpha}_0}\theta),
 \label{eq:rot-D7-3}
\end{equation}
similarly to $\Sigma_{b=1,2}$. This choice of the phase factor is to
ensure that the sixteen-SUSY-charge symmetric interaction 
in the superpotential,
\begin{equation}
  d^2 \theta \left( W = \sqrt{2}g_{U(6)} 2 \tr (\Sigma_2[\Sigma_3,\Sigma_1])
             \right),
\end{equation}
is invariant under the three independent rotations; indeed, we see that 
\begin{equation}
 \Sigma_1 \Sigma_2 \Sigma_3 (\theta) \mapsto 
 e^{i (\alpha_1+\alpha_2+\alpha_3)}\Sigma_1 \Sigma_2 \Sigma_3 
(e^{-i(\alpha_1 + \alpha_2 + \alpha_3)/2}\theta). 
\end{equation}

%% orbifold projection on D7
Let us now turn to the orbifold projection.
The orbifold projection associated with ${\bf
Z}_{12}\vev{\sigma}$ extracts only singlets of a ${\bf Z}_{12}$ symmetry 
generated by a $\sigma$-transformation of fields, and all other states 
are projected out of the theory.
Namely, the following condition is imposed:
\begin{equation}
\Sigma_a(x,{\bf y'},\theta)^k_{\;\; l} =  
\widetilde{\Sigma_a}(x,{\bf y'},\theta)^k_{\;\; l}, \quad {\rm where}\quad
\sigma :\Sigma_a(x,{\bf y'},\theta)^k_{\;\; l} \mapsto 
\widetilde{\Sigma_a}(x,{\bf y'},\theta)^k_{\;\; l}.
\label{eq:proj-D7-sigma} 
\end{equation}
The whole sector comprised of all singlets of a symmetry
consistently becomes a self-closed theory. 
(Singular points of the geometry should be treated carefully, 
since the field theories are not well defined there. 
This is the subject of the next subsection \ref{subsec:u2-anomaly}.) 
Here, the $\sigma$-transformation of fields is primarily determined by
the geometric rotation (\ref{eq:generator}). However, we also have one
degree of freedom in determining the $\sigma$-transformation of fields
on the geometry --- the geometric rotation of fields can be accompanied by
a non-trivial twist through a rigid gauge transformation
by $\tilde{\gamma}_{\sigma;7} \in$U(6).
Notice that the $\sigma$-transformation is still an exact symmetry 
of the unorbifolded theory.
Now, the $\sigma$-transformation is given as follows:
\begin{eqnarray}
 \sigma : &&(\Sigma_0)^k_{\;\; l}(x,{\bf y}',\theta) \mapsto  
 (\widetilde{\Sigma_0})^k_{\;\; l}(x,{\bf y}',\theta)
\equiv                \qquad    (\tilde{\gamma}_{\sigma;7})^k_{\;\; k'} 
(\Sigma_0)^{k'}_{\;\; l'}(x,\sigma^{-1} \cdot {\bf y}',\theta)
(\tilde{\gamma}^{-1}_{\sigma;7})^{l'}_{\;\; l}, \label{eq:trsf-D7-sigmaI}\\
 \sigma : &&(\Sigma_b)^k_{\;\; l}(x,{\bf y}',\theta) \mapsto  
 (\widetilde{\Sigma_b})^k_{\;\; l}(x,{\bf y}',\theta)
\equiv e^{2\pi i v_b }(\tilde{\gamma}_{\sigma;7})^k_{\;\; k'} 
(\Sigma_b)^{k'}_{\;\; l'}(x,\sigma^{-1} \cdot {\bf y}',\theta)
(\tilde{\gamma}^{-1}_{\sigma;7})^{l'}_{\;\; l}, \label{eq:trsf-D7-sigmaII}
\end{eqnarray}
where $\tilde{\alpha}_a = 2\pi \tilde{v}_a$ are substituted into 
Eqs. (\ref{eq:rot-D7-0})--(\ref{eq:rot-D7-3}) and 
we take the 6 by 6 unitary matrix\footnote{
This 6 by 6 unitary matrix $\tilde{\gamma}_{\sigma;7}$ is related to 
32 by 32 unitary matrices  $\gamma_{\sigma;D7}$ found in references such
as \cite{GP,DM,IIB-orbifold} through 
\begin{equation}
 \gamma_{\sigma;D7} = \tilde{\gamma}_{\sigma;7} \oplus 
                      \tilde{\gamma}_{\sigma;7}^{-1} 
\oplus (20 {\rm ~by~} 20 {\rm ~matrix}).
\label{eq:32forD7}
\end{equation}
The unitary matrix $\gamma_{\Omega R_{89};D7}$ associated 
to the projection condition of $\Omega R_{89}$ is expressed in this
basis as
\begin{equation}
 \gamma_{\Omega R_{89};D7} = {\bf 1}_{6 \times 6}\otimes 
\left(\begin{array}{cc}
  0 & 1 \\ 1 & 0  \end{array} \right) \oplus {\bf 1}_{10 \times 10}\otimes 
\left(\begin{array}{cc}
  0 & 1 \\ 1 & 0        \end{array} \right).
\end{equation}
} $\tilde{\gamma}_{\sigma;7}$ as  
\begin{equation}
 (\tilde{\gamma}_{\sigma;7})^k_{\;\; l} = \diag \left( 
\overbrace{ e^{-\frac{1}{12}\pi i},... , e^{-\frac{1}{12}\pi i} }^5,
          - e^{-\frac{1}{12}\pi i}\right).
\label{eq:D7-twist-u2}
\end{equation}
First of all, in the above equations,
Eqs. (\ref{eq:trsf-D7-sigmaI}) and (\ref{eq:trsf-D7-sigmaII}), 
the Grassmann variable $\theta$ of the four-dimensional ${\cal N}$ = 1
SUSY does not receive a phase rotation
because $\sigma \in$ SU(3) $\subset$ SU(4) $\simeq$ SO(6) 
(i.e. $\tilde{\alpha}_0 = 2\pi \tilde{v}_0 = 0$).
This ensures \cite{CHSW} that the four-dimensional ${\cal N}$ = 1 SUSY 
is preserved in the spectrum obtained from the orbifold projection conditions
(\ref{eq:proj-D7-sigma}).
Secondly, we add a twist in the orbifold projection 
through the U(6) rigid transformation 
by $\tilde{\gamma}_{\sigma;7}$, which causes U(6)-symmetry breaking.

%% Reason for choosing specific rigid U(6)-twist
The specific choice of $\tilde{\gamma}_{\sigma;7}$ in
Eq. (\ref{eq:D7-twist-u2}) is based on the following reasons. 
First, the twelve times repeated application of $\tilde{\gamma}_{\sigma;7}$ 
leads to an adjoint action by $\tilde{\gamma}_{\sigma^{12};7} \equiv 
(\tilde{\gamma}_{\sigma;7})^{12} = -{\bf 1} \propto {\bf 1}$, 
whose effect is trivial.
This is a required property\footnote{
The $\gamma_{\sigma;D7}$ given in (\ref{eq:32forD7}) satisfies 
an algebraic constraint 
\begin{equation}
(\gamma_{\sigma;D7}) (\gamma_{\Omega R_{89};D7}) (\gamma_{\sigma;D7})^{T} 
= (\gamma_{\Omega R_{89};D7}),
\end{equation}
which is also required in \cite{DM}.}, since $\sigma^{12}$ acts
trivially on the space ${\bf T}^6$.
Secondly, the sixth diagonal entry of the $\tilde{\gamma}_{\sigma;7}$ 
is chosen differently from the first five entries 
in order to break the U(6) gauge symmetry down to U(5)$\times$U(1). 
Then, it follows from the requirement 
$(\tilde{\gamma}_{\sigma;7})^{12}\propto {\bf 1}$ that only 
the allowed difference is the twelfth root of unity 
between the sixth and other entries.
Thirdly, we avoid the phase difference  
$\{e^{2 \pi i n/12}|n=0,1,4,5,7,8,11\}$, since we do not want
SU(5)-charged matter particles to appear on the D7-branes.
This is first because the elementary Higgs particles $H^i({\bf 5})$ and
$\bar{H}_i({\bf 5^*})$ are not necessary in the SU(5)$_{\rm GUT}
\times $U(2)$_{\rm H}$ model, and second because we give up trying 
to obtain whole matter contents of the quarks and leptons from
D-branes\footnote{It is quite a difficult subject to obtain 
at the same time 
(i) {\em three families of} quarks and leptons,\\
(ii) a SUSY SU(5) unified gauge group, and finally 
(iii) a sector that breaks the the unified symmetry,
from open strings on D-branes. 
There are some trials as follows, yet they are not satisfactory.
The non-SUSY standard model is obtained in \cite{Madrid}, 
the SUSY standard model is obtained in \cite{Penn} with exotic chiral
multiplets, and the flipped SU(5)-unified model is obtained 
in \cite{CERN} with some necessary particles missing. 
Note also that all these models listed above are constructed 
using the intersecting D6--D6 system \cite{D6-D6} 
in the Type IIA string theory. 
This framework is not in a simple T-dual to the D3--D7 system 
in the Type IIB string theory.}.
Fourthly, we do not use the phase difference $\{e^{2\pi i
n/12}|n=3,9\}$, since we want the U(6) symmetry to be restored, at least, 
at the $\sigma^6$-projection, as we explain later in this subsection 
\ref{subsec:u2-config}.
Finally, among the remaining candidates\footnote{The ${\bf Z}_6(1/6,1/6,-2/6)$
and ${\bf Z}'_6(1/6,-3/6,2/6)$, which have (seven+one)-dimensional fixed loci 
away from O7-planes, do not have such candidates. For the ${\bf Z}_6$, 
$\{e^{2\pi i n/6}|n=0,1,2,4,5\}$ are excluded from the third criterion
in the text, but the fourth criterion requires 
$\{e^{2\pi i n/6}|n=0,2,4\}$.
For the ${\bf Z}'_6$ orbifold, all possibilities of choosing a phase
difference in $\tilde{\gamma}$ are excluded by the third
criterion. In short, both orbifolds have too simple structure to
fulfil our requirements. The ${\bf Z}'_6$ orbifold is disfavoured also by
anomaly arguments, as mentioned briefly in subsection
\ref{subsec:u2-anomaly}.} $\{e^{2\pi i n/12}|n=2,6,10\}$, we use $n=6$,  
since two other possibilities are excluded by anomaly arguments
in subsection \ref{subsec:u2-anomaly}.
The overall phase $e^{-\pi i /12}$ in $\tilde{\gamma}_{\sigma;7}$ is not
important since the $\tilde{\gamma}_{\sigma;7}$ acts on $\Sigma_a$'s
through {\bf adj.} representation of the U(6). 
This choice is just to make the expression similar to the
conventions \cite{GP,DM,IIB-orbifold} among string theorists.

%For the projection condition associated with the generator $\Omega
%R_{89}$, we adopt the same projection condition as the one common in
%string theories.
%That is,
%%% \begin{equation}
%%%  \Omega R_{89} : (\Sigma_a)_{ij}(x,{\bf y}') \mapsto 
%%% - 
%%% (\gamma_{\Omega R_{89};7})_{ii'} (\Sigma_a)_{j'i'}(x,{\bf y}')
%%% (\gamma^{-1}_{\Omega R_{89};7})_{j'j} 
%%% \equiv (\widetilde{\Sigma_a})_{ij}(x,{\bf y}'),
%%% \end{equation} 
%%% where
%%% \begin{equation}
%%%  (\gamma_{\Omega R_{89};7})_{ii'} = \left( \begin{array}{cc}
%%%        {\bf 0}  & {\bf 1} \\ {\bf 1} & {\bf 0} 
%%%        \end{array}\right),
%%% \end{equation}
%with each ${\bf 0}$ or ${\bf 1}$ is a diagonal $7 \times 7$ matrix,
%and the following projection condition is imposed:
%%% \begin{equation}
%%%  (\Sigma_a)_{ij}(x,{\bf y}') = 
%%% (\widetilde{\Sigma_a})_{i j}(x,{\bf y}').
%%% \label{eq:proj-D7-omega}
%%% \end{equation}
%Since the fixed locus $z_3 = \frac{{\bf e}_8+2{\bf e}_9}{3}$ is moved to 
%the one at $z_3 = \frac{2{\bf e}_8+{\bf e}_9}{3}$ by $R_{89}$, D7-branes
%with the i=1,...,7 should be identified with the D7-branes with
%i=8,...,14, which is realized by the matrix $\gamma_{\Omega R_{89};7}$.
%Twice repeated action of the $\gamma_{\Omega R_{89};7}$ is unity, which
%is also a desirable feature because of $(\Omega R_{89})^2=1$.

%% Surviving KK o-modes, no quarks and leptons 
Only the U(5)$\times$U(1)$_6$ vector multiplet of the four-dimensional
${\cal N}$ = 1 SUSY survive these orbifold projection 
in the low-energy spectrum.
% No Higgs multiplet appears after the projection in this model.
We identify the SU(5) part of the U(5) $\simeq$ SU(5)$ \times$U(1)$_5$ 
gauge group with the SU(5)$_{\rm GUT}$ gauge group. 
Some linear combination of the U(1)$_5 \times$U(1)$_6$ is identified
with the fiveness\footnote{The fiveness is equivalent to B$-$L in the
standard model.} gauge symmetry in subsection \ref{subsec:u2-phen}.
There is no massless chiral multiplet arising from D7-branes. 
Although we do not need Higgs multiplets 
in the SU(5)$_{\rm GUT} \times$U(2)$_{\rm H}$ model,  
quarks and leptons (and right-handed neutrinos) are missing. 
%There are string based \cite{LPT} or non-string based 
%\cite{WY-anarchy} trials to understand the origin of the quarks and
%leptons in a set-up similar to the present model (i.e. D3--D7 system,
%not the intersecting D6-D6 system), but it is not easy 
%to find straightforward connection between the SU(5)$_{\rm GUT}$ 
%breaking models discussed here and those models discussing quarks and leptons.
It is not easy to accommodate all the three families of quarks and
leptons along with the model of SU(5)$_{\rm GUT}$ breaking we discuss. 
We consider that the three families of quarks and leptons reside at a
fixed point of the orbifold geometry, although we cannot specify their
origin.

%%%%%%%%%%%%%%%%%%%%%%%%%%%%%%%%%%%%%%%%%%%%%%
%%% D3--D7 system
%%  D3 should be put at N=2 fixed point
Let us now derive from the D3--D7 bound state 
the matter contents of the SU(5)$_{\rm GUT}$-breaking sector. 
D3-branes are put on a fixed point that preserves ${\cal N}$ = 2
SUSY \cite{WY-ss}. 
We refer such fixed points to the ${\cal N}$ = 2 fixed points.
There would be unwanted SU(2)$_{\rm H}$-{\bf adj.} chiral multiplets 
if the D3-branes were not put at a fixed point; however,
the multiplet structure of the ${\cal N}$ = 2 SUSY would be lost 
if they were put on a fixed point the ${\cal N}$ = 2 SUSY is not preserved.

%% N=2 fixed point is a condition on isotropy group
SUSY of local geometry at a given fixed point is determined by 
its isotropy group, that is a subgroup of the orbifold group 
that fixes the point.
The isotropy group determines the local geometry around the fixed point, 
and hence the SUSY.
Matter contents from D-branes located at that fixed point are also
determined by imposing orbifold projection conditions associated  
to the isotropy group.

%% isotropy group in SU(2)
The ${\cal N}$ = 2 fixed points should have isotropy group contained 
in a particular SU(2) subgroup of the SU(4) $\simeq$ SO(6) rotation.
This is because the local geometry, which is determined by the isotropy
group, should preserve the ${\cal N}$ = 2 SUSY of the D3--D7 system.
Since the SUSY charges of this system are in the 
$-\Gamma^{7654}=1$ eigenspace, the isotropy group should act only on 
the $-\Gamma^{7654} \neq 1$ eigenspace. 
It is the SU(2) rotation of the two complex planes ${\bf C}^2=\{
(z_1,z_2)|z_1,z_2 \in {\bf C} \}$ that acts on this eigenspace; 
indeed this SU(2) subgroup (one of SU(2)$\times$SU(2) 
$\simeq$ (SO(4) of ${\bf C}^2$)) is given by
\begin{equation}
 \exp \left( \phi_a \bar{\eta}^a_{m'n'}\Gamma^{m'n'}\right),
\end{equation}
where $\bar{\eta}^a_{m'n'}$ is the 't Hooft $\eta$-symbol \cite{eta-symbol} 
with $a=1,2,3;m',n'=4,5,6,7$, whose generators 
$\bar{\eta}^a_{m'n'}\Gamma^{m'n'}$ are trivial on the
$-\Gamma^{7654}=1$ eigenspace:
\begin{equation}
 \bar{\eta}^a_{m'n'}\Gamma^{m'n'} 
= \bar{\eta}^a_{m'n'}\Gamma^{m'n'}(-\Gamma^{7654}) 
= -  \bar{\eta}^a_{m'n'}\Gamma^{m'n'},
\end{equation}
since $\bar{\eta}^a_{m'n'}\epsilon_{m'n'k'l'}=- 2 \bar{\eta}^a_{k'l'}$.

%% $\sigma^4$- and $\sigma^6$-fixed are candidates
There are essentially two different candidates of the ${\cal N}$ = 2 fixed
points on the D7-branes: 
two points at which the isotropy group is ${\bf Z}_2\vev{\sigma^6}$ and
a point at which the isotropy group is ${\bf Z}_4\vev{\sigma^3}$.
All of them\footnote{Intersection of the two-dimensional singularity at
${\bf y}'$={\bf 0} with the D7-branes, however, has only ${\cal N}$ = 1
SUSY, because the isotropy group is ${\bf Z}_{12}\vev{\sigma}$ there.}
are at the intersections of the two-dimensional singularities in the
{\bf T}$^6$ with the D7-branes. 
Note that $\sigma^3,\sigma^6$ and $\sigma^9$ belong to 
the above-mentioned SU(2) subgroup, 
since it does not rotate the third complex plane spanned 
by ${\bf e}_8$ and ${\bf e}_9$ because of $v_3 \times 3 \in {\bf Z}$.

%% U(6)-fixed point
We put the D3--D7 bound state at an ${\cal N}$ = 2 fixed point where the
original U(6) symmetry is enhanced on the D7-branes. 
This is because the ${\cal N}$ = 2 hypermultiplet 
$(\bar{Q}^6_{\;\; \alpha},Q^\alpha_{\;\; 6})$ is necessary 
in the SU(5)$_{\rm GUT}$-breaking sector in addition to 
the hypermultiplet $(\bar{Q}^i_{\;\; \alpha},Q^\alpha_{\;\; i})$ ($i=1,...,5$).
The U(6) symmetry would be enhanced only at fixed points with isotropy group
${\bf Z}_3\vev{\sigma^4}$, where there is only ${\cal N}$ = 1 SUSY, 
if we were to adopt the phase difference 
$\{e^{2\pi i n/12}|n=3,9\}$ in (\ref{eq:D7-twist-u2}). 
Thus, there would be no fixed point where the ${\cal N}$ = 2 SUSY 
and the U(6) symmetry are simultaneously obtained. 
This is the reason why we reject the phase difference $\{e^{2\pi i n
/12}|n=3,9\}$ in the $\tilde{\gamma}_{\sigma;7}$.
On the contrary, when we take the phase difference 
$\{ e^{2\pi i n/12}|n=2,6,10\}$, the symmetry is enhanced up to U(6) 
at fixed points with isotropy group ${\bf Z}_{2}\vev{\sigma^6}$, 
where the ${\cal N}$ = 2 SUSY is also restored. 
%If the phase difference is $e^{2\pi i 6/12}$, as we take, 
%the U(6) is realized at fixed points whose isotropy group is contained 
%in ${\bf Z}_6\vev{\sigma^2}$.
%After all, the symmetry cannot be enhanced to U(6) on the ${\cal N}$ = 2 
%fixed point with isotropy group ${\bf Z}_4\vev{\sigma^3}$(, 
%though not in the case for those with isotropy group 
%${\bf Z}_2\vev{\sigma^6}$).

%% U(6)- and N=2 fixed point ... only $\sigma^6$-isotropy 
Therefore, we put two D3-branes on one of the fixed points 
where the isotropy group is ${\bf Z}_2\vev{\sigma^6}$.
There are two such fixed points on the 
${\bf T}^6/({\bf Z}_{12}\vev{\sigma}\times{\bf Z}_2\vev{\Omega
R_{89}})$, and the resulting phenomenology is 
different. However, we postpone choosing one from these two candidates
until subsection \ref{subsec:u2-phen}, since they make no difference 
in the theoretical construction of the models.
Such a fixed point, whichever one chooses, consists of twelve points in 
the covering space ${\bf T}^6$, 
which are identified under the $({\bf Z}_{12}\vev{\sigma} \times
{\bf Z}_2\vev{\Omega R_{89}})/{\bf Z}_2\vev{\sigma^6}$.
Thus, two D3-branes have to be introduced at each of these twelve points. 
Twenty-four D3-branes are necessary as a whole in the covering space
${\bf T}^6$.
The rest of the D3-branes can be used for other sectors, which are not
directly observed today.

%% Matter contents before orbifold projection
We concentrate on a set of fields at one of these twelve images
on which the orbifold projection associated with the isotropy group 
${\bf Z}_2\vev{\sigma^6}$ is imposed:
those fields are a U(2) vector multiplet $(X)^\alpha_{\;\;\beta}$ 
($\alpha,\beta=4,5$) of the four-dimensional ${\cal N}$ = 4 SUSY 
and a hypermultiplet of the four-dimensional ${\cal N}$ = 2 SUSY 
in the ({\bf 6},{\bf 2}$^*$) representation of the U(6)$\times$U(2) 
gauge group.
They are decomposed into irreducible multiplets of the ${\cal
N}$ = 1 SUSY: a U(2) vector multiplet 
$(X_0)^\alpha_{\;\;\beta}(x,\theta) \equiv 
({\cal W}^{\U(2)})(x,\theta)$ and three chiral multiplets 
$(X_b)^\alpha_{\;\;\beta}(x,\theta)$ ($b=1,2,3$) in the U(2)-{\bf adj.}
representation, chiral superfields $Q^\alpha_{\;\; k}(x,\theta)$ in the ({\bf
6}$^*$,{\bf 2}) representation and $\bar{Q}^k_{\;\; \alpha}(x,\theta)$ in the 
({\bf 6},{\bf 2}$^*$) representation.

%% Rotational property
Let us see how these fields transform under the three independent
rotational symmetries on the ${\bf C}^3$ : $z_b \mapsto
e^{i\alpha_b} z_b$ for $b=1,2,3$.
The four chiral multiplets of four-dimensional ${\cal N}$ = 1 SUSY $X_a$'s,
which are under the control of sixteen SUSY charges, transform in the same way
as $\Sigma_a$'s on the D7-branes (except for ${\bf y}'$-dependence):
\begin{eqnarray}
 X_0(x,\theta) &\mapsto &e^{i\tilde{\alpha}_0}
 X_0(x,e^{-i\tilde{\alpha}_0}\theta), \\
 X_{b}(x,\theta) &\mapsto & e^{i \alpha_b}
 X_{b}(x,e^{-i\tilde{\alpha}_0}\theta) \quad {\rm for~} 
b=1,2,3.
\end{eqnarray}
The AdS/CFT correspondence on the D3-branes in the large 't Hooft
coupling limit provides sufficient evidence for this determination 
\cite{AdS/CFT1}.
Transformation properties of $(\bar{Q},Q)$ are determined through
field theoretical arguments. 
The four-dimensional ${\cal N}$ = 2 SUSY interaction, 
\begin{equation}
 W = \sqrt{2}g_{\rm U(2)} 
 \bar{Q}^k_{\;\; \alpha} (X_3)^{\alpha}_{\;\;\beta} Q^\alpha_{\;\; k},
\label{eq:N=2super}
\end{equation}
requires that the product $\bar{Q}Q$ transforms as 
\begin{equation}
 \bar{Q}Q(x,\theta) \mapsto e^{i(\alpha_1+\alpha_2)}
 \bar{Q}Q(x,e^{-i\tilde{\alpha}_0}\theta).
\end{equation}
On the other hand, vanishing U(2)[rotation]$^2$ anomalies require
that $\bar{Q}$ and $Q$ have the same rotational charge. Thus 
\begin{eqnarray}
 \bar{Q}(x,\theta) & \mapsto & e^{i (\alpha_1+\alpha_2)/2}
\bar{Q}(x,e^{-i\tilde{\alpha}_0}\theta),\\
 Q(x,\theta) & \mapsto & e^{i (\alpha_1+\alpha_2)/2}
 Q(x,e^{-i\tilde{\alpha}_0}\theta).
\end{eqnarray}

%% Orbifold projection
These fields transform under the $\sigma^6$ as
\begin{eqnarray}
 \sigma^6 : && (X_0)^\alpha_{\;\;\beta}(x,\theta) \mapsto
   (\widetilde{X_0})^\alpha_{\;\;\beta}(x,\theta) \equiv  \qquad
(\tilde{\gamma}_{\sigma^6;3})^\alpha_{\;\;\alpha'}
(X_0)^{\alpha'}_{\;\;\beta'}(x,\theta)
  (\tilde{\gamma}_{\sigma^6;3}^{-1})^{\beta'}_{\;\;\beta}  ,  
                                    \label{eq:trsf-D3-XI-U2}  \\
 \sigma^6 : && (X_b)^\alpha_{\;\;\beta}(x,\theta) \mapsto 
    (\widetilde{X_b})^\alpha_{\;\;\beta}(x,\theta) \equiv  e^{2\pi i v_b 6}
  (\tilde{\gamma}_{\sigma^6;3})^\alpha_{\;\;\alpha'}
  (X_b)^{\alpha'}_{\;\;\beta'}(x,\theta)
  (\tilde{\gamma}_{\sigma^6;3}^{-1})^{\beta'}_{\;\;\beta}, 
                                    \label{eq:trsf-D3-XII-U2} 
\end{eqnarray}
\begin{eqnarray}
 \sigma^6 : &&Q^{\alpha}_{\;\;k}(x,\theta) \mapsto 
     \widetilde{Q}^\alpha_{\;\;k}(x,\theta)  \equiv e^{\pi i (v_1+v_2) 6}
(\tilde{\gamma}_{\sigma^6;3})^\alpha_{\;\;\alpha'} 
Q^{\alpha'}_{\;\; k'}(x,\theta)
(\tilde{\gamma}_{\sigma;7}^{-6})^{k'}_{\;\; k},  \label{eq:trsf-D3-D7-Q-U2} \\
 \sigma^6 : && \bar{Q}^{k}_{\;\;\alpha}(x,\theta) \mapsto 
   \widetilde{\bar{Q}}^k_{\;\;\alpha}(x,\theta) \equiv  e^{\pi i (v_1 + v_2)6}
(\tilde{\gamma}_{\sigma;7}^{6})^{k}_{\;\; k'} 
 \bar{Q}^{k'}_{\;\; \alpha'}(x,\theta)
(\tilde{\gamma}_{\sigma^6;3}^{-1})^{\alpha'}_{\;\;\alpha},
                                                 \label{eq:trsf-D3-D7-Qbar-U2}
\end{eqnarray}
where we take the 2 by 2 unitary matrix for the rigid U(2)
transformation\footnote{
The 32 by 32 matrix $\gamma_{\sigma;D3}$ in
Refs. \cite{GP,DM,IIB-orbifold} is expressed in terms of the 
$\tilde{\gamma}_{\sigma^6;3}$ as 
\begin{equation}
\gamma_{\sigma;D3} = \left( \begin{array}{cccccc}
                      & & & & & 1 \\ 1 & & & & & \\ & 1 & & & & \\ 
                      & & 1 & & & \\ & & & 1 & & \\ & & & & 1 & 
                            \end{array}\right) \otimes  
               (\tilde{\gamma}_{\sigma^6;3})^{\frac{1}{6}} \oplus 
               \left( \begin{array}{cccccc}
                     & 1 & & & & \\ & & 1 & & & \\ & & & 1 & & \\
                     & & & & 1 & \\ & & & & & 1 \\ 1 & & & & & 
                            \end{array}\right) \otimes  
               (\tilde{\gamma}_{\sigma^6;3})^{-\frac{1}{6}} \oplus 
                ({\rm 8~by~8~matrix}). 
\label{eq:32forD3}
\end{equation}.} 
$\tilde{\gamma}_{\sigma^6;3}$ as
\begin{equation}
 \tilde{\gamma}_{\sigma^6;3} = \diag ( 
e^{-\frac{1}{2}\pi i},e^{-\frac{1}{2}\pi i}).
\end{equation}
Here, $(\tilde{\gamma}_{\sigma^6;3})^2 = - {\bf 1} \propto {\bf 1}$ is
satisfied. Phase $e^{- \pi i /2}$ is chosen so that the $(\bar{Q},Q)$
survive the orbifold projection conditions  
\begin{equation}
 X_a = \widetilde{X_a}, \quad Q = \widetilde{Q}, \quad 
                         \bar{Q} = \widetilde{\bar{Q}},
\label{eq:proj-D3-XQQbar}
\end{equation}
with $a=0,1,2,3$.

%% Surviving KK 0-modes
Massless modes that survive the orbifold projection are 
a U(2) vector multiplet ($X_0$,$X_3$) of the four-dimensional ${\cal
N}$ = 2 SUSY, and a hypermultiplet ($Q^\alpha_{\;\; 
k}$,$\bar{Q}^{k}_{\;\; \alpha}$) ($k=1,...,6$).
This is exactly the matter content of the SU(5)$_{\rm GUT}$-breaking
sector in the product-group unification model based on the 
SU(5)$_{\rm GUT} \times$U(2)$_{\rm H}$ gauge group.
An unwanted ${\cal N}$ = 2 hypermultiplet in the U(2)-{\bf adj.}
representation ($X_1$,$X_2$) has been projected out.

\subsection{Anomaly Cancellation}
\label{subsec:u2-anomaly}

There is no anomaly in the ten-dimensional bulk. 
However, anomalies generally arise at orbifold singularities.
This subsection is devoted to the analysis of such anomalies.
%%% Only ${\bf Z}_{12}\vev{\sigma}$-projection is mainly treated.
We mainly consider the anomalies due to the ${\bf Z}_{12}\vev{\sigma}$ 
projection. Anomalies due to the ${\bf Z}_2\vev{\Omega R_{89}}$
projection are briefly touched upon in the course of the discussion.

\subsubsection{Triangle Anomalies}
%%%  Fixed point wise cancellation
%%   Triangle vanish as a whole, but not necessarily at each fixed point
The matter contents we obtained below the Kaluza--Klein scale is free
from triangle anomalies. However, this only implies that the total sum 
of triangle anomalies localized at all the fixed points 
cancel one another.
Thus, anomalies at each fixed point can be non-zero. 
If it is the case, then there will be violation of unitarity 
at an energy scale higher than the Kaluza--Klein scale, 
which means that the description using higher-dimensional field theories 
is no longer valid.

Kaluza--Klein particles of higher-dimensional massless fields  
play a crucial role in the determination of the triangle anomaly 
distribution at fixed points, 
although the total sum of the anomaly is determined only from 
the four-dimensional massless particles.
This is intuitively obvious from the fact that the anomaly distribution
only from Kaluza--Klein zero modes is homogeneous over the orbifold,
because of their homogeneous wave functions. 
It is the Kaluza--Klein towers that collect and redistribute 
to fixed points the anomalies carried by zero
modes (see below for a more concrete explanation). 
The resulting distribution depends on the Kaluza--Klein spectrum.

Higher-dimensional massive excitations above the fundamental scale, 
if they exist, also change the anomaly distribution. 
Their existence can results in replacement of the anomaly 
that is once localized at a fixed point by the Kaluza--Klein towers 
to another fixed point.
This effect will be described by the Chern--Simons terms 
on the D7-branes after those massive excitations are integrated out 
\cite{CSeffectively}.
Thus, in general, one can expect that triangle anomalies vanish at all
fixed points whenever the total sum of these anomalies vanish; 
the anomaly distribution determined by the Kaluza--Klein towers of
higher-dimensional massless fields can be gathered at a single
fixed point to vanish in the presence of unknown massive excitations.

% N=4 in effective theory (with assumption) rules out the above possibility
However, the general possibility described above does not work
straightforwardly %\footnote{
%There {\em is} a way to replace (non-Abelian) triangle anomalies from a 
%fixed point to another. 
%It is an intersecting D6-D6 system \cite{D6-D6} 
%with suitable background field, which
%is introduced to preserve the SUSY \cite{bg-field-SUSY}.
%It might be possible to preserve the SUSY by introducing background
%field, since the D5-D7 system is T-dual to the intersecting D6-D6
%system.
%The anomaly cancellation is discussed in \cite{Witten-G2} in the context
%of the $G_2$-holonomy compactification of the M-theory, which is
%considered to be dual to the intersecting D6-D6 system in the Type IIA
%string theory.} 
in the presence of highly extended SUSY in the
extra-dimensional space (on D7-branes).
Let us discuss this issue by taking the effective theory description 
(i.e. massive particles are integrated out and 
only massless fields of the Type IIB supergravity (including D-brane
fluctuations) are used). 
It is true that the effective Chern--Simons term on the D7-branes, 
\begin{equation}
 \int_{D7} (dC_{(2)})\wedge \tr (A FF + \cdots), 
\label{eq:CS}
\end{equation}
which is allowed by the ${\cal N}$ = 1 SUSY of eight-dimensional space
\cite{bwb},  
replaces \cite{Witten-5brane,inflow} triangle anomalies\footnote{
Discussed in Ref. \cite{Witten-5brane} is the replacement of box anomalies, 
which can be done without breaking SUSY.} 
from a fixed point to another, provided the $C_{(2)}$ field 
has a background configuration such that 
\begin{equation}
 d (dC_{(2)}) = \sum_{{\bf y}'_*\in \{{\rm fixed~ points}\} }
              n_{{\bf y}'_*}\delta^4({\bf y}'-{\bf y}'_*),
\end{equation}
where ${\bf y}'$ is the coordinate on the D7-branes ${\bf y}'=(z_1,z_2)$
and $\delta^4({\bf y}'-{\bf y}'_*)$ denotes, here, 
a delta-function-supported 4-form.
However, this condition on the background means that
\begin{equation}
 \int_{S_3} dC_{(2)} = \int_{B_4} d(dC_{(2)}) = n_{{\bf y}'_*}, 
\end{equation}
where the $S_3$ is a 3-sphere in D7-branes surrounding a fixed point
${\bf y}'_*$ and $\partial B_4 = S_3$.
In other words, there are $n_{{\bf y}'_*}$ 5-branes intersecting the
D7-branes at ${\bf y}'_*$.
Since the SUSY charges preserved in the presence of 5-branes alone are 
\begin{equation}
 {\cal Q} - \Gamma^{7654}{\cal Q}',
\end{equation}
these SUSY charges contain no common subset with the SUSY charges 
on the D7-branes (\ref{eq:SUSY-D7}).
Thus, it is in SUSY-violating vacua 
that the triangle anomalies on the D3--D7 system can be replaced 
through the interaction (\ref{eq:CS}).

%%% Calculation of anomaly distribution
%%  ACG is the simplest way
Therefore, we require that the triangle anomalies vanish at all of the
fixed points on the D7-branes, in the distribution determined 
from the Kaluza--Klein spectrum.
The simplest way to calculate the triangle anomaly distribution 
is given in \cite{ACG}.

%% Intuitive introduction to ACG, and mechanism of fixed point localization
In the case of the $S^1/{\bf Z}_2$ orbifold \cite{ACG}, zero modes have 
distribution function $1/2$, the $n$-th excited Kaluza--Klein modes 
have distribution functions $\cos^2 (n y/L)$ or $\sin^2 (n y/L)$, with
anomaly coefficient opposite in sign. The total summation of all these
contributions from all the Kaluza--Klein towers, 
\begin{equation}
 \frac{1}{2} + \sum_{n > 0} \left( \cos^2\left(\frac{n y}{L}\right)
                                 - \sin^2\left(\frac{n y}{L}\right)\right),
\end{equation}
leads to a delta-function distribution of the anomaly 
supported on the $S^1/{\bf Z}_2$ fixed points.

%% Generalized ACG, expression by def.
Now the generalization to higher-dimensional orbifold is
straightforward.
The anomaly distribution is calculated as 
the total summation of absolute square of each Kaluza--Klein wave
function weighted by its anomaly coefficients:
\begin{equation}
 \sum_{I,a} A_I \frac{1}{12}
    \sum_{{\bf p}'\in \Lambda_0'} |\psi_{I,a,{\bf p}'}({\bf y}')|^2.
\label{eq:ACG}
\end{equation}
Kaluza--Klein towers are labelled by $I$ and $a$, 
where $a=0,1,2,3$ runs over all components of the SO(3,1)-irreducible 
decomposition of eight-dimensional Weyl fermions 
(the fermion contents of the U(6)-{\bf adj.} ${\cal N}$ = 1 vector
multiplet of the eight-dimensional space-time), 
while the $I$ runs over irreducible representations 
$I \in \{ {\bf 24}^{(0,0)}, {\bf 1}^{(0,0)},
           {\bf 5}^{(1,-1)},{\bf 5}^{*(-1,1)} \}$ (which form U(6)-{\bf adj.})
of the unbroken gauge group SU(5)$_{\rm GUT} \times$U(1)$_5
\times$U(1)$_6$.
The anomaly coefficient $A_I$ of a given type 
(such as [SU(5)$_{\rm GUT}$]$^3$, U(1)$_5$[SU(5)$_{\rm GUT}$]$^2$,
etc.)
is determined as usual: 
\begin{equation}
 \tr (\{ t^b_I , t^c_I \} t^a_I) = 
A_I \tr ( \{ t^b_I , t^c_I \} t^a_I))|_{I={\bf fund.}
{\rm ~of~SU(5)}_{\rm GUT}}. 
\end{equation}
% which is common within the same representation $I$. 
Each Kaluza--Klein particle in a Kaluza--Klein tower is labelled by its 
Kaluza--Klein momentum ${\bf p}'$. In (\ref{eq:ACG}) the $\Lambda'_0$  
denotes the dual lattice of the four-dimensional space lattice
$\Gamma'_0$ spanned by 
${\bf e}_4$, ${\bf e}_5$, ${\bf e}_6$, ${\bf e}_7$, over which the
D7-branes are stretched.
The Kaluza--Klein wave function on the ${\bf Z}_{12}\vev{\sigma}$
orbifold is given by
\begin{equation}
 \psi_{I,a,{\bf p}'} = \frac{1}{ \sqrt{12 {\rm vol}({\bf T}^4)}} 
     \sum_{g \in{\bf Z}_{12}\vev{\sigma}} \rho_{a}(g)
     \rho_I(g)
     e^{- i {\bf p}'\cdot (g \cdot {\bf y}')} 
\end{equation}
with phase factors $\rho_{a}(g)$ and $\rho_I(g)$ being fixed by the local
Lorentz rotation and by the U(6) rigid transformation associated with the
orbifold transformation $g\in{\bf Z}_{12}$, respectively. 
These two phase factors are explicitly given by 
\begin{equation}
 \rho_{a}(\sigma^k)  =  
      (e^{i \pi \tilde{v}_a})|_{a=0,1,2,3}
\label{eq:localLorentz4Dirr}
\end{equation}
for four Weyl fermions of SO(3,1); 
$\rho_I(\sigma^k)=(-1)^k$ for 
$I \in \{{\bf 5}^{(1,-1)},{\bf 5}^{*(-1,1)}\}$   
and $\rho_I(\sigma^k)=1$ for 
$I \in \{ {\bf 24}^{(0,0)}, {\bf 1}^{(0,0)}\}$.

%% Group element decomposition through generalized ACG
The distribution function of triangle anomalies (\ref{eq:ACG}) is
rewritten as 
\begin{equation}
 \frac{1}{12}\sum_{g \in {\bf Z}_{12}\vev{\sigma}} 
 \left( \frac{1}{{\rm vol}({\bf T}^4)}\sum_{{\bf p}'\in \Lambda'_0}
         e^{- i {\bf p}'\cdot (g \cdot {\bf y}'-{\bf y}')}\right)
 \sum_I A_I \rho_I(g) \sum_{a=0}^3 \rho_a(g),
\label{eq:decomposition}
\end{equation}
where algebraic relations 
$\rho_a(g')^* \rho_a(g) = \rho_a(g'^{-1}\cdot g)$ and 
$\rho_I(g')^* \rho_I(g) = \rho_I(g'^{-1}\cdot g)$ are used.
The Poisson formula tells us that 
\begin{equation}
   \frac{1}{{\rm vol}({\bf T}^4)} \sum_{{\bf p}'\in \Lambda_0'} 
                   e^{ -i (g^{-1}\cdot {\bf p}'- {\bf p}')\cdot {\bf y}'} 
 =  \frac{1}{|\Gamma'_g : \Gamma'_0|} \sum_{{\bf y}'_* \in \Gamma'_g} 
     \delta^{4}({\bf y}'-{\bf y}'_*).     
\label{eq:Poisson}
\end{equation}
Here, a momentum-space lattice 
$\Lambda'_g \equiv \{g^{-1} \cdot {\bf p}' - {\bf p}'|{\bf p}' 
\in \Lambda'_0\}$ is a superlattice of $\Lambda'_0$, and $\Gamma'_g$ in
the above expression denotes the sublattice of $\Gamma'_0$ 
dual to $\Lambda'_g$. This $\Gamma'_g$ is also characterized as 
a set of $g$-fixed points on the D7-branes. 
$|\Gamma'_g : \Gamma'_0|$ is the number of the
$g$-fixed points within the covering space ${\bf T}^4$, 
which is the world volume of the D7-branes before the orbifold projection. 
Thus, the integration of the (\ref{eq:Poisson}) over a
single unit cell of ${\bf T}^4$ yields unity.
Therefore, the triangle-anomaly distribution over the orbifold geometry 
is decomposed into parts, each of which corresponds 
to an element $\sigma^k $ of the orbifold group and is 
supported on the $\sigma^k$-fixed points.
The total amount of anomaly carried by this component is 
\begin{equation}
 \frac{1}{12} i \sum_{a=0}^3 \rho_a(\sigma^k)\sum_I (-i A_I \rho_I(\sigma^k)) 
= \frac{1}{12} 
   4\left(\prod_{b=1}^3 \sin (\pi v_b k) + 
         i \prod_{b=1}^3 \cos (\pi v_b k)
    \right)  \sum_I (-i A_I \rho_I(\sigma^k)) . 
\label{eq:triangle-coeff-k}
\end{equation}
$1/|\Gamma'_{\sigma^k}:\Gamma'_0|$ of (\ref{eq:triangle-coeff-k}) is
distributed at each $\sigma^k$-fixed point through the $\sigma^k$-component.
The anomaly localized at a fixed point is given by the sum of all such
$g$-component contribution, where the fixed point is $g$-fixed
The cosine part in this expression cancels with that of
the $\sigma^{12-k}$-component, and hence only the sine part is of importance.

%% Vanishing triangles at all fixed points in the case of $e^{2\pi i (n=6)/12}$
Now it is easy to see that all triangle anomalies vanish at all fixed
points on the D7-branes in the SU(5)$_{\rm GUT} \times$ U(2)$_{\rm H}$ model 
described in the previous
subsection \ref{subsec:u2-config}. It is sufficient, though not necessary, 
to see that all $(-i \sum_I A_I \rho_I(\sigma^k))$ vanish for $k=0,...,11$.
It is indeed the case, since 
$\rho_{I={\bf 5}^{(1,-1)}}(\sigma^k) = \rho_{I={\bf
5}^{*(-1,1)}}(\sigma^k) = (-1)^k$, and 
$A_{I={\bf 5}^{(1,-1)}} = - A_{I={\bf 5}^{*(-1,1)}}$ for all types of
triangle anomalies between SU(5)$_{\rm GUT}$, U(1)$_5$, U(1)$_6$ and gravity.
It is extremely encouraging that there is a set of geometry and 
U(6)-twisting matrix $\tilde{\gamma}_{\sigma;7}$ where the triangle
anomalies vanish at all fixed points. 

%% The reason why we rejected $n=\pm 2$ above
If we were to take $e^{2\pi i (n=\pm 2)/12}$ as the phase difference 
in $\tilde{\gamma}_{\sigma;7}$, rather than $e^{2\pi i (n=6)/12}$ as in the
text, it turns out that [SU(5)$_{\rm GUT}$]$^3$ triangle anomalies 
are distributed at all fixed points contained in $\Gamma'_{\sigma^{\pm 4}}$ 
and $\Gamma'_{\sigma^{\pm 1}} = \Gamma'_{\sigma^{\pm 2}} 
= \Gamma'_{\sigma^{\pm 5}}$ by a fractional amount that cannot be 
cancelled by introducing new particles at fixed points.
%the $\sigma^4$- and $\sigma^8$-components do not vanish.
%Since the $\sigma^4$-fixed points are fixed only by
%$\{\sigma^k|k=0,4,8\}$, we can conclude in these cases 
%that there are fixed points where non-vanishing triangle anomalies appear.
This is the reason why we do not adopt $e^{2\pi i
(n = \pm 2)/12}$ as the phase difference in (\ref{eq:D7-twist-u2}).

\subsubsection{More Anomalies}
\label{subsubsec:u2-anomaly-more}

%%% Two issues, yet they are UV dependent
Now that the triangle anomalies vanish at all of the fixed points, the next
question is whether they are the only ones that we have to care about or not.
We point out in the rest of this subsection \ref{subsec:u2-anomaly} 
that there are two other issues we have to discuss
--- higher-dimensional anomalies and ``anomalies in internal dimensions''. 
In the end, however, it is concluded that two consistency conditions that
come from those two issues depend on UV physics, and hence there is a
chance that these conditions would be satisfied by choosing suitable UV
physics. Therefore, they are not necessary conditions  for physics below the
cut-off scale.

%%% Higher-dimensional anomalies
Let us first discuss higher-dimensional anomalies.
Although the ten-dimensional space-time is directly compactified to
the four-dimensional space-time, singular loci, on which anomalies may be
localized, are not necessarily (three+one)-dimensional sub-space-time.
The ${\bf T}^6/{\bf Z}_{12}\vev{\sigma}$-compactified ten-dimensional
space-time possesses a couple of (five+one)-dimensional  singularities, 
on which box anomalies may appear.

%%% We have to care about higher-dimensional anomalies
First, we show that the box anomalies {\em do} appear 
at those  singularities. Indeed, six-dimensional (box) 
anomalies arise on the same footing as the triangle anomalies  
in the Fujikawa method \cite{Fujikawa} extended to
orbifold-compactified geometry \cite{BCCRS}.
Massless particles in the bulk and massless particles on D-branes are
already specified, and hence it is a well-defined question 
how the box anomalies arise from these particles. 
%Those anomalies are calculated, (which in turn provides a constraint 
%to other unknown sectors that also contribute those anomalies.)
%%%%%  %%%  Fujikawa method
%%%%%  %%   Only pure gauge anomalies are treated in Fujikawa method
%% We discuss pure gauge anomalies as examples to see how box anomalies
%% arise in the Fujikawa method. 
We first show that the distribution of the triangle anomalies 
are re-obtained in the Fujikawa method and then discuss 
how the box anomalies arise. 
We use non-gravitational (pure gauge) anomalies as examples,  
just for illustration.
Non-gravitational anomalies only arise from gauge-charged Weyl fermions, 
and hence it is easy to calculate them.
Anomalies involving gravity can also be treated in a similar way 
by regarding the gravity as local Lorentz gauge theory. 
Actual calculation of gravitational anomaly is presented later.

%% source of gauge anomalies
Gauge fermions on the D7-branes are the only source of pure gauge
anomalies, since fields on the D3--D7 bound state 
propagate only in four-dimensional
space-time and since they keep a vector-like structure 
even after the orbifold projection.
We calculate the anomalies at each fixed point that come from 
the fermions of the U(6) vector multiplet, which propagates 
in the eight-dimensional space-time. 

%% Fujikawa anomaly by def
Anomalies are understood as anomalous variations of the functional
measure \cite{Fujikawa}.
The daAnomalous variation of an action due to a chiral transformation 
is given by
\begin{equation}
 \delta S = {\rm Tr}\left( \delta^4 (\tilde{x}-x) \gamma_5 \right)
\end{equation}
from the measure of a four-dimensional Weyl fermion,
where the trace is taken over space-time coordinates $\tilde{x}=x$, 
spinor indices and gauge indices.
One can think of anomaly to a more general symmetry transformation by
replacing $\gamma_5$ to the generator $t^a$ of that symmetry transformation.
A straightforward extension to the higher-dimensional space-time 
compactified over orbifold geometry is simply given
\cite{BCCRS} by
\begin{equation}
 \delta S = {\rm Tr}_{co,sp,\omega,I} \left(
      \left(\frac{1}{12}\sum_{g \in {\bf Z}_{12}\vev{\sigma}}
      \delta^4(\tilde{{\bf y}'}-g\cdot {\bf y}')\delta^4(\tilde{x} - x)
      \rho_{sp}(g)\rho_{I}(g)
      \right)
      t^a_I     \right), 
\label{eq:Fujikawa}
\end{equation}
where $ t^a_{I}$ is a suitable representation of the generator $ t^a$ of
the transformation of which we consider the anomaly; 
$(\delta^4(\tilde{x} - x))$ in four dimensions is replaced 
by its orbifold analogue; the delta
function on the orbifold geometry is given by
\begin{equation}
 \frac{1}{12}\sum_{g \in {\bf Z}_{12}\vev{\sigma}}
      \delta^4(\tilde{{\bf y}'}-g\cdot {\bf y}')\delta^4(\tilde{x} - x)
      \rho_{sp}(g)\rho_{I}(g).
\end{equation}
Note that the gauge fermions of interest propagate 
in eight-dimensional space-time. Coordinates of four extra dimensions 
are denoted by ${\bf y}',\tilde{{\bf y}'} \in 
{\bf T}^4\equiv{\bf C}^2/\Gamma'_0$. %  as in the preceding sections.
Here, the $\Gamma'_0$ denotes the lattice spanned by ${\bf e}_4$, ${\bf e}_5$, 
${\bf e}_6$ and ${\bf e}_7$; 
${\bf y}'$ and $g\cdot {\bf y}'$($g\in {\bf Z}_{12}\vev{\sigma}$) are
the same point on the orbifold ${\bf T}^4/{\bf Z}_{12}\vev{\sigma}$.
Fields on these two points are identified up to two internal transformations,
$\rho_{sp}(g)$, which denotes the effect of local Lorentz rotation under
$g\in {\bf Z}_{12}\vev{\sigma}$, and $\rho_I(g)$, which denotes the effect of 
the rigid U(6) transformation given by $\tilde{\gamma}_{\sigma;7}$, as before.
Here, $\rho_{sp}(g)$ plays the same role as $\rho_a(g)$ 
in (\ref{eq:localLorentz4Dirr}), which is now given by
\begin{equation}
\rho_{sp}(\sigma^k) \equiv \prod_{b=1}^3 \left(\cos (\pi v_b k) + \sin
(\pi v_b k) \Gamma^{(2b+2)(2b+3)}\right). 
\label{eq:space-twist-Gamma}
\end{equation}
%U(6) rigid transformation results in multiplying phase factor 
%$\rho_I(g)$ for each irreducible representation in 
%$I \in \{ {\bf 24}^{(0,0)}, {\bf 1}^{(0,0)},
%           {\bf 5}^{(1,-1)},{\bf 5}^{*(-1,1)} \}$ of 
%the unbroken gauge group SU(5)$_{\rm GUT} \times$U(1)$_5 \times$U(1)$_6$
%--- $\rho_I(\sigma^k)=1$ for ${\bf 24}^{(0,0)}, {\bf 1}^{(0,0)}$ and 
%$\rho_I(\sigma^k)=(-1)^k$ for ${\bf 5}^{(1,-1)},{\bf 5}^{*(-1,1)}$.

%% technical supplement to the prev. paragraph
The trace is taken over the space-time coordinates 
(i.e. summation over $x$ and ${\bf y}'$ with $\tilde{x} = x$ and 
$\tilde{{\bf y}'}={\bf y'}$ imposed), over eight spinor indices 
(since we consider a Weyl fermion of eight-dimensional space-time), 
over weights ($\omega$) in a single irreducible representation of 
the unbroken gauge groups, and finally over different irreducible 
representations 
$I \in \{ {\bf 24}^{(0,0)}, {\bf 1}^{(0,0)},
           {\bf 5}^{(1,-1)}, \\ {\bf 5}^{*(-1,1)} \}$ .
%All (Kaluza--Klein towers of) four irreducible representations 
%of the unbroken gauge group, which formerly form a single U(6)-{\bf
%adj.} representation, contribute to the anomaly distribution, 
%although the total amount of anomalies integrated
%over the extra-dimensional manifold can be calculated only from the
%Kaluza--Klein zero modes surviving the orbifold projection conditions.

%% contribution from the CPT conjugate
The CPT conjugate of an eight-dimensional Weyl fermion 
gives a hermitian conjugate of (\ref{eq:Fujikawa}).
The (\ref{eq:Fujikawa}) and its hermitian conjugate as a whole are
expressed just through taking the trace over the whole sixteen spinor 
components of the eight dimensions.
This is because the CPT-conjugate of a Weyl fermion of eight
dimensions has the chirality opposite to the original one, 
and also because the whole relevant fermions form the vector-like 
({\bf adj.}) representation of the U(6).
%%% and also because the twisting used in the orbifold
%   projection acts as an adjoint on the U(6)-{\bf Adj.}. 
The $\Gamma^{89}$ in (\ref{eq:space-twist-Gamma}) is understood as 16 by
16 part (which acts on the U(6) gauge fermion) of the 32 by 32 matrix of
$\Gamma^{89}$.

%% In momentum frame
The trace over the coordinate in the $(x,{\bf y}')$-base can be 
rewritten in the momentum-space base as
\begin{eqnarray}
 \delta S &=& \int \left(\frac{dp}{2\pi}\right)^4 
            \frac{1}{{\rm vol}({\bf T}^4)}\sum_{{\bf p}'\in \Lambda_0'}
            \int d^4\tilde{x} d^4 \tilde{{\bf y}'}  \int  d^4 x d^4 {\bf y}'
             \delta^4(\tilde{x} - x) \\
  & & \qquad \quad \frac{1}{12}
      \sum_{g \in {\bf Z}_{12}\vev{\sigma}}
        \tr{}_{sp,\omega,I} \left( 
                     e^{-i (p \cdot \tilde{x}+{\bf p}'\cdot \tilde{{\bf y}'})} 
                     \delta^4(\tilde{{\bf y}'}-g\cdot {\bf y}')
                     \rho_{sp}(g)\rho_{I}(g)
                     e^{i (p \cdot x+{\bf p}'\cdot {\bf y}')}
                     t^a_I 
                            \right),  \nonumber  \\
 & =& \int d^4 x d^4 {\bf y}'
            \int \!\!\! \left(\frac{dp}{2\pi}\right)^4 
            \frac{1}{{\rm vol}({\bf T}^4)} \!\! \sum_{{\bf p}'\in \Lambda_0'} 
            \! \frac{1}{12} \!\!\! \sum_{g \in {\bf Z}_{12}\vev{\sigma}}
               \!\!\! \tr{}_{sp,\omega,I} \left( 
                     e^{- i {\bf p}'\cdot (g \cdot {\bf y}'- {\bf y}')} 
                 \rho_{sp}(g) \rho_{I}(g)  t^a_I \right)\! \! ,    
\end{eqnarray}
where now the trace runs only over the spinor and gauge indices. 
%the $\Lambda_0'$ is the dual lattice of the $\Gamma'_0$.
%Kaluza--Klein momenta run over this dual lattice.

%% Gauge invariant regularization 
%%  decomposition to group elements / basic expression
This anomalous variation of the functional measure is regularized
in a gauge-invariant way by inserting $e^{\Dsl\cdot \Dsl/2M^2}$, where $M$ is
an energy scale of the regulator and $\Dsl$ the Dirac operator of
eight-dimensional space-time. 
We adopt, here, the SO(7,1)-symmetric regularization.
Then,
\begin{eqnarray}
 \delta S &=& \frac{1}{12}\sum_{g \in {\bf Z}_{12}\vev{\sigma}}
            \int d^4 x d^4 {\bf y}'
            \;\; \frac{1}{{\rm vol}({\bf T}^4)} \sum_{{\bf p}'\in \Lambda_0'} 
                   e^{ - i (g^{-1} \cdot {\bf p}' - {\bf p}') \cdot {\bf y}'}
        \label{eq:decomp-Fujikawa}\\  
& & \qquad \qquad \lim_{M \rightarrow \infty}
           i \left(\frac{M^2}{2\pi}\right)^{2}
                   \tr{}_{sp,\omega,I} \left( 
                e^{-\frac{{\bf p}'\cdot {\bf p}'+\frac{i}{2}F_{AB}\Gamma^{AB}}
                            {2 M^2}
                      }
             \rho_{sp}(g)\rho_{I}(g)  t^a_I \right) \nonumber .
\end{eqnarray}
The space-time indices $A$ and $B$ run from 0 to 7.
Here, we can see a structure similar to the anomaly distribution
obtained in Eq. (\ref{eq:decomposition}).
The anomalies are decomposed into parts, each of which corresponds 
to each element $g$ of the orbifold group ${\bf Z}_{12}\vev{\sigma}$. 
Each component has its own distribution function determined by $g$. 
The major difference from Eq. (\ref{eq:decomposition}) 
is that the expression for the anomaly of each component 
\begin{equation}
 \frac{1}{12} \times 
  \lim_{M \rightarrow \infty} i \left(\frac{M^2}{2\pi}\right)^2 
       e^{-\frac{{\bf p}'\cdot {\bf p}'}{2M^2}}
       \tr{}_{sp,\omega,I}\left( e^{-i \frac{\Gamma \cdot F}{4M^2}}
       \rho_{sp}(g)\rho_I(g) t^a_I
\right),
\label{eq:basic-expression}
\end{equation}
contains not only the triangle anomalies but also more information, 
including box anomalies as we see below.

%% Triangle anomalies are re-obtained
The triangle anomalies are obtained 
from (\ref{eq:basic-expression}) in the following way. 
The $\rho_{sp}(\sigma^k)$ in the above expression 
contains a term $\prod_{b=1}^3 \sin (\pi v_b k) \Gamma^{456789}$. 
Then, the trace over spinor indices becomes non-vanishing 
when the regulator $e^{-i (\Gamma \cdot F)/(4M^2)}$ provides 
$-(\Gamma \cdot F)^2/(2(4M^2)^2)$. The decoupling limit of the regulator
$M \rightarrow \infty$ leaves convergent and (generally) non-vanishing 
quantities. 
The $\sigma^k$-component includes triangle anomalies as 
\begin{eqnarray}
&& 
  \frac{1}{12}
 i \left( \frac{M^2}{2\pi}\right)^2 
   \left( \prod_{b=1}^3 \sin (\pi v_b k)\right)
   \sum_I \rho_I(g) \tr{}_{sp,\omega} \left( 
    t^a_I\frac{-1}{2(4M^2)}(\Gamma^{MN}F_{MN})^2\Gamma^{456789}\right) 
                          \label{eq:anomaly-distr} \\
&=& \frac{\tr{}_{\omega}(t^a\{ t^b,t^c \})|_{I={\bf fund.}}}{32\pi^2}
     (F^b \cdot \tilde{F}^c)|_{4D}
     \frac{4}{12} 
     \left( \prod_{b=1}^3 \sin (\pi v_b k)\right) \sum_I (-i\rho_I(g) A_I)
     \frac{\tr{}_{sp}\left(\Gamma^{0123456789}_{16{\rm ~by~}16}\right) }{16},
                               \nonumber   
\end{eqnarray}
which coincides with (\ref{eq:triangle-coeff-k}).

%% Fujikawa method to six-dimensional singularity

Let us now turn our attention to the box anomalies.
The six-dimensional singularities are associated with elements of the
orbifold group that do not rotate one of three complex planes.
All the six-dimensional singularities on the ${\bf T}^6/{\bf
Z}_{12}\vev{\sigma}$ orbifold extend in the third complex plane labeled
by $z_3$.
The isotropy groups at these singularites are generically
${\bf Z}_4\vev{\sigma^3}$ or ${\bf Z}_2\vev{\sigma^6}$, 
as explained in section \ref{sec:geometry}.
The $\sigma^{3k}$($k=0,1,2,3$) do not rotate this third complex plane. 
In other words, the loci of $\sigma^{3k}$-fixed points are no longer points in
${\bf T}^6$, but rather fixed two-dimensional planes in ${\bf T}^6$, 
and hence they are (five+one)-dimensional singularities.

Since the gauge fields do not propagate in two dimensions among 
the six dimensions of those singularities, we do not have to care about
the pure gauge box anomalies on these six-dimensional singularities. 
This is one of the benefits of the ${\bf Z}_{12}\vev{\sigma}$ orbifold,
and one of the reasons why we adopt the ${\bf Z}_{12}$ orbifold.
When the ${\bf Z}'_6$ orbifold is adopted, where
$(v_b)|_{b=1,2,3}=(1/6,-3/6,2/6)$, six-dimensional singularities develop
in the directions to which the D7-branes are stretched. 
In this case, one has to take care of the pure gauge box anomalies.

However, since the gravitational field propagates in all ten
dimensions, pure gravitational box anomalies may be localized on those
six-dimensional singularities.
We require that all of them be cancelled out on each singularity. 
The distribution function in the first line of Eq. (\ref{eq:decomp-Fujikawa})
is now given by
\begin{equation}
 \frac{1}{{\rm vol}({\bf T}^6)} \sum_{{\bf p} \in \Lambda_0}
 e^{-i (g^{-1} \cdot {\bf p} - {\bf p}) \cdot {\bf y}}, 
\end{equation}
where $\Lambda_0$ is the dual lattice of the $\Gamma_0$.
This distribution function is independent of $z_3$, 
when $g = \sigma^{3k}$ does not rotate the third complex plane. 
Thus, the $\sigma^{3k}$-component of anomalies are localized on these
six-dimensional singularities.
At the same time, (\ref{eq:basic-expression}) yields box anomalies for
$\sigma^{3k}$. Indeed, the $\rho_{sp}(\sigma^{3k})$ do not contain 
the $\Gamma^{456789}$ term, since $\sin (\pi v_3 (3k) ) = 0$, but rather 
% $\rho_{sp}(\sigma^{3k})$ include 
another term 
\begin{equation}
\prod_{b=1}^2 (\sin (\pi v_b (3k))) \cos(\pi v_3 (3k)) \Gamma^{4567} .
\end{equation}
The trace over the spinor indices
becomes non-trivial when the regulator provides $i (\Gamma
\cdot F)^3/(3!(4M^2)^3)$; the $F$ is understood as the field strength
$R$ of the local Lorentz symmetry. 
One can see that the decoupling limit of the regulator $M \rightarrow
\infty$ leaves convergent and (generally) non-vanishing results 
$ \propto \tr{}_{sp} ( (R \wedge R \wedge R)|_{6D} t^a)|_{I}$.

%% actual calculation of pure gravitational box anomalies

The box anomalies consist of only $\sigma^{3k}$-components ($0,1,2,3$).
All other transformations in the orbifold group are irrelevant.
This implies that the box anomalies can be calculated by assuming 
only the ${\bf Z}_4\vev{\sigma^3}$-orbifold projection. 
This is also intuitively reasonable. Since anomalies are, in some sense,
a UV phenomenon, they can be determined by looking at only local geometry.
Then, since the local geometry around the six-dimensional singularities
of ${\bf T}^6/{\bf Z}_{12}\vev{\sigma}$ is the same as that in ${\bf
C} \times ({\bf T}^4/{\bf Z}_4\vev{\sigma^3})$, box anomalies calculated
in the latter geometry should be the same locally as those calculated in
the former one.
Now, the ten-dimensional space-time is compactified on the ${\bf T}^4/{\bf
Z}_4\vev{\sigma^3}$ orbifold, and we have Kaluza--Klein towers of
six-dimensional particles. 
Thus, we can calculate box anomalies by
summing up the absolute square of Kaluza--Klein wave functions,  
just as we have done for the triangle anomalies at the beginning 
of this subsection \ref{subsec:u2-anomaly}.
The actual calculation is much easier in this way than having to calculate
the anomalies with various representations of SO(9,1) 
in the Fujikawa method extended for orbifold geometry.
The anomaly on the $\Gamma'_{\sigma^{3k}}$ is again given by 
\begin{equation}
 \frac{1}{\# {\bf Z}_4\vev{\sigma^3}} 
   \sum_{I} A_I^{\rm box} \rho_{I}(\sigma^{3k}).
\end{equation}

The supergravity multiplet of the Type IIB supergravity has the
following fields that contribute to the pure gravitational anomalies:
two Weyl and Majorana gravitinos (two times fifty-six on-shell states), 
two Weyl and Majorana fermions (two times eight on-shell states) and 
real self-dual 4-form fields (thirty-five on-shell states).
These fields are irreducible representations of ten-dimensional
space-time, but they are decomposed into various fields in the
six-dimensional space-time after the Kaluza--Klein reduction, which we
label $I$.
The term $\rho_I(g)$ denotes the phase factor due to the local Lorentz
transformation associated with $g \in {\bf Z}_4\vev{\sigma^3}$.

A pair of the Weyl and Majorana gravitinos and Weyl and Majorana fermions
is an ${\bf 8}_v \otimes {\bf 8}_s$ representation of the little group 
SO(8) in ten dimensions, where {\bf 8}$_v$ is the vector representaion
of the SO(8) and {\bf 8}$_s$ the spinor representation. 
This is decomposed into Kaluza--Klein towers of
six-dimensional fields 
$({\bf 4}_v + {\bf 2}_{scl+} + {\bf 2}_{scl-}) \otimes 
({\bf 4}_{-} + {\bf 4}_+)$, where
${\bf 4}_v$ is vector representation, ${\bf 2}_{scl\pm}$ are complex
scalars, and ${\bf 4}_{\pm}$ are Dirac spinors with opposite 
$\Gamma^7$-chiralities.
All these towers contribute to the pure gravitational box anomalies.
The transformation $(z_1,z_2) \mapsto (e^{i\alpha}z_1,e^{-i\alpha}z_2)$,
to which the ${\bf Z}_4\vev{\sigma^3}$ belong, acts on each
Kaluza--Klein tower through
${\bf 4}_v,{\bf 4}_- \mapsto {\bf 4}_v,{\bf 4}_-$,
${\bf 2}_{scl \pm} \mapsto e^{\pm i\alpha} {\bf 2}_{scl \pm}$ and 
${\bf 4}_+ \mapsto e^{i\alpha}{\bf 4}_+$.  
The phase factor $\rho_{I}$ is calculated from these transformations for
each Kaluza--Klein tower $I$.

The self-dual 4-form is decomposed into Kaluza--Klein towers 
of two scalars, four vector fields, one 2-form field, 
one complex rank-2 tensor contained in ${\bf 4}_+\otimes {\bf 4}_+$
and one complex rank-2 tensor in ${\bf 4}_- \otimes {\bf 4}_-$.
Only the last two of them contribute to the box anomalies. 
The tensor in ${\bf 4}_+\otimes {\bf 4}_+$ is multiplied by
$e^{2i\alpha}$, while the tensor in ${\bf 4}_- \otimes {\bf 4}_-$ is
multiplied by 1 under the transformation described in the previous paragraph. 

Let us calculate the irreducible part of the box anomalies.
The coefficient $A_I^{\rm box,irr}$ of the 
irreducible part of the pure gravitational box anomalies is given as follows:
a gravitino in ${\bf 4}_v \otimes {\bf 4}_-$ has 
$A_I^{\rm box,irr} = 245/360$, 
$A_I^{\rm box,irr}=1/360$ for a Dirac fermion ${\bf 4}_-$, and
$A_I^{\rm box,irr}=-56/360$ for a complex rank-2 tensor in ${\bf 4}_+
\otimes {\bf 4}_+$. Fields of opposite chirality have $A_I^{\rm box,irr}$
with opposite sign. 

The coefficient on the $\Gamma'_{\sigma^{3k}}|_{k=0,1,2,3}$ lattice is
calculated as
\begin{equation}
 \frac{1}{4}\frac{1}{360}(2 \times (245 (1-i^k) -1 (1-i^k) 
                                  + 2 (i^k - i^{2k} + i^{-k} - 1)) 
                          - 56 (i^{2k} - 1 )),
\end{equation}
where $e^{i\alpha}=i$ for $\sigma^3$ has been used.
The coefficients do not vanish for $k=1,2,3$. 
Thus, we examine how much anomaly is distributed to each six-dimensional 
singularity. 
Sixteen  points in $\Gamma'_{\sigma^6}$ receive $1/16$ of
$240/360$ from the $\sigma^6$-component, and four of them, 
which are also in $\Gamma'_{\sigma^3}=\Gamma'_{\sigma^9}$,
also receive $1/4$ of 
$(1/360)\times ((122(1-i)+28) + (122(1+i)+28))=300/360$ from the
$\sigma^3$- and $\sigma^9$-components of the anomaly.
Thus, the twelve loci of $\sigma^6$-fixed points in the covering space 
${\bf T}^6$ have $15/360$ pure gravitational irreducible anomaly, 
and the remaining four loci of $\sigma^3$-fixed points in the ${\bf T}^6$ 
have $90/360$ anomaly.
Those sixteen singularities in the covering space ${\bf T}^6$ 
correspond to four distinct two-dimensional singularities 
of the ${\bf T}^6/{\bf Z}_{12}\vev{\sigma}$.
90/360, 270/360, 90/360 and 90/360 of anomalies are localized on the four
singularities, respectively.
%Thus, the twelve $\sigma^6$-fixed loci becomes two singularities,
%on each of which $90/360$ anomaly is localized.
%The three of among four $\sigma^3$-fixed loci become a single
%singularity, which has $270/360$ anomaly.
%Finally, the remaining one $\sigma^3$-fixed loci has $90/360$ anomaly.

One can, in general, cancel these irreducible pure gravitational
anomalies, by introducing gauge singlet fields at these singularities.
However, the situation is not so simple in our case of interest.
It is because the matter contents that can be introduced on these
singularities are quite limited, since there is extended SUSY there.
Since SUSY is broken at these singularities only by 
${\bf Z}_2\vev{\sigma^6}$- or ${\bf Z}_4\vev{\sigma^3}$-orbifold, 
there are sixteen SUSY charges, which form (0,2) SUSY of the six-dimensional
space-time (e.g. \cite{Witten-5brane}).  
The minimal SUSY multiplet of the (0,2) SUSY theories is a tensor
multiplet. 
Other SUSY multiplets cannot be introduced because
they always include fields of spin more than 1.
The tensor multiplet consists of five real scalars, one real 
rank-2 tensor in ${\bf 4}_+ \otimes {\bf 4}_+$, and two Dirac fermions
in ${\bf 4}_+$. 
Therefore, a single tensor multiplet of the (0,2) SUSY contributes
to the irreducible part of the pure gravitational box anomaly by
$(-28-2\times 1)/360=-30/360$ (e.g. \cite{Witten-5brane}).

The irreducible part of the pure gravitational box anomaly can be cancelled 
by introducing the tensor multiplets of the (0,2) SUSY 
at each of four six-dimensional singularities.
This is because the amount of anomaly happens to be an integral multiple 
of $-30/360$, with opposite sign at each singularity.
All the box anomalies are cancelled out if 
the number of tensor multiplets at those singularities is 
3, 9, 3 and 3, respectively.
This is also a miraculous result.

There will be non-vanishing reducible part of pure gravitational
anomalies, but they can be cancelled by the Green--Schwarz mechanism
\cite{Witten-5brane}. 

Finally, the ${\bf Z}_2\vev{\Omega R_{89}}$-projection does not 
give rise to anomalies since the theory obtained by 
gauging only $\Omega R_{89}$ is nothing but the Type I$'$ theory known to 
be consistent.

We have discussed the box anomaly cancellation on
six-dimensional singularities, and the pentagonal anomalies on
eight dimensions. In particular, we have shown that the irreducible 
part of the pure gravitational anomalies can be cancelled 
by introducing suitable massless fields in the six-dimensional singularities.
However, the introduction of new massless fields might not be
necessary in a situation such as the following.
When there are an infinite number of massive excitations on D3-branes,
D7-branes or four-dimensional fixed points, those particles sometimes 
give rise to higher-dimensional anomalies. 
Winding modes on D-branes in the Type IIB string theory are good examples.
Since those particles can contribute to the pure gravitational box 
anomalies, it does not make sense, in principle, to discuss
higher-dimensional anomalies without specifying a spectrum of infinite
massive particles on four-dimensional space-time\footnote{
For a related discussion, see subsubsection
\ref{subsubsec:u3-anomaly-rel}.}\raisebox{1mm}{$\!\!$,}\footnote{
We expect that things are much the same for global anomalies,  
which we do not discuss in this paper.}. 
We show above that it is possible to cancel the anomaly in a genuinely 
field-theoretical manner (cancelling box anomalies through six-dimensional
massless fields). We do not consider that this is the only way, but 
rather we claim that there is at least one way of cancelling the anomaly.
 
%%%%%%%%%%%%%%%%%%%%%%%%%%%%%%%%%%%%%%%%%%%%%%%%%%%%%%%%%%%
%% Other anomalies at (three+one)-dim fixed points

We have examined so far anomalies over untwisted space-time.
We have discussed triangle anomalies at (three+one)-dimensional singularities 
and box anomalies at (five+one)-dimensional singularities. 
They are anomalies over space-time that extend in untwisted directions.
However, there is another class of ``anomalies'' that arise from
Eq. (\ref{eq:basic-expression}), and these are over space 
in the twisted directions. This is the issue discussed in the rest of
this subsection \ref{subsec:u2-anomaly}.
We only treat such anomalies on D7-branes, but the following discussion 
can easily be extended to include the gravity in the bulk.

Equation (\ref{eq:basic-expression}) implies that there are other
``anomalies''. 
Terms in $\rho_{sp}(g)$ proportional to ${\bf 1}$, 
$\Gamma^{45}$, $\Gamma^{67}$ and $\Gamma^{4567}$ give rise to
non-vanishing values after the trace over the spinor index is taken 
by extracting terms proportional to themselves
from the regulator $e^{ - ({\bf p}'\cdot {\bf p}'+ \frac{i}{2}\Gamma_{AB}
\cdot F^{AB})/2M^2}$ we adopt. We call them ``anomalies in internal 
space dimensions''.
The $\sigma^k$-component of these anomalies is given by
\begin{eqnarray}
 && \frac{4 M^4}{\pi^2} 
      \left(\prod_{b=1}^{3}\cos (\pi v_b k)\right) 
      \left(\sum_I i \rho_I(\sigma^k) \tr{_\omega}(t^a_I) \right)  
\label{eq:int-anomaly-1}\\
 &+&   \frac{2M^2}{\pi^2} 
      \sin (\pi v_1 k) \left(\prod_{b=2}^{3}\cos (\pi v_b k)\right) 
      \left(\sum_I - \rho_I(\sigma^k) 
                   \tr{}_{\omega}(t^a_I t^b_I)F^b_{45}\right) 
\label{eq:int-anomaly-2}\\
 &+&   \frac{2M^2}{\pi^2} 
      \sin (\pi v_2 k) \left(\prod_{b=3}^{1}\cos (\pi v_b k)\right) 
      \left(\sum_I - \rho_I(\sigma^k) 
                   \tr{}_{\omega}(t^a_I t^b_I)F^b_{67}\right) 
\label{eq:int-anomaly-3}\\
 &+&  \frac{\tr{}_\omega(t^a\{t^b,t^c\})|_{I={\bf fund.}}}{32\pi^2} 
      (F^b \tilde{F}^c)|_{\rm{on~4567\mbox{-}th~plane}}     
\label{eq:int-anomaly-4}\\
 & & \qquad \qquad \qquad 
       4 \left(\prod_{b=1}^{2}\sin (\pi v_b k)\right)\cos(\pi v_3 k) 
      \left(\sum_I -i \rho_I(\sigma^k) A_I \right).  \nonumber 
\end{eqnarray}
Now, these expressions explicitly depend on the regulator mass $M$.
This is in sharp contrast with the ordinary anomalies discussed before.
This suggests that these ``anomalies in internal dimensions'' 
are regularization-dependent, and their values can be changed 
by UV physics ( $\simeq$ regularization).
Thus, the ``anomalies'' (\ref{eq:int-anomaly-1}) -- (\ref{eq:int-anomaly-3})
do not lead to reliable constraints on low-energy physics.
 
The ``anomaly'' (\ref{eq:int-anomaly-4}), which is proportional to 
$(F \wedge F)|_{\rm {on~4567\mbox{-}th~plane}}$, 
seems to be reg-ularization-independent, but it is not the case either. 
We have only discussed so far the consequences of adopting the regulator 
$e^{\Dsl \cdot \Dsl /2M^2}$. 
This regulator fully respects the whole SO(7,1) Lorentz symmetry.
However, it does not have to do so, since the geometry 
of the orbifold already breaks this symmetry.
It is true that there exists the SO(7,1) local Lorentz symmetry on the
D7-brane world volume away from singularities. 
It is also true that there exists the SO(5,1) local Lorentz symmetry around
the six-dimensional singularities and that there exists the SO(3,1) local
Lorentz symmetry at any point in the orbifold, even around the
four-dimensional singularities.
However, the regulator does not have to respect broken SO(7,1)/SO(3,1)
symmetry (SO(7,1)/SO(5,1) symmetry) at four-dimensional
(six-dimensional) singularities, respectively. 
Then, regulators can include 
$(F \wedge F)|_{\rm{on~4567\mbox{-}th~plane}}/M^4$ explicitly 
at singularities, without being accompanied by $\Gamma^{4567}$, 
although it is not easy to write down explicitly 
such regularization in the momentum-diagonal base. 
Notice that the gauge transformation around the singularities cannot be
topologically non-trivial\footnote{One can understand this statement by
considering the gauge symmetry on S$_1$-compactified and
S$_1$/{\bf Z}$_2$-compactified five dimensions.} 
in the twisted ${\bf e}_{4,5,6,7}$ directions; 
hence the explicit $(F \wedge F)|_{\rm{on~4567\mbox{-}th~plane}}/M^4$ 
(at singularities) in the regulator is not forbidden 
by the remaining gauge symmetry. 
The $\left(\prod_{b=1}^{3}\cos (\pi v_b k)\right)$ term 
in (\ref{eq:space-twist-Gamma}), thus, can also give rise to the 
``anomaly in internal dimensions'' proportional to 
$(F \wedge F)|_{\rm{on~4567\mbox{-}th~plane}}$.
This contribution depends on the regularization.
Therefore, the ``anomaly'' (\ref{eq:int-anomaly-4}) also depends on 
regularization (more specifically, on UV physics at singularities), 
and does not lead to a constraint only on low-energy physics.

It is clear from the above argument that the anomaly in the untwisted
directions is not susceptible to these variation in the regulators.
Regulators do not contain $F_{01}$, $F_{23}$ or 
$(F \wedge F)|_{\rm{on~0123\mbox{-}th~plane}}$ because of the unbroken 
local SO(3,1) symmetry and topologically non-trivial gauge-symmetry
transformation. 
Thus, terms in $\rho_{sp}(g)$ that are not proportional to
$\Gamma^{456789}$ % or $\Gamma^{4567}$ 
do not give rise to the anomalies proportional to 
$(F \wedge F)|_{\rm{on~0123\mbox{-}th~plane}}$.
The situation is the same for the box anomalies 
$(R \wedge R \wedge R)|_{\rm{on~012389\mbox{-}th~plane}}$ in the untwisted
directions.

We conclude that all the ``anomalies in internal dimensions'' are not
constraints on low-energy physics.
Although they might be constraints on UV physics, especially 
on UV physics at singularities, we do not discuss this issue further.

\subsection{Discrete R Symmetry}

Orbifold geometry preserves a discrete rotational symmetry, 
which is a subgroup of the SO(6) rotational symmetry of the ${\bf C}^3$.
This subsection is devoted to consequences of this symmetry.

Now that all the particles in the SU(5)$_{\rm GUT}$-breaking sector 
are obtained from D-branes, we know how those fields transform under
the discrete rotational symmetry of the orbifold geometry.
On the other hand, such a rotational symmetry is regarded as 
an internal symmetry at energies below the Kaluza--Klein scale. 
This symmetry, in general, rotates SUSY charges 
since these are in the spinor representation 
of the space rotational symmetry.
Thus, it becomes an R symmetry below the
Kaluza--Klein scale.
Therefore, we can figure out how those fields transform 
under the discrete R symmetry.

The R symmetry obtained in this way is a gauged symmetry. 
Indeed the rotation is nothing but the combined action of a general
coordinate transformation and a local Lorentz symmetry, both of which are
gauged.
Thus, the discrete gauge R symmetry (obtained in this way) is exact
unless it is spontaneously broken. 
This is quite important because the (mod 4)-R symmetry
of the product-group unification models should be preserved
at the $10^{-14}$ level to keep the two Higgs doublets almost massless.

The (mod 4)-R symmetry is expected to be spontaneously broken 
by vacuum condensation of the superpotential, which is related to the
spontaneous breaking of ${\cal N}$ = 1 SUSY.
Spontaneous symmetry breaking of geometry is related to the spontaneous
breaking of SUSY also in string theories \cite{NS-R-SUSY}; 
the SUSY breaking causes tadpoles of NS--NS fields, leading to 
instabilities of the geometry. 
Deformations of geometry due to this instability (due to SUSY breaking) 
will lead to the spontaneous breaking of the discrete R symmetry. 
One might expect that a similar thing could happen in a model based on
supergravity. 

The transformation properties of various fields under the rotation of
extra dimensions have been already given in subsection \ref{subsec:u2-config}.
One can easily see that the R charge is properly assigned 
for all the particles in the SU(5)$_{\rm GUT}$-breaking sector 
when the R symmetry is identified\footnote{The required R-charge
assignment is properly obtained as long as $\alpha_1 = - \alpha_2$. 
We put $\alpha_1=\alpha_2=0 $ in the text just because of its simplicity.
The unbroken subgroup discussed in the next paragraph is the mod 4 subgroup 
whenever $(\alpha_1 = -\alpha_2) \in \alpha_3 {\bf Z}$.} with 
the rotational symmetry of the third complex plane 
$z_3 \mapsto e^{i\alpha}z_3$.

The toroidal compactification breaks the SO(6) rotational symmetry of
the ${\bf C}^3$ down to a discrete subgroup of the SO(6).
Discrete rotation that corresponds to the mod 4 part of the R symmetry 
should be preserved by the geometry; otherwise the R symmetry, 
although gauged, is spontaneously broken at the Kaluza--Klein scale 
and does not play any role in phenomenology.
Notice that the mod 4 discrete subgroup {\em is} naturally 
preserved by geometry 
since that the subgroup corresponds to rotation of the third complex
plane $z_3 \in ({\bf C}/({\bf Z}_{12}\vev{\sigma}\times {\bf Z}_2\vev{\Omega
R_{89}})$ by an angle $\pi$. 

The supergravity multiplet of the Type IIB supergravity also provides
Kaluza--Klein zero modes that survive the orbifold projection 
associated to ${\bf Z}_{12}\vev{\sigma} \times {\bf Z}_2\vev{\Omega
R_{89}}$.
Those matter contents (all are gauge singlets) consist
\cite{IIB-orbifold} of one supergravity
multiplet of  four-dimensional ${\cal N}$ = 1 supergravity and four chiral
multiplets.
Three of the four chiral multiplets correspond to the metric and the 2-form 
in three different complex planes, and the other one corresponds to
the dilaton and an axion.
R charges of all these chiral multiplets are 0.
Therefore, these moduli are massless without (mod 4)-R symmetry breaking
unless particles appear at singularities whose R charges are 2.

Since we do not specify the origin of the quarks and leptons, it is
impossible to determine the R charges of those particles. 
We just expect that their R charges are determined 
as those given in Table \ref{tab:u2-R4}.

It is discovered in \cite{KMY-da} that the (mod 4)-R symmetry has 
a vanishing anomaly with the SU(2)$_{\rm H}$ gauge group, and can 
be anomaly-free with SU(5)$_{\rm GUT}$ gauge group, 
if there is an extra pair of $({\bf 5},{\bf 1})$ and 
$({\bf 5}^*,{\bf 1})$ chiral multiplets of the 
SU(5)$_{\rm GUT} \times$U(2)$_{\rm H}$ gauge group.
The higher-dimensional construction motivates the existence 
of this SU(5)$_{\rm GUT}$-charged vector-like pair at the TeV scale.
The vector-like pair cannot have mass unless (mod 4)-R symmetry is broken.
Notice, however, that the (mod 4 R)[SU(5)$_{\rm GUT}$]$^2$ 
can be cancelled also by the generalized Green--Schwarz mechanism.

Although we examined whether triangle anomalies are cancelled at all fixed
points, we do not have to check that the (mod 4)-R symmetry has vanishing
anomalies at each fixed point. % not just as a whole sum from all fixed points.
This is because the angle-$\pi$ rotation that we are interested in 
is a rigid rotation of the whole orbifold. 
We are not interested in a space rotation with the angle changing point by
point.
 
\subsection{Toward a Realistic Model}
\label{subsec:u2-phen}

The most successful feature of this higher-dimensional construction is
that the superpotential of ${\cal N}$ = 2 SUSY (the first and 
the second lines of (\ref{eq:super2})) is automatically obtained 
from (\ref{eq:N=2super}).
The approximate ${\cal N}$ = 2 SUSY relation is naturally expected as a
result of the extended SUSY in the UV physics.
The third line of (\ref{eq:super2}) is also allowed by the ${\cal N}$ = 2
SUSY and the (mod 4)-R symmetry (the Fayet--Iliopoulos F-term).

Let us now discuss what is needed to make the model realistic beyond the
orbifold construction obtained so far.
The first issue to be discussed is the necessary particles that we could
not obtain from the orbifold construction.

Quarks and leptons, SU(5)$_{\rm GUT}$-{\bf 10} and -${\bf 5}^*$, are not
obtained in our orbifold construction. 
There is no model using D-brane construction based on string theories
that has succeeded in obtaining all of (i) the three families 
of quarks and leptons, (ii) a unified gauge group and 
(iii) a sector to break that symmetry; 
this is not a difficulty limited to our construction.
We consider that the SU(5)$_{\rm GUT}$-breaking sector with ${\cal N}$ = 2 SUSY
is a strong indication of the structure of higher-dimensional
space, and we, therefore, construct the model 
so that this structure is manifestly realized.
The orbifold geometry thus obtained provides a good description of the
structure of the SU(5)$_{\rm GUT}$-breaking sector, but not of
the quarks and leptons; 
these are described as particles put by hand at a fixed point. 
The fixed point where they reside should preserve neither 
the U(6) symmetry nor ${\cal N}$ = 2 SUSY.
There is only one candidate for such a fixed point: 
${\bf y}'={\bf 0}$ on the D7-branes.
  
As we have discussed in subsection \ref{subsec:u2-anomaly}, the pure
gravitational anomalies localized on six-dimensional singularities
requires that some particles be newly introduced.
The tensor multiplet of (0,2) SUSY of the six dimensions with the number
specified in subsection \ref{subsec:u2-anomaly} is one of the possibilities.
Towers of infinite massive particles on D3-branes or D7-branes, if they exist, 
may also contribute to the anomaly cancellation. 
 
The gauge symmetry of our model is now SU(5)$_{\rm GUT} \times$U(1)$_5
\times$U(1)$_6 \times $U(2)$_{\rm H}$. 
However, when quarks and leptons are introduced at the ${\bf y}'={\bf 0}$
fixed point, it is probable that only one linear combination of U(1)$_5$
and U(1)$_6$ remains free from  mixed anomaly with the SU(5)$_{\rm GUT}$
gauge group; the other combination will be anomalous, 
but its anomaly will be cancelled by the generalized Green--Schwarz
mechanism \cite{DSW}.
The mixed-anomaly free combination must be what is
called the ``fiveness'' U(1) symmetry\footnote{The ``fiveness'' U(1)
symmetry is given by a linear combination of the U(1) B$-$L symmetry and
U(1)$_Y$ of the standard model that commutes with all generators in
the SU(5)$_{\rm GUT}$.}, since this is the unique assignment that leads to
a vanishing mixed anomaly. However, there are still non-vanishing
U(1)$_{\rm fiveness} $[gravity]$^2$ and [U(1)$_{\rm fiveness}$]$^3$
triangle anomalies.
These can be cancelled (i) by the generalized
Green--Schwarz mechanism or (ii) by introducing another particle.
Three families of right-handed neutrinos are sufficient to cancel these
two anomalies simultaneously.
In case (ii), the U(1)$_{\rm fiveness}$ symmetry is not broken at
the Kaluza--Klein scale, and it should be spontaneously broken at the
intermediate scale to explain the small neutrino masses via seesaw
mechanism \cite{seesaw}. 

The ${\bf T}^6/({\bf Z}_{12}\vev{\sigma}\times {\bf Z}_2\vev{\Omega
R_{89}})$ geometry would have more discrete symmetries in addition to the
(mod 4)-R symmetry discussed above. 
We consider that they may be broken by condensations of gauge-singlet
fields introduced at singularities, although their geometrical
interpretations are not clear.
Therefore, we do not take such discrete symmetries seriously 
as a symmetry that determines the low-energy physics.

Now finally, at the end of this section \ref{sec:u2}, we discuss the
low-energy superpotential.
Yukawa interactions of quarks and leptons are 
\begin{equation}
 W = c {\bf 10} \; {\bf 10} \; 
         \frac{(\bar{Q}^i_{\;\; \alpha}Q^\alpha_{\;\; 6})}{M_*}+
     c'\frac{(\bar{Q}^6_{\;\; \alpha}Q^\alpha_{\;\; i})}{M_*}
        \cdot {\bf 10}^{ij} 
        \cdot {\bf 5}^*_j 
\end{equation}
in the SU(5)$_{\rm GUT} \times$U(2)$_{\rm H}$ model. 
These interactions can be induced (i) by massive particle exchange
and (ii) by unknown non-perturbative effects.

The wave-function renormalization of composite states 
$(\bar{Q}^i_{\;\; \alpha}Q^\alpha_{\;\; 6})$ and 
$(\bar{Q}^6_{\;\; \alpha}Q^\alpha_{\;\; i})$
in the K\"{a}hler potential may be\footnote{The hyper-K\"{a}hler metric 
is not renormalized in general ${\cal N}$ = 2 SUSY gauge theories 
\cite{hyperKahler}. However, 
our case of interest is asymptotic non-free, and hence it is not
obvious that it is indeed the case in the present model.}
protected from the strong U(2)$_{\rm H}$ couplings by the approximate 
${\cal N}$ = 2 SUSY \cite{hyperKahler}.
The SU(5)$_{\rm GUT}$ gauge interactions, which do not preserve ${\cal
N}$ = 2 SUSY, do not lead to a sizeable wave-function renormalization.
Thus, the effective Yukawa coupling may include a $\vev{Q}/M_* \simeq v/M_*
\simeq 10^{-1}$ suppression factor in both cases (i) and (ii).

The effective coefficients $c$ and $c'$ include 
an exponential suppression factor 
if they are generated by massive particle exchanges (in case (i)).
This is because these interactions can be generated only by
particles whose masses are of the order of the fundamental scale $M_*$
and because the SU(5)$_{\rm GUT}$-breaking sector, to which the
$Q,\bar{Q}$ belong, and quarks and leptons reside at different fixed
points.
The Kaluza--Klein modes of the U(6) vector multiplet cannot induce the
Yukawa couplings, 
since those particles that have suitable gauge charges do not have 
non-vanishing wave functions at the fixed where quarks and leptons reside.
The exponential suppression factor is 
$e^{- M_* L_4/2}$ or $e^{- M_* L_4}$,  depending 
on which $\sigma^6$-fixed point the SU(5)$_{\rm GUT}$-breaking sector 
resides. Note that we have not yet chosen one from two candidates of 
the $\sigma^6$-fixed points in subsection \ref{subsec:u2-config}. 
The former choice is preferred because of its moderate suppression factor.
The effective Yukawa coupling includes 
$e^{-M_* L_4/2}\times (v/M_*)\sim 10^{-2}$ suppression factor as a
whole.
However, there is no way to estimate the effective coupling 
between the massive particle in the extra dimensions and 
the quarks and leptons, since their origins are not known.
Thus, we see that the suppression of the order of $10^{-2}$ is 
marginal\footnote{Although figures larger that 1 are certainly required
other than the exponential suppression factor in $c$ and $c'$, the effective
coefficients $c$ and $c'$ themselves do not exceed the bound from Born
unitarity below the cut-off scale $M_*$ as long as $c,c' \lsim 4\pi$. } 
to obtain the top Yukawa coupling of order 1. 

The disunification between strange quarks and muons is obtained
through an operator
\begin{equation}
 W = \frac{c''}{M_*^3} (\bar{Q}Q)^6_{\;\; i}{\bf 10}^{ij} 
               (\bar{Q}Q)^{k}_{\;\; j}{\bf 5}^*_k.
\end{equation} 
Thus, the effective coefficient would involve 
extra $(v/M_*)^2 \sim 10^{-2}$ 
suppression with respect to the Yukawa couplings of bottom quarks 
and tau leptons.

When there are non-perturbative contributions to the Yukawa couplings
(in  case (ii)), it is impossible to determine how the effective couplings
$c$ and $c'$ are suppressed. 
Yukawa couplings may be obtained without an extra suppression factor
if they are generated by  non-perturbative effects.
But the study of such effects is beyond the scope of this paper.

%%%%
%%%%\include{u3}
%%%%
%%%%%%%%%%%%%%%%%%%%%%%%%%%%%%%%%%%%%%%%%%%%%%%%%%%%%%%%%%%%%%%%%%%%
\section{SU(5)$_{\rm GUT} \times$U(3)$_{\rm H}$ Model}
\label{sec:u3}
%%%%%%%%%%%%%%%%%%%%%%%%%%%%%%%%%%%%%%%%%%%%%%%%%%%%%%%%%%%%%%%%%%%%

\subsection{D-brane Configuration and Orbifold Projection}
\label{subsec:u3-config}

Let us now describe how the SU(5)$_{\rm GUT}$-breaking sector of the  
SU(5)$_{\rm GUT} \times$U(3)$_{\rm H}$ model is derived.
We adopt the same geometry, 
${\bf T}^6/({\bf Z}_{12}\vev{\sigma}\times {\bf Z}_2\vev{\Omega
R_{89}})$, as in the previous model.
We basically assume the same massless field contents on the D-branes 
at the beginning, i.e. U($N$) vector multiplet on $N$ 
coincident D-branes, etc. 
What is different between the construction of the two models is the  
D-brane configuration, and also the choice of unitary matrices
$\tilde{\gamma}_{\sigma^k;7}$ and $\tilde{\gamma}_{\sigma^k;3}$ that appear 
in the orbifold projection conditions.

We put seven D7-branes at the two ${\bf Z}_{12}\vev{\sigma}$-fixed loci 
$z_3=\frac{{\bf e}_8+2{\bf e}_9}{3}$ and 
$z_3=-\frac{{\bf e}_8+2{\bf e}_9}{3}$. 
The ${\bf Z}_2\vev{\Omega R_{89}}$ only identifies the fields on both
fixed loci, and the identified fields are subject only to the 
orbifold projection conditions of the ${\bf Z}_{12}\vev{\sigma}$.
Projection conditions are written in the same way as in 
Eqs.
(\ref{eq:trsf-D7-sigmaI}), (\ref{eq:trsf-D7-sigmaII}) and 
(\ref{eq:proj-D7-sigma}).
The only difference is that the vector multiplet 
$(\Sigma)^k_{\;\; l}$ ($k,l=1,...,7$) is of the U(7) gauge group 
rather than of the U(6). 
The 7 by 7 unitary matrix $\tilde{\gamma}_{\sigma;7}$ is now chosen
as
\begin{equation}
 (\tilde{\gamma}_{\sigma;7})^k_{\;\; l} = 
  \diag \left(\overbrace{e^{-\frac{1}{12}\pi i},...,e^{-\frac{1}{12}\pi i}}^5,
               e^{-\frac{9}{12}\pi i},e^{-\frac{11}{12}\pi i}\right), 
\label{eq:twist-D7-3}
%,
%             \overbrace{e^{\frac{1}{12}\pi i},...,e^{\frac{1}{12}\pi i}}^5,
%              e^{\frac{9}{12}\pi i},e^{\frac{11}{12}\pi i}\right).
\end{equation}
instead of (\ref{eq:D7-twist-u2}).

The sixth diagonal entry and seventh diagonal entry are chosen differently
from the first five entries, so that the U(7) gauge symmetry is broken down
to U(5)$\times$U(1)$_6 \times$U(1)$_7$.
The phase difference between the first five entries and the sixth 
is chosen as $e^{-2\pi i 4/12}$ for the following reasons.
Since we require that the U(6) gauge symmetry be restored at
the $\sigma^3$-projection, for a reason that is explained later, the phase
difference should be the third root of unity. The phase difference can be
$e^{+ 2\pi i 4/12}$, but the model is not essentially different from the
model with the phase difference $e^{-2\pi i4/12}$.

No matter which phase difference we adopt, there exists a
U(5)-{\bf (anti-)fund.} 
chiral multiplet that survives the orbifold projection. 
On the other hand, we observe in the next subsection
\ref{subsec:u3-anomaly} that it is hard to cancel the triangle anomalies
at each fixed point unless we take a construction such that 
the U(5)-conjugate particle also survives the orbifold projection.
This is the reason why we start from U(7) vector multiplet.
As a result, the SU(5)$_{\rm GUT} \times $U(3)$_{\rm H}$ model
inevitably includes SU(5)-({\bf 5}+{\bf 5}$^*$) pair in the 
Kaluza--Klein zero modes from D7-branes.
The phase difference between the first five entries and the seventh entry is
chosen so that U(5)-conjugate matter appears in the low-energy spectrum.
There are only two possibilities --- $e^{-2\pi i 5/12}$ and $e^{2\pi
i/12}$. Both possibilities essentially lead to the same physics.
Under the choice of the phase difference in (\ref{eq:twist-D7-3}), 
i.e. $e^{-2\pi i 5/12}$, the U(7) symmetry is not enhanced 
at the $\sigma^3$-projection. 
If the U(7) symmetry were to remain at the $\sigma^3$-fixed point on the
D7-branes, then the unwanted hypermultiplet 
$(\bar{Q}^7_{\;\; \alpha},Q^\alpha_{\;\; 7})$ would appear in the
SU(5)$_{\rm GUT}$-breaking sector. 

The Kaluza--Klein zero modes that survive the orbifold projection in 
Eq. (\ref{eq:proj-D7-sigma}) are as follows:
${\cal N}$ = 1 vector multiplets of U(5)$\times$U(1)$_6 \times$U(1)$_7$, 
where the SU(5) subgroup of the U(5) $\simeq$ SU(5)$\times$U(1)$_5$ 
is identified with the SU(5)$_{\rm GUT}$ gauge group, and 
chiral multiplets, $(\Sigma_3)^6_{\;\;i}$ ,
$(\Sigma_2)^i_{\;\; 7}$ and $(\Sigma_1)^7_{\;\; 6}$, 
which transform ({\bf 5}$^*$)$^{(-1,1,0)}$, ({\bf 5})$^{(1,0,-1)}$ and
({\bf 1})$^{(0,-1,1)}$ under the gauge group 
SU(5)$_{\rm GUT} \times$U(1)$_5 \times$U(1)$_6 \times $U(1)$_7$.
The index ``$i$'' now runs from 1 to 5, not to 7.
Anomalies of the above gauge group are discussed in the next subsection
\ref{subsec:u3-anomaly}. 
We identify $(\Sigma_3)^6_{\;\; i}$ and $(\Sigma_2)^i_{\;\; 7}$
with Higgs multiplets $\bar{H}_i({\bf 5}^*)$ 
and $H^i$({\bf 5}) in subsections \ref{subsec:u3-phen} and 
\ref{subsec:u3-discreteR}, respectively.
  
%%%%%%%%%%%%%%%%%%%%%%%%%%%%
%% putting D3-branes
Three D3-branes are put on a fixed point on the D7-branes 
where the isotropy group\footnote{
The SU(5)$_{\rm GUT} \times $U(3)$_{\rm H}$ model cannot be constructed 
by putting the D3-branes at a fixed point where the isotropy group is 
${\bf Z}_2\vev{\sigma^6}$. 
This is because thirty-six D3-branes (twelve images of three D3-branes) 
are necessary within the covering space ${\bf T}^6$ in this case.
Thirty-six D3-branes are too many to cancel their 3-brane charges by negative
charges of O3-planes.  } is ${\bf Z}_4\vev{\sigma^3}$.
There is only one such a candidate in the ${\bf T}^6/({\bf
Z}_{12}\vev{\sigma}\times {\bf Z}_2\vev{\Omega R_{89}})$ geometry.
These three D3-branes provide the U(3)$_{\rm H}$ 
gauge group in the SU(5)$_{\rm GUT}$-breaking sector. 
Eighteen D3-branes (six images of three D3-branes) are necessary 
as a whole within the covering space ${\bf T}^6$, 
yet fourteen remaining D3-branes can be used for other sectors.

Fields are restricted only under the orbifold projection by 
the isotropy group ${\bf Z}_{4}\vev{\sigma^3}$.
Thus, the U(6) symmetry, which is required to be restored in the SU(5)$_{\rm
GUT}$-breaking sector, is required to be restored on D7-branes 
under the $\sigma^3$-projection. This is the major reason why we take  
the phase difference between the first five diagonal entries and 
the sixth diagonal entry as the third root of unity in (\ref{eq:twist-D7-3}).
Fields on the D3-branes transform under the ${\bf Z}_4\vev{\sigma^3}$ as
\begin{eqnarray}
 \sigma^3 : && (X_0)^\alpha_{\;\;\beta} \mapsto
   (\widetilde{X_0})^\alpha_{\;\;\beta} \equiv  \qquad
(\tilde{\gamma}_{\sigma^3;3})^\alpha_{\;\;\alpha'}(X_0)^{\alpha'}_{\;\;\beta'}
  (\tilde{\gamma}_{\sigma^3;3}^{-1})^{\beta'}_{\;\;\beta}  ,  
                                    \label{eq:trsf-D3-XI-U3}  \\
 \sigma^3 : && (X_b)^\alpha_{\;\;\beta} \mapsto 
    (\widetilde{X_b})^\alpha_{\;\;\beta} \equiv e^{2\pi i \tilde{v}_b 3}
  (\tilde{\gamma}_{\sigma^3;3})^\alpha_{\;\;\alpha'}(X_b)^{\alpha'}_{\;\;\beta'}
  (\tilde{\gamma}_{\sigma^3;3}^{-1})^{\beta'}_{\;\;\beta}, 
                                    \label{eq:trsf-D3-XII-U3} 
\end{eqnarray}
\begin{eqnarray}
 \sigma^3 : &&Q^{\alpha}_{\;\;k} \mapsto 
     \widetilde{Q}^\alpha_{\;\;k}  \equiv e^{\pi i (v_1+v_2)3}
(\tilde{\gamma}_{\sigma^3;3})^\alpha_{\;\;\alpha'} Q^{\alpha'}_{\;\; k'}
(\tilde{\gamma}_{\sigma;7}^{-3})^{k'}_{\;\; k},  \label{eq:trsf-D3-D7-Q-U3} \\
 \sigma^3 : && \bar{Q}^{k}_{\;\;\alpha} \mapsto 
     \widetilde{\bar{Q}}^k_{\;\;\alpha} \equiv  e^{\pi i (v_1 + v_2)3}
(\tilde{\gamma}_{\sigma;7}^{3})^{k}_{\;\; k'} \bar{Q}^{k'}_{\;\; \alpha'}
(\tilde{\gamma}_{\sigma^3;3}^{-1})^{\alpha'}_{\;\;\alpha},
                                                 \label{eq:trsf-D3-D7-Qbar-U3}
\end{eqnarray}
where the $(X_a)$'s form a U(3) vector multiplet of the four-dimensional ${\cal
N}$ = 4 SUSY and the $(\bar{Q},Q)$ is a hypermultiplet of the
four-dimensional ${\cal N}$ = 2 SUSY in the ({\bf 7},{\bf 3}$^*$) 
representation of the U(7)$\times$U(3) gauge group.
The 3 by 3 unitary matrix $\tilde{\gamma}_{\sigma^3;3}$ is chosen as
\begin{equation}
 \tilde{\gamma}_{\sigma^3;3} = \diag ( 
e^{\frac{3}{4}\pi i},e^{\frac{3}{4}\pi i},e^{\frac{3}{4}\pi i}),
\label{eq:twist-D3}
\end{equation}
so that the hypermultiplets $(\bar{Q}^k_{\;\; \alpha},Q^\alpha_{\;\;
k})$ ($k=1,...,6$; $\alpha=1,2,3$) survive the following orbifold projection
conditions. 
The orbifold projection imposes Eq. (\ref{eq:proj-D3-XQQbar}).
Projected out by these conditions are 
the ${\cal N}$ = 2 hypermultiplet ($X_1,X_2$) in the U(2)-{\bf adj.}
representation and ($\bar{Q}^7_{\;\;\alpha}$,$Q^\alpha_{\;\; 7}$).
What is left is exactly the matter contents of 
the SU(5)$_{\rm GUT}$-breaking sector.
It is very encouraging that the full multiplets for 
the SU(5)$_{\rm GUT}$-breaking sector are obtained with 
the ${\cal N}$ = 2 SUSY structure.

\subsection{Anomaly Cancellation and Tadpole Cancellation}
\label{subsec:u3-anomaly}

\subsubsection{Anomaly Cancellation}
\label{subsubsec:u3-anomaly-tri}
The SU(5)$_{\rm GUT} \times$U(3)$_{\rm H}$ model discussed in this
section differs from the SU(5)$_{\rm GUT} \times$U(2)$_{\rm H}$ model in
the previous section only in the D-brane configuration and the unitary
matrices $\tilde{\gamma}_{\sigma^k;7}$ and $\tilde{\gamma}_{\sigma^k;3}$.
Therefore, the discussion in subsubsection 
\ref{subsubsec:u2-anomaly-more} exactly holds also in this model.
In particular, the pure gravitational box anomalies are cancelled
because only the supergravity multiplet in the bulk is relevant, whose
Kaluza--Klein spectrum is exactly the same as in the previous model.
Anomalies in internal dimensions are not the conditions on low energy
physics either in this model, just because of the ambiguity of
regularization at singularities.   
However, triangle anomalies, which are related to gauge fields 
on D7-branes, should be examined in this model again.

%%%%%% 

Let us discuss triangle anomalies between the unbroken gauge groups
SU(5)$_{\rm GUT} \times$U(1)$_5 \times$ U(1)$_6 \times$U(1)$_7$ and gravity.
We first discuss the [SU(5)$_{\rm GUT}$]$^3$ anomaly cancellation at
each fixed point because this type of anomaly cannot be cancelled out by
incorporating the generalized Green--Schwarz mechanism.

Formulae (\ref{eq:decomposition}) with (\ref{eq:triangle-coeff-k}) (or
equivalently, (\ref{eq:decomp-Fujikawa}) with (\ref{eq:anomaly-distr}))
are applicable also to the SU(5)$_{\rm GUT} \times$U(3)$_{\rm H}$ model
discussed in the previous subsection \ref{subsec:u3-config}. 
What is obvious from the expression (\ref{eq:anomaly-distr}) is that
no anomaly arises on a fixed-point lattice $\Gamma'_{\sigma^k}$ in which 
${}^\exists v_b k \in {\bf Z}$.
In particular, the fixed points with ${\cal N}=2$ SUSY, which are only
on one of the $\Gamma'_{\sigma^{3k}}$'s, do not have any kind of triangle
anomalies.
This is because a vector-like structure is still kept at those fixed points. 
Triangle anomalies are carried by $\sigma^k$-components with 
$k=1,2,4,5,7,8,10$ and 11.
These eight components have only two independent distribution functions.
One is the $\sigma^4$-fixed lattice,
$\Gamma'_{\sigma^4}$ $(=\Gamma'_{\sigma^8})$, which contains nine points 
within a torus ${\bf T}^4 \equiv {\bf C}^2/\Gamma'_0$.
The other is the $\Gamma'_{\sigma^{1,2,5,7,10,11}}$, all of which
consist of only one and the same fixed point, ${\bf y'}$ = {\bf 0}.
Thus, the anomaly in the $\sigma^{\pm 4}$-components
and the total anomaly that comes from the 
$\sigma^{\pm 1}$, $\sigma^{\pm 2}$ and $\sigma^{\pm 5}$-components 
should separately vanish, so that the triangle anomalies vanish at all
the fixed points.

The [SU(5)$_{\rm GUT}$]$^3$ anomaly of the 
SU(5)$_{\rm GUT} \times $U(3)$_{\rm H}$ model 
on the $\Gamma'_{\sigma^4}$ vanishes because 
\begin{equation}
\sigma^4 + \sigma^8 : 
   \frac{8}{12} \times \left(-\frac{3\sqrt{3}}{8}\right) \times 
   2 {\rm Im} (e^{2\pi i \frac{4}{3}} + e^{2\pi i \frac{5}{3}})=0.
\end{equation}
Now we do not have to make more calculations to arrive at the
conclusion that there is no [SU(5)$_{\rm GUT}$]$^3$ anomaly 
on the $\Gamma'_{\sigma^{1,2,5,7,10,11}}$ lattice (i.e. fixed point 
${\bf y}'$ = {\bf 0}) either, and hence at any fixed points.
This is because the total amount of the [SU(5)$_{\rm GUT}$]$^3$ anomaly 
integrated over the whole compact space vanishes (due to the anomaly-free
spectrum of the zero modes), and also because 
this total anomaly consists of anomalies only on the
$\Gamma'_{\sigma^{\pm 4}}$ and on ${\bf y}'={\bf 0}$.

We also show, however, the explicit calculation of the anomaly 
at the fixed point ${\bf y'}$ = {\bf 0} just as a preparation 
for discussion in the next subsubsection \ref{subsubsec:u3-anomaly-rel}. 
The [SU(5)$_{\rm GUT}$]$^3$ anomalies from
$\sigma^{\pm 1}$, $\sigma^{\pm 2}$ and $\sigma^{\pm 5}$-components are
\begin{eqnarray}
 \sigma^1 + \sigma^{11}: && 
       \frac{8}{12} \times \left(-\frac{\sqrt{3}}{8}\right)
       \times 
       2 {\rm Im}\left(e^{2\pi i \frac{4}{12}}+e^{2\pi i \frac{5}{12}}\right),
  \label{eq:sigma1-comp}\\
 \sigma^5 + \sigma^{7} : && 
       \frac{8}{12} \times \left(\frac{\sqrt{3}}{8}\right)
       \times 
  2 {\rm Im}\left(e^{2\pi i 5 \frac{4}{12}}+e^{2\pi i 5 \frac{5}{12}}\right),
   \label{eq:sigma5-comp}\\
 \sigma^2 + \sigma^{10} : && 
       \frac{8}{12} \times \left(-\frac{\sqrt{3}}{8}\right)
       \times 
  2 {\rm Im} \left(e^{2\pi i 2 \frac{4}{12}}+e^{2\pi i 2 \frac{5}{12}}\right),
   \label{eq:sigma2-comp}
\end{eqnarray}
and the sum of all these three contributions vanishes, though each
component does not.

If we were to start from the U(6) vector multiplet rather than the U(7), 
then the $\sigma^{\pm 4}$-components and the sum of $\sigma^{\pm 1}$,
$\sigma^{\pm 2}$ and $\sigma^{\pm 5}$-components are separately
non-vanishing. The former has $-3/4$ and the latter has $-1/4$ times the
anomaly of the SU(5)$_{\rm GUT}$-{\bf anti-fund.} representation.
These anomalies are distributed into four different ${\cal N}$ = 1 SUSY 
fixed points, % on the ${\bf T}^4/{\bf Z}_{12}\vev{\sigma}$, 
and the amount of anomaly at each fixed
point is a fractional number. These anomalies cannot be
cancelled even by introducing SU(5)$_{\rm GUT}$ charged particles\footnote{
In the previous paper \cite{WY-ss}, there is a mistake in this calculation.
The model described there is also valid if it is possible to replace
triangle anomalies from one fixed point to another, though.}.

Let us now discuss the anomalies of the U(1) gauge groups.
First, the U(1)$_{5+6+7}$ gauge field decouples from all the matter
contents on the D7-branes and hence there is no anomaly associated to
this gauge group. 
Second, it is easy to see, from the same calculation as in the case of 
the [SU(5)$_{\rm GUT}$]$^3$ anomaly,  
that the U(1)$_5$ gauge group does not have 
U(1)$_5$[SU(5)$_{\rm GUT}$]$^2$, U(1)$_5$[gravity]$^2$
and [U(1)$_5$]$^3$ anomalies.
Finally, the U(1)$_{6-7}$ gauge group has a number of anomalies at
various fixed points, and hence these anomalies should be cancelled by 
the generalized Green--Schwarz mechanism at all these fixed points.
Thus, the U(1)$_{6-7}$ symmetry is spontaneously broken.

\subsubsection{Relation to the Ramond--Ramond Tadpole Cancellation}
\label{subsubsec:u3-anomaly-rel}

Here, we clarify the relation between the Ramond--Ramond tadpole
cancellation in string theories and the anomaly cancellation discussed
in subsection \ref{subsec:u2-anomaly} and in subsubsection
\ref{subsubsec:u3-anomaly-tri}. 
This part is not necessary for the rest of this paper.
The following discussion is basically along the line of
Ref. \cite{RR-tadpoleII}. It is argued there that both 
conditions are {\em generically} equivalent, while sometimes
different.

The Ramond--Ramond tadpole cancellation conditions 
for the ${\bf Z}_{12}\vev{\sigma}$-orbifold is given by
\cite{IIB-orbifold,RR-tadpoleIII} 
\begin{eqnarray}
\tr(\tilde{\gamma}_{\sigma^k;7}\oplus \tilde{\gamma}_{\sigma^k;7}^{-1}) - 
 \tr{}_{{\bf y}'=0}
(\tilde{\gamma}_{\sigma^k;3}\oplus \tilde{\gamma}_{\sigma^k;3}^{-1}) = 0  
&{\rm for} & k=1,2,5,7,10,11,
\label{eq:RR-1}\\
 \tr(\tilde{\gamma}_{\sigma^k;7}\oplus \tilde{\gamma}_{\sigma^k;7}^{-1}) + 
3 \tr{}_{{\bf y}' \in {\rm Eq.}(\ref{eq:sigma4-fixed})}
(\tilde{\gamma}_{\sigma^k;3}\oplus \tilde{\gamma}_{\sigma^k;3}^{-1})=0 
& {\rm for} & k=4,8,
\label{eq:RR-4}
\end{eqnarray}
and 
\begin{eqnarray}
 \tr(\gamma_{\sigma^k;7}) + 
2 \tr{}_{{\bf y}'\in {\rm Eq.}(\ref{eq:sigma3-fixed})}(\gamma_{\sigma^k;3}) 
= 0 & {\rm for} & k=3,9, 
\label{eq:RR-3}\\
 \tr(\gamma_{\sigma^k;7}) + 
4 \tr{}_{{\bf y}'\in {\rm Eq.}(\ref{eq:sigma6-fixed})}(\gamma_{\sigma^k;3}) 
= 0 & {\rm for} & k=6,
\label{eq:RR-6}
\end{eqnarray}
with $\tr(\gamma_{\sigma^0;7})=32$ and $\tr (\gamma_{\sigma^0;3})=32$.
Here, the $\gamma_{\sigma^k;7}$ denotes a 32 by 32 unitary matrix 
$(\gamma_{\sigma;D7})^k$ obtained from (\ref{eq:32forD7}) 
and the 
$\gamma_{\sigma^k;3}|_{{\bf y}'\in {\rm Eq.}(\ref{eq:sigma3-fixed})}$ or 
$\gamma_{\sigma^k;3}|_{{\bf y}'\in {\rm Eq.}(\ref{eq:sigma6-fixed})}$ 
a diagonal block of $(\gamma_{\sigma;D3})^k$
obtained from (\ref{eq:32forD3}) that corresponds to D3-branes at
six-dimensional singularities whose {\bf y}$'$ coordinates 
are one of Eq. (\ref{eq:sigma3-fixed}) or Eq. (\ref{eq:sigma6-fixed}).
These matrices are replaced by corresponding ones when the 
SU(5)$_{\rm GUT} \times$U(3)$_{\rm H}$ model is considered. 
Equations (\ref{eq:RR-4}), (\ref{eq:RR-3}) and (\ref{eq:RR-6}) are imposed 
at each of their fixed points separately, e.g. Eq. (\ref{eq:RR-4}) for $k=4$
contains nine equations for nine fixed points given 
in Eq. (\ref{eq:sigma4-fixed}).
There are more Ramond--Ramond tadpole cancellation conditions, 
which, however, do not restrict the projection on the D7-branes and 
D3-branes in our model.
Thus, we do not list them here.

Let us first show that Eqs. (\ref{eq:RR-1}) and (\ref{eq:RR-4}) are derived
{\em generically} from the cancellation of non-Abelian triangle
anomalies on the D7-branes. 
Let us take the $\tilde{\gamma}_{\sigma;7}$ as 
\begin{equation}
 \tilde{\gamma}_{\sigma;7}= \diag (
e^{2\pi i \frac{1}{24}} {\bf 1}_{n_1 \times n_1}, \ldots, 
e^{2\pi i \frac{2j-1}{24}} {\bf 1}_{n_j \times n_j}, \ldots,
e^{2\pi i \frac{23}{24}} {\bf 1}_{n_{12} \times n_{12}}).
\end{equation}
This is a generalization of (\ref{eq:D7-twist-u2}) and
(\ref{eq:twist-D7-3}).
Then, the gauge group from these D7-branes is $\prod_{j=1}^{12}U(n_j)$.
Now the $\sigma^k$-component of the [SU($n_j$)]$^3$ anomaly is
proportional to 
\begin{eqnarray}
 &&\frac{4}{12} \left(\prod_{b=1}^{3} \sin (\pi v_b k)\right)
 (-i \sum_I A_I \rho_I(\sigma^k))  \nonumber \\
&=& \frac{4}{12} \left(\prod_{b=1}^{3} \sin (\pi v_b k)\right) (-i)
\left(e^{2\pi i \frac{k}{24}(2j-1)} \tr (\tilde{\gamma}_{\sigma^k;7}^{*})
    - e^{-2\pi i \frac{k}{24}(2j-1)} \tr (\tilde{\gamma}_{\sigma^k;7})\right) 
                              \nonumber \\
&=& \frac{4}{12} \left(\prod_{b=1}^{3} \sin (\pi v_b k)\right) (-2){\rm Im} 
  \left( e^{-2\pi i \frac{k}{24}(2j-1)} 
   \tr (\tilde{\gamma}_{\sigma^k;7})\right).
\label{eq:each-comp}
\end{eqnarray}
When it is required that all the $\sigma^k$-components
separately\footnote{
The Ref. \cite{RR-tadpoleII} does not require that each component
of triangle anomalies separately vanish. 
Requiring only vanishing total triangle anomalies integrated over the
orbifold is sufficient to derive the Eq. (\ref{eq:my-tadpole}) 
as far as the triangle anomalies are concerned.} vanish
for all [SU($n_j$)]$^3$ anomalies ($j=1,...,12$), then
\begin{equation}
 \tr (\tilde{\gamma}_{\sigma^k;7}) = 0 \qquad {\rm for~} k=1,2,5,7,10,11,4,8
\label{eq:my-tadpole}
\end{equation}
follows. 
No condition follows from the triangle anomaly cancellation 
for $k=0,3,6,9$, because $\sin(\pi v_3 k)=0$.
Equations (\ref{eq:my-tadpole}) are the same conditions 
as Eqs. (\ref{eq:RR-1}) and (\ref{eq:RR-4}) 
in the absence of the D3-branes at those fixed points. 
Notice that it is the case in our construction, since D3-branes are put
only at $\Gamma'_{\sigma^6}$ in the SU(5)$_{\rm GUT} \times$U(2)$_{\rm
H}$ model or at $\Gamma'_{\sigma^{\pm 3}}$ in the SU(5)$_{\rm GUT}
\times$U(3)$_{\rm H}$ model, not at $\Gamma'_{\sigma^{\pm 1,2,5}}$ or 
$\Gamma'_{\sigma^{\pm 4}}$ .

The triangle anomaly cancellation 
(for the SU(5)$_{\rm GUT} \times$U(3)$_{\rm H}$ model 
discussed in this subsection) does not require
that each $\sigma^k$-component (\ref{eq:each-comp}) separately vanishes,
but rather, it is sufficient to require vanishing sum of components that
have the same distribution function. 
In particular, the sum of
(\ref{eq:sigma1-comp}), (\ref{eq:sigma5-comp}) and (\ref{eq:sigma2-comp})
vanishes, but not separately.
This is one of the differences between the triangle anomaly cancellation 
and the Ramond--Ramond tadpole cancellation. 
The other difference is that we do not have to impose a non-Abelian
triangle anomaly cancellation when $n_j < 3$. 
In particular, the [SU($n_{12}=5$)]$^3$ anomaly cancellation is imposed, 
while no other ``[SU($n_{j \neq 12}$)]$^3$ anomaly'' does not have to be
cancelled, since $n_{7,8}=1$ and all other $n_j$'s are 0.

The origin of quarks and leptons is not identified, but they reside on
the ${\bf y}'={\bf 0}$ fixed point, which is exactly the
$\Gamma'_{\sigma^1}=\Gamma'_{\sigma^5}=\Gamma'_{\sigma^2}$.
Therefore, they can also give certain contributions\footnote{
The authors thank M.~Cvetic for useful discussion.} to each of 
(\ref{eq:sigma1-comp}), (\ref{eq:sigma5-comp}) and
(\ref{eq:sigma2-comp}), although we cannot calculate each contribution 
in terms of $\tr{}_{{\bf y}'=0} (\tilde{\gamma}_{\sigma^k;3})$.
In particular, there is a possibility that each
$\sigma^k$-component vanishes separately owing to contributions from
particles whose origins are not well-specified yet.

The Ramond--Ramond tadpole cancellation conditions for $k=0,3,6,9$, 
Eqs. (\ref{eq:RR-3}) and (\ref{eq:RR-6}) are not obtained from the triangle
anomaly cancellation \cite{RR-tadpoleII}. 
However, Eq. (\ref{eq:RR-3})\footnote{
Equation (\ref{eq:RR-6}) is trivially satisfied as long as one takes 
the $\gamma_{\sigma;D7}$ and $\gamma_{\sigma;D3}$ as in
(\ref{eq:32forD7}) and (\ref{eq:32forD3}).} is a
condition for pure gauge box anomaly cancellation, 
assuming the massive spectrum in the Type IIB string theory.
It will be easily guessed from discussion 
in subsection \ref{subsec:u2-anomaly}.
Since the infinite towers of string excitations winding in the 
$z_3$-direction on D7-branes and D3-branes behave 
as Kaluza--Klein towers through T-duality, gauge theories on those branes
effectively  extend in ten-dimensional and six-dimensional space-time,
respectively. 
Thus, one has to consider the cancellation of 
(the irreducible part of) pure gauge box anomalies. 
Equation (\ref{eq:RR-3}) ensures that we can cancel that anomaly 
using the anomaly inflow to the singularities.   
In the above situation, where the winding modes in the $z_3$-direction play
an important role, all the fields on D7-branes and D3-branes, whatever
the $z_3$-coordinates are, contribute 
to the same box anomalies on the six-dimensional singularities;
$z_3$-coordinates of initial points and end points of strings are
no longer important in the sense of local field theories 
when they can wind around the torus of the $z_3$-direction. 
Therefore, all thirty-two D7-branes and all the D3-branes 
in a given six-dimensional singularity contribute 
to Eq. (\ref{eq:RR-3}).

However, this condition depends highly on the spectrum above the cut-off
scale (including the existence of the winding modes), and on the
configuration of D7- and D3-branes that are away from the fixed loci at 
$z_3=\pm ({\bf e}_8 + 2{\bf e}_9)/3$. 
This is the reason why we do not take this condition,
Eq. (\ref{eq:RR-3}), seriously in our generic study based on supergravity. 
The same thing is expressed in another way also in subsection
\ref{subsubsec:u2-anomaly-more}.

\subsection{Discrete R Symmetry}
\label{subsec:u3-discreteR}

The matter contents obtained from D-branes are the whole SU(5)$_{\rm
GUT}$-breaking sector and three chiral multiplets 
$(\Sigma_1)^7_{\;\; 6}$, $(\Sigma_2)^i_{\;\; 7}$ and 
$(\Sigma_3)^6_{\;\; i}$.
We regard the rotational symmetry $z_b \mapsto e^{i\alpha_b} z_b$ 
of ${\bf C}^3$ with $\alpha_1 = - \alpha_2 = \alpha_3$ 
as the principal origin of the (mod 4)-R symmetry. 
Note that the (mod 4)-subgroup is preserved by the geometry, 
since it is generated by the rotation of three complex planes 
by an angle $\pi$.
The zero modes have the following R charges under this rotation:
2 for $(X_3)^\alpha_{\;\;\beta}$, 
0 for $\bar{Q}^k_{\;\; \alpha}$ and $Q^\alpha_{\;\; k}$,
2 for $(\Sigma_1)^7_{\;\; 6}$, $-2$ for $(\Sigma_2)^i_{\;\; 7}$ and 
2 for $(\Sigma_3)^6_{\;\; i}$.

As shown in the next subsection \ref{subsec:u3-phen}, it is reasonable to
identify the chiral multiplet $(\Sigma_3)^6_{\;\; i}$ with one of the Higgs
multiplets $\bar{H}_i({\bf 5}^*)$. 
Then, the R charge of the $(\Sigma_3)^6_{\;\; i}$ should be 0 mod 4,
while the charge obtained from the rotation is 2. Thus, we consider
that the (mod 4)-R symmetry is a suitable linear combination 
of the rotational symmetry and anomaly-free U(1)$_{6+7}$
symmetry\footnote{The (mod 4)-R symmetry can be a linear combination of
the U(1)$_{\rm H}$ and the U(1)$_5$ symmetry in addition to the
geometric rotation and the U(1)$_{6+7}$. 
We do not exclude this possibility. The choice made in the text is just 
to simplify the description.}, 
so that the $(\Sigma_3)^6_{\;\; i}$ has R charge 0. Then, it follows
that $(\Sigma_2)^i_{\;\; 7}$ also has R charge 0. 
The $(\Sigma_2)^i_{\;\; 7}$ has the same SU(5)$_{\rm
GUT}$ charge and the same R charge as the Higgs $H^i({\bf 5})$.
Therefore, we identify the $(\Sigma_2)^i_{\;\; 7}$ with the Higgs
multiplet $H^i({\bf 5})$.
The R charges (mod 4) of all the zero modes are now obtained exactly as
in Table \ref{tab:u3-R4}, including those of the 
$\bar{Q}^6_{\;\; \alpha}$ and $Q^\alpha_{\;\; 6}$.
We also note here that the SU(5)$_{\rm GUT}$-singlet 
$(\Sigma_1)^7_{\;\; 6}$ has R charge 2.

\subsection{Toward a Realistic Model}
\label{subsec:u3-phen}

We have obtained the SU(5)$_{\rm GUT} \times $U(1)$_5
\times$U(1)$_{6+7}$ vector multiplet and three chiral multiplets 
$S \equiv (\Sigma_1)^7_{\;\; 6}$, 
$\bar{H}_i({\bf 5}^*) \equiv (\Sigma_3)^6_{\;\; i}$ and 
$H^i({\bf 5}) \equiv (\Sigma_2)^i_{\;\; 7}$ from the D7-branes.
The SU(5)$_{\rm GUT}$-breaking sector is exactly obtained on the
D3-branes.
Interactions determined by extended SUSY provide tree-level interactions
of these (Kaluza--Klein zero mode) fields. 
Some of them are written in the superpotential as:
\begin{eqnarray}
 W &=& \sqrt{2}g_{\rm H} 
  \bar{Q}^i_{\;\;\alpha} (X_3)^{\alpha}_{\;\;\beta} Q^\alpha_{\;\; i} 
   +\sqrt{2}g_{\rm H} 
  \bar{Q}^6_{\;\;\alpha} (X_3)^{\alpha}_{\;\;\beta} Q^\alpha_{\;\; 6} 
                                     \label{eq:N=2-gauge-int} \\
  && + \sqrt{2}g_{\rm GUT} 
   Q^\alpha_{\;\; 6} (\Sigma_3)^6_{\;\; i} \bar{Q}^i_{\;\; \alpha}
   + \sqrt{2}g_{\rm GUT}
   (\Sigma_1)^7_{\;\; 6} (\Sigma_3)^6_{\;\; i} (\Sigma_2)^i_{\;\; 7}.
\end{eqnarray}
%The first line comes from the U(3) gauge interactions and the second
%line from the U(7) gauge interactions.  
The first line is the ${\cal N}$ = 2 SUSY interaction in (\ref{eq:super}), 
whose natural explanation is one of the main purposes 
of our higher-dimensional construction. 
We identify the $(\Sigma_3)^6_{\;\; i}$ as one of the Higgs multiplets 
$\bar{H}_i({\bf 5}^*)$, because the first term in the second line
gives the first term of the fourth line of the superpotential
(\ref{eq:super}). 
The last term implies that there exists a trilinear term
\begin{equation}
 W= \sqrt{2} g_{\rm GUT} S \bar{H}_i H^i.
\label{eq:trilinear}
\end{equation}

All particle contents have been obtained, except for three families of
quarks and leptons, {\bf 5}$^*$+{\bf 10}.
They are introduced at the fixed point ${\bf y}'={\bf 0}$, just as 
in the SU(5)$_{\rm GUT} \times$U(2)$_{\rm H}$ model. 
Only one linear combination of the U(1)$_5$ and the U(1)$_{6+7}$ gauge
groups is expected to be free from mixed anomaly 
with the SU(5)$_{\rm GUT}$ gauge group 
in the presence of quarks and leptons. 
It should be the U(1)$_{\rm fiveness}$.
The other candidate of the anomaly-free gauge symmetry is the (mod 4)-R 
symmetry discussed in the previous subsection \ref{subsec:u3-discreteR}.
This symmetry is a linear combination of the rotational symmetry of the
extra dimensions and U(1)$_{6+7}$.
It was discovered in Ref. \cite{KMY-da} that this symmetry has vanishing
mixed anomalies, not only with the SU(3)$_{\rm H}$ gauge group but also
with the SU(5)$_{\rm GUT}$ gauge group, provided there is an extra pair
of SU(5)$_{\rm GUT}$-({\bf 5}+{\bf 5}$^*$) chiral multiplets. 
In the presence of these extra particles, this anomaly-free discrete gauge
R symmetry can be kept unbroken at low energies, until the
vacuum condensation of the superpotential breaks it.
We consider that other linear combinations are anomalous, and that their 
anomalies will be cancelled by the generalized Green--Schwrz mechanism or
rather simply spontaneously broken.
Thus, those symmetries are not preserved at low energies.
In the absence of the extra pair of SU(5)$_{\rm GUT}$-({\bf 5}+{\bf
5}$^*$), the (mod 4)-R symmetry is also anomalous, whose anomaly is
also cancelled by the generalized Green--Schwarz mechanism.
%However, it is impossible to consider why this particular 
%linear combination becomes anomaly free, 
%since this question ((is/may be)) deeply related to the
%dynamics that generates quarks, leptons and the extra {\bf 5}+{\bf 5}$^*$.

Yukawa couplings of quarks and leptons and a coloured Higgs mass term
$W = h \bar{Q}^6_{\;\; \alpha}Q^{\alpha}_{\;\; i}H^i$ are expected to be
generated through non-perturbative effects.
We cannot estimate the Yukawa couplings since we do not know the
dynamics that generates these couplings.
We expect all terms allowed by symmetries, namely 
the SU(5)$_{\rm GUT} \times$U(3)$_{\rm H}$
gauge symmetry, the (mod 4)-R symmetry, and U(1)$_{\rm fiveness}$ (which is
assumed to be spontaneously broken at some intermediate scale), 
are generated dynamically.

The symmetries listed in the previous paragraph allows a superpotential
\begin{equation}
 W= \lambda S^3 + m^2 S.
\end{equation}
However, the order of magnitude of $m$ does not allow any expectation
since it highly depends on UV the cut-off. 

Although both Higgs multiplets $\bar{H}_i({\bf 5}^*)$ and $H^i({\bf 5})$
originate from higher-dimensional polarizations of the U(7) gauge fields,
the U(7) gauge transformation, which causes inhomogeneous shifts 
also to the higher-dimensional polarizations, does not prevent
the Yukawa couplings from being generated. 
Both Higgs multiplets do not transform inhomogeneously under the U(7)
gauge transformation, since they are zero modes, although Kaluza--Klein
excitations do.
Therefore, the Yukawa couplings can be generated and can be finite in
the effective action below the compactification scale, where the
spontaneous breaking of higher-dimensional Lorentz symmetry 
is already taken into account.

\section{Summary and Phenomenological Consequences}
\label{sec:fine}

The product-group unification constructed in 
four-dimensional space-time has been proposed to solve the
doublet-triplet mass splitting problem in SUSY GUT's, which 
has a number of interesting features. 
Models use product group as a ``unified gauge group'' 
with strong gauge coupling constants for extra gauge groups. 
The ${\cal N}$ = 2 SUSY is necessary to maintain the strong coupling, and
the structures of the SU(5)$_{\rm GUT}$-breaking sectors of these models
accommodate the ${\cal N}$ = 2 SUSY. The cut-off scale of the models
should lie somewhat lower than the Planck scale. Finally, the symmetry
principle of these models, the (mod 4) R symmetry, can be a discrete
gauge symmetry, shedding some light on the required $10^{-14}$ precision
to keep light Higgs doublets at the TeV scale.

All these features can be naturally explained when these models are
embedded into an extra-dimensional space with extended SUSY, where the
SU(5)$_{\rm GUT}$-breaking sector is expected to be localized on a 
point in the extra dimensions.

We have considered in this paper that the above localization mechanism of 
the SUSY gauge theories is realized on solitonic solutions of the
ten-dimensional Type IIB supergravity. 
Although the localization of a particular SUSY gauge
theory predicted by the Type IIB string theory is not perfectly proved
within the Type IIB supergravity, 
we assume that the same massless contents are realized on the D3--D7
system as in the Type IIB string theory. 
The D3--D7 system preserves ${\cal N}$ = 2 SUSY before the orbifold
projection condition is imposed, which is necessary in the models.

We have pursued the basic idea that the SU(5)$_{\rm GUT}$-breaking sector
with ${\cal N}$ = 2 SUSY is realized on the D3--D7 system.
We have shown that the whole of the sector is obtained 
from D-brane fluctuations together with the ${\cal N}$ = 2 SUSY, 
while the whole system has only ${\cal N}$ = 1 SUSY;
${\bf T}^6/({\bf Z}_{12}\vev{\sigma}\times {\bf Z}_2\vev{\Omega
R_{89}})$ is adopted as the compactified manifold.
Quarks and leptons are assumed to reside on one of the fixed points, since
they are not obtained from D-brane fluctuations.
Anomalies are suitably cancelled within the framework of field theories.
The R charges are suitably obtained for particles that are
identified with the D-brane fluctuations.

We finally summarize a couple of phenomenological consequences 
of these models. 
The first issue is the proton decay. The analysis of
Refs. \cite{FW-ss,FW-ss2} has made two assumptions.
One is the approximate ${\cal N}$ = 2 SUSY relations in 
Eqs. (\ref{eq:N=2relation}) and (\ref{eq:N=2relation2}), and 
the other is the absence of a tree-level contribution that involves 
SU(5)$_{\rm GUT}$-breaking VEV such as the second term
in (\ref{eq:tree-SU(5)breaking}). 
Both assumptions are justified in our construction, 
because the ${\cal N}$ = 2 SUSY is a symmetry at short distances, 
and the second term in (\ref{eq:tree-SU(5)breaking}) 
has an extra suppression factor of $10^{-2}$ relative to the first term 
in (\ref{eq:tree-SU(5)breaking}) because the first term comes from the
whole D7-branes, while the second term is only on D3-branes.

Threshold corrections to the MSSM gauge coupling constants 
arising from particles in the SU(5)$_{\rm GUT}$-breaking sector almost cancel
each other in both the SU(5)$_{\rm GUT} \times$U(2)$_{\rm H}$ 
and the SU(5)$_{\rm GUT} \times$U(3)$_{\rm H}$ models.
This is because of the approximate ${\cal N}$ = 2 SUSY relation.
Cancellation of the threshold corrections enables one to estimate the
GUT gauge boson mass, leading to a prediction of the lifetime of the proton.
Typically $\tau (p \rightarrow e^+ \pi^0) \simeq (3 - 10)\times 10^{34}$
yrs is the prediction common to both models \cite{FW-ss,FW-ss2}\footnote{
The analysis in \cite{FW-ss,FW-ss2} is based on models in
four-dimensional space-time. Although the higher-dimensional effects would
not change the prediction very much, a detailed analysis of their effects 
will be given elsewhere \cite{FIW-ss}. }, 
which is a fairly short lifetime compared with the typical prediction 
for ordinary grand unified theories, $\tau \simeq 10^{36}$yrs. 
Although there is an SU(5) unification model \cite{HN,HN2} 
that also predicts a short lifetime of the proton 
(typically $\tau \sim 10^{34}$ yrs), their model and the present models
can be distinguished experimentally because 
all the decay modes $p \rightarrow e^+\pi^0,\mu^+ \pi^0,e^+ K^0,\mu^+
K^0,\pi^+\bar{\nu},K^+ \bar{\nu}$ can have sizeable branching ratios 
in \cite{HN2}, while the standard decay mode 
$p \rightarrow e^+ \pi^0$ is the dominant one in the models we discuss
in this paper.

The second issue is the gaugino mass. 
This mass does not necessarily satisfy the SU(5) GUT relation \cite{AM}, 
since there are contributions from masses of 
SU(2)$_{\rm H} \times$U(1)$_{\rm H}$ (or
SU(3)$_{\rm H} \times$ U(1)$_{\rm H}$) gauginos.
We cannot determine the gaugino masses without fixing how the SUSY
breaking is mediated, however.
Contact interaction between the U(2)$_{\rm H}$ (U(3)$_{\rm H}$) vector
multiplet and chiral multiplets carrying the SUSY-breaking F-term VEV 
is, in general, forbidden by local ${\cal N}$ = 2 SUSY, and then gaugino masses
only come from the SU(5)$_{\rm GUT}$, %($\times$ various U(1)'s), 
and the SU(5) GUT relation is almost satisfied. 
However, such an ${\cal N}$ = 2 SUSY-violating interaction can
be generated in an effective action below the Kaluza--Klein scale, and
hence there is no definite prediction.

The third issue is the discrete gauge R symmetry. 
Now the (mod 4)-R symmetry is a gauged symmetry. 
Although it has vanishing mixed anomaly with SU(2)$_{\rm H}$ or
SU(3)$_{\rm H}$ gauge group, the mixed anomaly 
(mod 4 R)[SU(5)$_{\rm GUT}$]$^2$ does not vanish.
This anomaly might be cancelled through the generalized Green--Schwarz
mechanism, or otherwise, new SU(5)$_{\rm GUT}$ charged particles are
required. Those particles do not have masses without SUSY breaking,
which breaks the (mod 4)-R symmetry down to R parity, and hence they are
expected, if they exist, around the TeV scale \cite{KMY-da}.

Finally, there are possibilities that gauge-singlet particles exist
(moduli) with masses of the order of the TeV scale. 
The Kaluza--Klein zero modes that survive the orbifold projection 
become moduli fields, unless they have mass partners whose R charges are 2.
Those particles may have interesting implications in the thermal history 
of the Universe.
Another possibility is the gauge singlet $S \equiv (\Sigma_1)^7_{\;\; 6}$
in the SU(5)$_{\rm GUT} \times$U(3)$_{\rm H}$ model,  
which is characterized by its trilinear coupling with the two Higgs
doublets in (\ref{eq:trilinear}). 
This particle remains in the low-energy spectrum as long as
there is no mass partner having R charge 0.
If the tadpole term is not generated for the gauge singlet field $S$,
then the model becomes the so-called next-to-minimal SSM \cite{NMSSM}.

%%%%%%%%%%%%%%%%%%%%%%%%%%%%%%%%%%%%%%%%%%%%%%%%%%%%%%%%%%%%%%%%%%%%%%
\section*{Acknowledgement}
%%%%%%%%%%%%%%%%%%%%%%%%%%%%%%%%%%%%%%%%%%%%%%%%%%%%%%%%%%%%%%%%%%%%%%
%
This work was supported by the Japan Society for the Promotion of
Science (T.W.) and Grant-in-Aid for Scientific Research (S) 14102004 (T.Y.).
%
%%%%%%%%%%%%%%%%%%%%%%%%%%%%%%%%%%%%%%%%%%%%%%%%%%%%%%%%%%%%%%%%%%%%%
%%%%%%%%%%%%%%%%%%%%%%%%%%%%%%%%%%%%%%%%%%%%%%%%%%%%%%%%%%%%%%%%%%%%%

%

\begin{table}[ht]
\begin{center}
\begin{tabular}{c|cccc}
Fields            & $Q,\bar{U},\bar{E}$ & $\bar{D},L$ & $H_u$ & $\bar{H}_d$ \\
\hline 
R charges (mod 4) &   $ 1 - n/5 $       & $1+3n/5 $ & $0+2n/5$ & $0-2n/5$  \\
\end{tabular}
\caption{R charges (mod 4) of the fields in the MSSM are given
 here. $n$ is an arbitrary integer.}
\label{tab:mssm-R4}
\end{center}
\end{table}
\begin{table}[ht]
\begin{center}
\begin{tabular}{c|cccccccc}
Fields    & ${\bf 10}^{ij}$ & ${\bf 5}^*_i$ 
          & $H({\bf 5})^i$ & $\bar{H}({\bf 5}^*)_i $ 
          & $X$ & $Q_i,\bar{Q}^i$ & $Q_6$ & $\bar{Q}^6$ \\
\hline 
R charges & 1 & 1 & 0 & 0 & 2 & 0 & 2 & $-2$  \\
\end{tabular}
\caption{R charges (mod 4) of the fields in the 
SU(5)$_{\rm GUT}\times$U(3)$_{\rm H}$ model are given here. }
\label{tab:u3-R4}
\end{center}
\end{table}
\begin{table}[ht]
\begin{center}
\begin{tabular}{c|ccccc}
Fields    & ${\bf 10}^{ij}$ & ${\bf 5}^*_i$ 
          & $X$ & $Q_i,\bar{Q}^i$ & $Q_6$,$\bar{Q}^6$ \\
\hline 
R charges & 1 & 1 & 2 & 0 & 0  \\
\end{tabular}
\caption{R charges (mod 4) of the fields in the 
SU(5)$_{\rm GUT}\times$U(2)$_{\rm H}$ model are given here. }
\label{tab:u2-R4}
\end{center}
\end{table}

\end{document}